\documentclass[apj]{emulateapj} 
\usepackage{graphicx,color}
\usepackage[usenames,dvipsnames]{xcolor} 
\usepackage{amssymb,amsmath,mathtools,mathrsfs,bm} 
\usepackage{epsfig,subfigure,placeins,float} 
\usepackage{booktabs,longtable,ctable,multirow} 
\usepackage{exscale,relsize} 
\usepackage[normalem]{ulem} 
\usepackage{enumerate}

\newcommand{\corr}{}

\newcommand{\StageOne}{Stage I}
\newcommand{\StageTwo}{Stage II}
\newcommand{\StageThree}{Facility}

\newcommand{\be}{\begin{equation}}
\newcommand{\ee}{\end{equation}}
\newcommand{\bea}{\begin{eqnarray}}
\newcommand{\eea}{\end{eqnarray}}

\usepackage[stable]{footmisc}

\begin{document}

\setlength{\skip\footins}{1.5em}

\title{Late-time cosmology with 21cm intensity mapping experiments}

\author{Philip Bull}
\email{p.j.bull@astro.uio.no}
\affil{Institute of Theoretical Astrophysics, University of Oslo, P.O. Box 1029 Blindern, N-0315 Oslo, Norway}

\author{Pedro G. Ferreira}
\affil{Astrophysics, University of Oxford, DWB, Keble Road, Oxford OX1 3RH, UK}

\author{Prina Patel} 
\affil{Department of Physics, University of Western Cape, Cape Town 7535, South Africa}
\affil{SKA SA, 3rd Floor, The Park, Park Road, Pinelands, 7405, South Africa}
\affil{Astrophysics, Cosmology Gravity Centre, and Department of Mathematics and Applied Mathematics, University of Cape Town, Cape Town, 7701, South Africa.}

\author{M\'{a}rio G. Santos} 
\affil{Department of Physics, University of Western Cape, Cape Town 7535, South Africa}
\affil{SKA SA, 3rd Floor, The Park, Park Road, Pinelands, 7405, South Africa}
\affil{CENTRA, Instituto Superior T\'{e}cnico, Universidade de Lisboa, Lisboa 1049-001, Portugal}


\begin{abstract}
We present a framework for forecasting cosmological constraints from future neutral hydrogen intensity mapping experiments at low to intermediate redshifts. In the process, we establish a simple way of comparing such surveys with optical galaxy redshift surveys. We explore a wide range of experimental configurations and assess how well a number of cosmological observables (the expansion rate, growth rate, and angular diameter distance) and parameters (the densities of dark energy and dark matter, spatial curvature, the dark energy equation of state, etc.) will be measured by an extensive roster of upcoming experiments. A number of potential contaminants and systematic effects are also studied in detail. The overall picture is encouraging -- {\corr if autocorrelation calibration can be controlled to a sufficient level,} Phase I of the SKA should be able to constrain the dark energy equation of state about as well as a DETF Stage IV galaxy redshift survey like Euclid, in roughly the same timeframe.
\end{abstract}




\section{Introduction}

As the drive towards ever-greater cosmological precision continues, it becomes necessary to survey progressively larger volumes of the Universe in order to stay ahead of the fundamental limits on measurement accuracy set by cosmic variance. In principle, the best we can ever do is to map out the structure of the whole of the observable Universe, and on large scales at least, this possibility may soon be within reach. The tracer of choice is likely to be neutral hydrogen (HI), which pervades space from the time of recombination up to the present day. HI is thought to be a (biased) tracer of the underlying dark matter distribution, and has a characteristic line emission at around 21cm -- well into the radio -- that is mostly immune to obscuration by intervening matter. The redshifting of this line additionally gives a measure of cosmic distance, making it possible to reconstruct the three-dimensional matter density field over a wide range of redshifts and scales.

At late times the Universe has reionised, and so the bulk of the neutral hydrogen is thought to reside in comparatively dense gas clouds (damped Lyman-alpha systems) embedded in galaxies, where it is shielded from ionising UV photons. HI is therefore essentially a tracer of the galaxy distribution. Detecting sufficient numbers of HI-emitting galaxies to do precision cosmology would be a mammoth task, but fortunately this is not necessary; one can instead simply measure the total HI intensity over comparatively large angular scales, without needing to resolve individual galaxies. The result is a map of large-scale fluctuations in 21cm intensity, similar to a CMB map, except now the signal is also a function of redshift. Combined with the high frequency (and thus redshift) resolution of modern radio telescopes, this {\it intensity mapping} (IM) methodology makes it possible to efficiently survey extremely large volumes \citep{Battye:2004re, 2006ApJ...653..815M, Chang:2007xk,2008PhRvD..78b3529M, 2008PhRvL.100p1301L, 2008PhRvD..78j3511P, 2008MNRAS.383..606W, 2008MNRAS.383.1195W, 2009astro2010S.234P,Bagla:2009jy,Seo:2009fq,2011ApJ...741...70L,2012A&A...540A.129A, 2013MNRAS.434.1239B}.

As with the CMB, the 21cm signal is contaminated by a host of foreground emission sources, such as our own galaxy and extragalactic point sources, that are orders of magnitude stronger. The hope is that the spectra of the foreground sources are sufficiently smooth that, with a clever cleaning algorithm, it should be possible to suppress them to such a level that the cosmological signal can be extracted in an unbiased way \citep{2003MNRAS.346..871O, 2005ApJ...624L..65B, 2005ApJ...625..575S,  2006ApJ...648..767M, 2006ApJ...650..529W, 2008MNRAS.391..383G, 2008MNRAS.389.1319J, 2009MNRAS.398..401L, 2011MNRAS.413.2103P, 2011PhRvD..83j3006L, 2012ApJ...756..165P, 2012ApJ...752..137M, 2014arXiv1404.2596L}. First attempts at using intensity mapping have been promising, but have highlighted the challenge of calibration and foreground subtraction. The Effelsberg-Bonn survey \citep{2011AN....332..637K} has produced a data cube covering redshifts out to $z=0.07$, while the Green Bank Telescope (GBT) has produced the first (tentative) detection of the cosmological signal through IM by cross-correlating with the WiggleZ redshift survey \citep{Chang:2010jp,Switzer:2013ewa,2013ApJ...763L..20M}. As probes to constrain cosmological parameters these measurements are, as yet, ineffective, but they do point the way to a promising future.

The purpose of this paper is two-fold. First of all, we develop a self-consistent forecasting formalism, rooted in the mapping of two-dimensional diffuse emission, but which can easily be compared to (and even interpreted as) 3D redshift surveys of discrete sources. It is approximate -- using the ``flat-sky'' approximation, and slicing up the full dataset into approximately uncorrelated redshift bins -- but the formalism is remarkably effective in forecasting constraints for a diverse portfolio of cosmological parameters. We can also use it to discuss the impact of different experimental configurations, as well as the effectiveness of foreground subtraction on our results.

With this formalism in hand, we then use it to explore the observational campaigns that are planned up to, and including, Phase I of the Square Kilometre Array (SKA), which is due to see first light around 2020. One of the key results of this work is the prediction that cosmological constraints from forthcoming 21cm IM surveys will be able to compete with, and perhaps even surpass, those from traditional probes of large scale structure within the decade, even when future high-precision experiments such as Euclid and LSST are taken into account. {\corr This finding relies on being able to use large dish arrays like the SKA in an autocorrelation mode, rather than as a (more standard) interferometer -- a requirement that brings with it a number of calibration and data analysis challenges that have yet to be solved.}

The paper is structured as follows. We first present a mathematical model for the IM signal, and use it to construct the Fisher Matrix for a general set of cosmological parameters. In doing so, we discuss the approximations that we are making in modelling the cosmological signal, experimental set-up, and foreground subtraction. We then discuss the structure of our formalism, comparing it to what one obtains when forecasting for redshift surveys. The concept of ``effective volume'' becomes a useful way of discussing the strengths and weaknesses of IM. We then make a first pass at forecasting for cosmological parameters, focusing on the detectability of the Baryon Acoustic Oscillation (BAO) feature and the information that can be gleaned from it, including the Hubble rate and angular diameter distance, and the growth rate from redshift space distortions. Next, we turn our attention to forecasting for the canonical set of cosmological parameters, including various fractional energy densities and the equation of state of dark energy. The potential to constrain theories of modified gravity is also examined, and the importance of bias and the neutral hydrogen fraction is assessed. We then discuss the importance of survey design and foreground subtraction, and establish a set of goals that subtraction methods will need to satisfy if IM experiments are to be successful. We finally conclude with a series of desiderata for the future of cosmology with HI intensity mapping.

Throughout this paper we use the Planck best-fit $\Lambda$CDM model \citep{2013arXiv1303.5076P} for our fiducial cosmology,
\bea
& & h = 0.67,\,\,\, \Omega_\Lambda = 0.684,\,\,\, \Omega_K = 0,\,\,\, \Omega_b = 0.049, \nonumber \\
& & w = -1,\,\,\, n_s = 0.962,\,\,\, \sigma_8 = 0.834,\,\,\, N_\mathrm{eff} = 3.046, \nonumber
\eea
and all distances and scales are expressed in physical, rather than $h^{-1}$, units.

\section{Fisher forecast formalism}

We base our forecasting formalism on the {\it Fisher matrix} technique, which assumes that all parameters of interest can be approximated as being Gaussian-distributed, and that observations are unbiased. While it has its drawbacks \citep{2012MNRAS.424....2H,2012JCAP...09..009W}, Fisher forecasting is an effective way of getting an idea of how constraining a given experimental setup is likely to be without requiring detailed experiment-specific simulations.

We begin by defining our data model. The observed brightness temperature is $T_\mathrm{obs} = \overline{T} (1 + \delta T)$, where the total fluctuation in an individual ``voxel'' (volume element) is given by
\be
\delta T({\bm \theta}_p,\nu_p) = \delta T^S({\bm\theta}_p,\nu_p)+\delta T^N({\bm\theta}_p,\nu_p)+\delta T^F({\bm\theta}_p,\nu_p) \nonumber
\ee
with $p$ labelling the voxel given by a 2D angular direction, ${\bm\theta}_p$, and frequency, $\nu_p$. The total fluctuation consists of the cosmological signal ($S$), instrumental and atmospheric noise ($N$), and residual astrophysical foregrounds ($F$). Detailed models for each component will be set out in subsequent sections, but for now we need only note that they are all stochastic, and will be modelled as Gaussian-distributed, with mean zero, in each voxel.

It is usual to expand the different components of the data vector in terms of spherical harmonics. This is the preferred strategy for accurate forecasting on the largest scales -- for example, when testing for scale-dependent bias due to to non-Gaussianity \citep{2013PhRvL.111q1302C} or effects due to GR/modified gravity \citep{Hall:2012wd}. In this paper, however, we will work in the flat-sky limit, and describe the signal in terms of the comoving 3D power spectrum, $P({\bf k})$. This mimics what is used when forecasting for galaxy redshift surveys, and will be useful when we define an equivalence between redshift surveys and IM experiments.

In this limit, the mapping between the observed voxel and its comoving-space location is\footnote{We use the definition of transverse comoving distance from \citet{1999astro.ph..5116H}, which reduces to $r(z) = \int_0^z\frac{cdz}{H(z)}$ for $\Omega_K = 0$, and take $r_\parallel \approx \frac{dr}{dz}dz$.}
\bea
{\bf r}_\perp &\approx& r(z_i)({\bm\theta}_p - {\bm\theta}_i) \nonumber \\
r_\parallel &\approx& \frac{c(1+z_i)^2}{H(z_i)} ({\tilde \nu}_p - {\tilde \nu_i}) \equiv r_\nu(z_i) ({\tilde \nu}_p - {\tilde\nu}_i), \nonumber
\eea
where we have centred the survey on $({\bm \theta}_i, \nu_i)$, corresponding to a redshift bin centred at $z_i$, and have defined the dimensionless frequency ${\tilde \nu}\equiv\nu/\nu_{21} = (1+z)^{-1}$. We will predominantly work in observational coordinates, with the Fourier transform convention
\bea
\delta T({\bm q},y) = \int \delta T({\bm \theta},\nu) \, e^{i({\bm \theta}\cdot{\bm q}+y\cdot{\tilde \nu})} d^2{\theta} \, d {\tilde \nu}. \nonumber
\eea
The (dimensionless) observation-space Fourier variables are related to the comoving variables by $\mathbf{q} = \mathbf{k}_\perp r$ and $y = k_\parallel r_\nu$.

To construct the Fisher matrix we now need to define the covariance for each of the components. For a component $X$, this is defined as
\bea
 & &\langle \delta T^{X*} ({\bf q},y)\delta T^{X'} ({\bf q}',y')\rangle=\nonumber \nonumber \\ & & \ \ \ \ (2\pi)^3C^{X}({\bf q},y)
\delta^2({\bf q} - {\bf q}')\delta(y-y')\delta_{XX'}. \label{eqn-covariance-def}
\eea
We will make a number of approximations here. First of all, we assume that the signal, noise and foregrounds are uncorrelated with one other, and that the resulting covariance matrices are diagonal (i.e. they are statistically homogeneous and isotropic). The former is not true in practice, as the process by which foregrounds are removed from the data will introduce correlations, as discussed in Section \ref{sect-fg}. The latter is also not strictly true, as the cosmological signal is only diagonal in the flat-sky limit, and the main foreground that we need to correct for -- the Galaxy -- is anisotropic and will also introduce off-diagonal terms. Nevertheless, given the conservative modelling choices we make in coming sections, we believe our results will be close enough to the real situation.

A further approximation is that evolution can be neglected within each redshift bin, so that evolving cosmological functions are fixed to their values at the central redshift of the bin. This is a good approximation for sufficiently narrow bins, as most of the relevant functions (e.g. $H(z)$, $r(z)$) vary slowly with $z$. We have verified that our results are robust to the choice of bin width (which is chosen as $\Delta \nu = 60$ MHz for all experiments).

\subsection{Signal model}

Radio telescopes measure flux density -- the integral of the source intensity, $I_\nu$, over the solid angle of the telescope beam. We derive an expression for the HI line intensity in Appendix \ref{app-HI-line}. In the Rayleigh-Jeans limit, this can be converted into an effective HI brightness temperature, $T_b = c^2 I_\nu / 2 k_B \nu^2$, that can be split into a homogeneous part and a fluctuating part, $T_b = \overline{T}_b (1 + \delta_\mathrm{HI})$, where (from Appendix \ref{app-HI-line})
\be
\overline{T}_b = \frac{3}{32 \pi} \frac{h c^3 A_{10}}{k_B m_p \nu^2_{21}} \frac{(1+z)^2}{H(z)} \Omega_\mathrm{HI}(z) \rho_{c,0}.
\ee
The fluctuations are the quantity of interest here, and so we identify the cosmological signal as
\be
\delta T^S({\bm \theta}_p,\nu_p) = \overline{T}_b(z) \delta_{\rm HI}({\bf r}_p, z). \nonumber
\ee
At late times, most of the neutral hydrogen content of the Universe is expected to be localised to dense gas clouds within galaxies, where it is shielded from ionising photons. We therefore expect HI to be a biased tracer of the dark matter distribution, just as galaxies are. This allows us to write the HI density contrast as $\delta_{\rm HI} = b_{\rm HI} \star \delta_M$ (where $\delta_M$ is the total matter density perturbation, and $\star$ denotes convolution, accounting for the possibility of scale- and time-dependent biasing).

Because the HI intensity is measured as a function of frequency (and thus redshift) rather than comoving distance, we must also account for redshift space distortions (RSDs) caused by the peculiar velocities of the clouds and the galaxies in which they reside. Following \citet{Kaiser:1987qv}, we write the (Fourier-transformed) redshift-space HI contrast as
\be
\delta_\mathrm{HI}(\bm{k}) = (b_\mathrm{HI} + f \mu^2) \exp \left ( - k^2 \mu^2 \sigma^2_\mathrm{NL}/2 \right ) \delta_M(\bm{k}), \label{eqn-deltaHI}
\ee
where $\mu \equiv {k}_\parallel/k$ and the flat-sky approximation has been used again. We have assumed that the HI velocities are unbiased. The linear growth factor, $f$, is a key observable, telling us much about the growth of structure on linear scales; we will study it in detail in Section \ref{sect-f}. The exponential term accounts for the ``Fingers of God'' effect due to uncorrelated velocities on small scales\footnote{Alternatively, we could have used a Lorentzian instead of an exponential, or a slightly more complex exponential term that models non-linear smoothing of the BAO as well \citep{Samushia:2011cs}.}, which washes out structure in the radial direction past a cutoff scale parametrised by the non-linear dispersion, $\sigma_\mathrm{NL}$.

Substituting (\ref{eqn-deltaHI}) into (\ref{eqn-covariance-def}) and making use of the definition of the isotropic matter power spectrum,
\bea
\langle \delta^*_{M}({\bf k}) \delta_{M}({\bf k}')\rangle\equiv(2\pi)^2P(k)\delta^3({\bf k}-{\bf k}'), \nonumber
\eea
we can write the signal covariance as
\bea
C^S({\bf q},y) &=& T^2_b(z_i)\frac{P_{\rm tot}(z_i,\frac{\bf q}{r},\frac{y}{r_\nu})}{r^2r_\nu}, \label{sigcov}
\eea
where the factor of $r^2 r_\nu$ is from the conversion into observational Fourier coordinates, $(\mathbf{q}, y)$, and we have defined
\bea
P_{\rm tot}(z_i,{\bf k}_\perp,k_\parallel) &=& F_{\rm RSD}({\bf k}_\perp,k_\parallel)
D^2(z) P(k, z=0) \nonumber \\
F_{\rm RSD}({\bf k}_\perp,k_\parallel)&=&(b_{\rm HI} + f\mu^2)^2\exp(-k^2\mu^2\sigma_{\rm NL}^2). \nonumber
\eea
The redshift dependence of the matter power spectrum has been factored out into the linear growth factor, $D$, which is normalised to $D(z\!\!=\!\!0) \!=\! 1$. This is related to the growth rate by $f = d \log D / d \log a$. Strictly, the growth factor should be scale-dependent on small scales, but for simplicity we neglect this possibility here as we will mostly be concerned with larger scales.

It is straightforward to calculate fiducial values for the cosmological functions in (\ref{sigcov}). We use CAMB \citep{Lewis:1999bs} to calculate $P(k)$ at $z=0$ for our chosen fiducial cosmological parameters, and use the simple parametrisation $f(z) = \Omega^\gamma_M(z)$ for the linear growth rate \citep{peebles1980, Linder:2005in}, where $\Omega_M(z) = \Omega_M (1 + z)^3 H_0^2 / H^2(z)$, and $\gamma \approx 0.55$ for $\Lambda$CDM. For the other functions in (\ref{sigcov}), however, there is considerably more uncertainty in the choice of fiducial model.

One key uncertainty is the behaviour of the HI bias, $b_{\rm HI}$. The bias depends on the size of host dark matter haloes; if a halo is too small, gas clouds would be unable to gain sufficient density to shield themselves and keep the hydrogen neutral. The halo dependence can be modelled using the halo mass function with an appropriate lower mass cutoff (or lower rotation velocity); see \citep{Bagla:2009jy} for example. There are a few candidate models for the evolution of the bias as a function of redshift  that fit the current constraints from observations \citep{Switzer:2013ewa} or are calibrated against simulations \citep{2008MNRAS.388.1335W}, but there is considerable disagreement between them. In Section \ref{sect-HI-evol} and Appendix \ref{app-HI-signal}, we discuss the impact of the uncertain bias evolution. Unless stated otherwise, we will use a linear bias model for the rest of the paper, and -- rather conservatively -- marginalise over the value of $b_\mathrm{HI}$ separately in each redshift bin.

Another major uncertainty is in the HI density fraction, $\Omega_\mathrm{HI} = \rho_\mathrm{HI} / \rho_{c,0}$. This enters the signal covariance through $\overline{T}_b(z)$, since $\overline{T}_b \propto \Omega_\mathrm{HI}$. The current best constraints on the HI fraction come from \citet{Switzer:2013ewa}, who find 
\bea
\Omega_{\rm HI}b_{\rm HI} = 4.3\pm{1.1} \times 10^{-4} \nonumber
\eea
at the $68\%$ confidence level at $z=0.8$. This constitutes a relatively large uncertainty in the overall amplitude of the HI signal and, correspondingly, the signal-to-noise that can be achieved by a given experiment. We investigate the impact of this uncertainty in Section \ref{sect-HI-evol}, but for the rest of the paper we will adopt a fiducial value of $\Omega_{\mathrm{HI},0} = 4.86\times 10^{-4}$.

The non-linear dispersion scale, $\sigma_\mathrm{NL}$, is yet another source of uncertainty. Recent values from the literature vary between $\sigma_\mathrm{NL} \approx 4 - 10$ Mpc (e.g. \citet{Li:2007rpa, Reid:2011ar, Reid:2012sw, Contreras:2013bol}); for our fiducial model, we choose a middling value of $\sigma_\mathrm{NL} = 7$ Mpc \citep{Li:2007rpa}, which corresponds to a non-linear scale of $k_\mathrm{NL} \sim 0.14$ Mpc$^{-1}$ (or a velocity dispersion of $\sim 500$ km/s). We check the sensitivity of our results to this choice in Section \ref{sect-nonlinear}. IM experiments can independently constrain $\sigma_\mathrm{NL}$, so we leave it free, marginalising over it as a nuisance parameter in all forecasts.

\subsection{Noise model and effective beams} \label{sect-cn}

The noise covariance models the instrumental and sky noise for a given experiment, but we will also use it to include the effects of instrumental beams. For radio telescopes, assuming uncorrelated Gaussian noise, the noise covariance has the standard form
\bea
C^{N}({\bf q},y)=\frac{T^2_{\rm sys}}{t_{\rm tot} \Delta\nu} U_{\rm bin}\, {\cal I} B^{-2}_\perp B^{-1}_\parallel, \label{eqn-noise}
\eea
where $T_{\rm sys}$ is the system temperature, $t_{\rm tot}$ is the total integration time, $U_{\rm bin}=S_{\rm area} \,\Delta{\tilde \nu}$ is the volume of an individual redshift bin, and $S_{\rm area}$ and $\Delta{\tilde \nu}$ are the survey area and (dimensionless) bandwidth within the redshift bin respectively. The factors of $\mathcal{I}$ and $B$ describe the number (or number density) of receivers and their corresponding frequency and angular responses.

The system temperature has two main components: the instrument temperature, $T_\mathrm{inst}$, which depends on the hardware design, {\corr and $T_\mathrm{sky} \approx 60 \,\mathrm{K} \times (\nu / 300 \, \mathrm{MHz})^{-2.5}$, which} accounts for atmospheric and background radio emission, to give a total $T_\mathrm{sys} = T_\mathrm{inst} + T_\mathrm{sky}$. Values for $T_\mathrm{inst}$ are quoted in the design specifications for a given experiment, and are typically a few tens of Kelvin.

The survey area and total integration time are not intrinsic to the design of the instrument, but are instead chosen as part of the survey strategy. We will systematically examine the effects of varying these parameters in Section \ref{sect-survey-design}. One of the advantages of intensity mapping with radio telescopes is that substantial fractions of the sky can be surveyed to a useful depth over the course of only a year or so. This is thanks in part to the relative cheapness of low noise multiple-receiver systems, and the $\sim\,$degree-scale primary beams of dishes in most arrays, both of which act to substantially improve survey speed. For much of what follows, we will assume that all experiments can perform between $10$ to $25,000$ sq. deg. surveys over 10,000 hours total observing time, which is reasonable for a dedicated survey telescope.

We now turn to the effective beam terms. The instrumental resolution in the radial direction is limited by the bandwidth of an individual frequency channel, $\delta\nu$. We approximate the channel bandpass by a Gaussian, which gives an effective beam in the parallel direction,
\be
B_\parallel(y) = \exp \left ( -\frac{(y\delta {\nu / \nu_{21}})^2}{16\ln 2} \right ); \nonumber
\ee
the numerical factor comes from the definition of the full width at half maximum (FWHM), $\sigma = \theta_\mathrm{FWHM}\sqrt{8 \ln 2}$. Modern radio receivers can typically be built with narrow channel bandwidths of around 100 kHz or less. Narrow channels allow for more precise removal of artificial radio interference (RFI), for example, but also increase the data rate, which may require expensive increases in correlator performance for interferometers. In practise, the channel bandwidth is not the limiting factor in the radial resolution, as for realistic $\delta \nu$ this is instead determined by the non-linear dispersion scale, $\sigma_\mathrm{NL}$.

The expressions for the dish multiplicity factor, $\mathcal{I}$, and transverse effective beam, $B_\perp$, depend on whether the array is used as an interferometer or a collection of independent single-dishes, i.e. whether the signals from individual dishes are correlated with one another or not. For {\it single-dish} experiments that only use the autocorrelation of the signal from each dish, we have
\bea
{\cal I} &=& \frac{f(\nu)}{N_{\rm b}N_{\rm d}} \nonumber \\
B_{\perp}({\bf q}) &=& \exp \left ( - \frac{(q \theta_{\rm B})^2}{16 \ln 2} \right ), \nonumber
\eea
where $\theta_{\rm b} \approx \lambda / D_\mathrm{dish}$ is the FWHM of the beam of an individual dish of diameter $D_\mathrm{dish}$ at some wavelength $\lambda$, and $N_{\rm d}$ is the number of dishes in the array. $N_{\rm b}$ is the number of beams, which differs from unity if the experiment has multiple pixels or phased array feeds (PAFs) per dish. Each dish/beam will typically survey a different patch of the sky, thus increasing the survey speed. {\corr Any additional frequency dependence of the sensitivity (e.g. due to beam overlap for PAF receivers) can be accounted for by the $f(\nu)$ factor; specific forms of the noise expression appropriate for different types of receiver are given in Appendix \ref{app:noise}.}

For interferometers, in the case where we assume multiple pointings, i.e. $S_\mathrm{area} > \mathrm{FOV}$, we have
\bea
{\cal I} B^{-2}_\perp &=& \frac {\rm FOV} {n(u \!=\! q/2\pi)}, \nonumber
\eea
where $n(u)$ is the number density of samples in the $uv$ plane as a function of $|u|$, and $\mathrm{FOV} \approx (\lambda / D_\mathrm{dish})^2$ is the field of view. Each configuration of baselines will lead to a different $n(u)$, although if one assumes constant sampling in $uv$, it can be approximated by
\be
n(u) = \frac{N_d (N_d - 1)}{2 \pi (u^2_\mathrm{max} - u^2_\mathrm{min})}, \label{eqn-nu-const}
\ee
where $u_\mathrm{max} = D_\mathrm{max} / \lambda$, $u_\mathrm{min} = D_\mathrm{min} / \lambda$, and $D_\mathrm{max}$, $D_\mathrm{min}$ are the lengths of the longest/shortest baselines. The effective beam in the transverse direction is determined by $n(u)$, and so does not need to be defined separately.

For interferometric experiments where the baseline distribution is available, we calculate $n(u)$ specifically for that distribution; otherwise, we use the approximation in Eq. (\ref{eqn-nu-const}). For the former, one has to assume a declination of observation as well as a baseline distribution. As the sky drifts, a set of tracks will be mapped out onto the $uv$ plane. These are typically split into bins of size $(\Delta u)^2 \sim 1 / \mathrm{FOV}$, which can be taken to be independent. It is then possible to construct a simple model for $n(u)$ that is good enough for our forecasts (see Appendix \ref{app-baselines}).


\subsection{Foreground model} \label{forecast-dg}

Foreground contamination from the galaxy and extragalactic point sources dwarfs the cosmological HI signal. A number of different methods have been proposed for modelling and subtracting the foregrounds \citep{2003MNRAS.346..871O,2005ApJ...624L..65B,2005ApJ...625..575S,2006ApJ...648..767M, 2006ApJ...650..529W,2008MNRAS.391..383G,2008MNRAS.389.1319J, 2009MNRAS.398..401L,2011MNRAS.413.2103P}, but in this paper we will simply assume that some sort of method has been applied that removes them, leaving behind some residual contamination whose variance can be modelled as a sum of smooth power spectra. (One could also model instrumental calibration residuals in this way.) Our model for the residual foreground is
\bea
C^F(\mathbf{q},y) = \epsilon^2_{\rm FG} \sum_{X}A_X\left(\frac{l_{\rm p}}{2\pi q}\right)^{n_X}\left(\frac{{\nu}_p}{\nu_{\rm i}}\right)^{m_X}.
\eea
The amplitude and index parameters for four representative foregrounds are given in Table \ref{tab:fiducial}, following \citep{2005ApJ...625..575S}.

Subtraction algorithms also introduce a minimum wave number below which cosmological information cannot be extracted. This is because the smooth variation of the foregrounds in frequency is difficult to separate from cosmological modes on scales comparable to the total survey bandwidth, $k_\mathrm{FG} \sim 1 / (r_\nu \Delta {\tilde \nu}_\mathrm{tot})$. A subtraction method that relies on fitting and subtracting smoothly-varying functions will necessarily remove some power from radial cosmological modes larger than this scale as well.

\begin{table}[b]
\begin{center}
{\renewcommand{\arraystretch}{1.4} \begin{tabular}{|l|c|c|c|}
\hline 
{\bf Foreground} &$A_X [{\rm mK}^2]$ & ~$n_X$~ & ~$m_X$~ \\
\hline 
Extragalactic point sources~ &  57.0  & 1.1 &  2.07 \\ 
Extragalactic free-free  &  0.014  & 1.0 &  2.10 \\ 
Galactic synchrotron  &  700  & 2.4 &  2.80 \\ 
Galactic free-free  &  0.088  & 3.0 &  2.15 \\
\hline
\end{tabular} }
\end{center}
\vspace{-1em}\caption{Foreground model parameters at $\ell_p = 1000$ and $\nu_p = 130$ MHz, taken from \citet{2005ApJ...625..575S}.}
\label{tab:fiducial}
\end{table}

We have introduced an overall scaling, $\epsilon_{\rm FG}$, to parametrise the efficiency of the foreground removal process: $\epsilon_{\rm FG}=1$ corresponds to no foreground removal, while we will probably need $\epsilon_{\rm FG} \lesssim 10^{-5}$ to be able to extract the cosmological signal. One can also interpret $\epsilon_{\rm FG}$ as a measure of how smooth (or correlated) foregrounds are in frequency, and thus how well they can be modelled by smooth deterministic functions. For example, if we assume a Gaussian correlation function along the frequency direction, we have that $\epsilon_{\rm FG} \sim \exp[-(\Delta \nu/\xi)^2]$, where $\Delta \nu$ is the bandwidth of the redshift bin we are probing and $\xi$ is the correlation length in frequency. We have neglected cross-correlations between different frequencies here, although including them would not be difficult.

This treatment of foreground subtraction is necessarily simplified. For example, we are assuming that the various contributions to the total signal remain uncorrelated, yet all of the subtraction methods proposed so far remove modes that receive contributions from all components. Although the foregrounds will be the dominant part of the subtracted signal by far, this process will nevertheless induce cross-correlations between whatever is left in the residual. Practical experience of foreground subtraction from real data is quite limited so far (see \citep{Chang:2010jp, Switzer:2013ewa, 2013ApJ...763L..20M} for some initial attempts), and so we do not have a good picture of how important various aspects of the foreground problem are yet. For the time being, we believe that our model captures the essential elements of the foregrounds sufficiently well to be of use. We will return to the issue of foreground modelling in Sect. \ref{sect-fg}.

\subsection{The Fisher Matrix}

We are now in a position to construct the full Fisher matrix. To do this, we need to sum over the number of independent modes in ${\bf q}$ and $y$. If $S_{\it area}$ is the survey area, $\mathrm{FOV}$ is the field of view of a single array element, and $\Delta{\tilde \nu}$ is the (dimensionless) bandwidth in the redshift bin, we have $\Delta y = 2\pi/\Delta{\tilde \nu}$, and $\Delta q = 2\pi/\sqrt{S_{\it area}}$ (single-dish) or $\Delta q = 2\pi/\sqrt{\mathrm{FOV}}$ (interferometer). The sum over modes is then given by
\bea
\int\frac{d^2qdy}{(\Delta q)^2 \Delta y} \rightarrow U_{\rm bin}\int\frac{d^2qdy}{(2\pi)^3}, \nonumber
\eea
where $U_{\rm bin}=S_{\it area}\times \Delta{\tilde \nu}$. If we define $C^T=C^S+C^N+C^F$, the Fisher matrix for a set of cosmological parameters $\{p_i\}$ is given by
\bea
F^{\rm IM}_{ij}=\frac{1}{2} U_{\rm bin} \int\frac{ d^2 qdy}{(2\pi)^3}[\partial_i \ln C^T({\bf q},y)\partial_j \ln C^T({\bf q},y)], \,\,\,\,\,\, \label{fishqy}
\eea
where the derivatives will only act on $C^S$, since that is the only term containing parameters of interest.

We can rewrite (\ref{fishqy}) in terms of physical wave numbers by using the following dictionary: $V_{\rm bin} = U_{\rm bin} r^2 r_\nu$, ${\bf q}={\bf k}_\perp r$, and $y=k_\parallel r_{\nu}$, where $k_{\perp}=k\sqrt{1-\mu^2}$ and $k_\parallel=k\mu$. We then apply the substitutions
\begin{itemize}
\item $U_{\rm bin}\rightarrow  V_{\rm bin}$
\item $\int dq^2 dy \rightarrow 2\pi \int_{-1}^{+1}d\mu \int_{k_{\rm min}}^\infty k^2dk$
\item $q=kr\sqrt{1-\mu^2}$ and $y=kr_\nu\mu$.
\end{itemize}
We can now express the Fisher matrix in a familiar form to those working on galaxy redshift surveys -- a comparison we will pursue in Section \ref{sect-zsurvey}.

\subsection{Experimental configurations} \label{sect-experiments}

Our focus in this paper is on the lead-up to Phase I of the SKA. We consider a portfolio of planned experimental configurations, with the aim of exploring how they will impact constraints on cosmological parameters. Our approach is ecumenical -- we try to include as many proposed configurations as possible, although we are limited by what information has been made publicly available.

We will first of all consider three illustrative experimental setups, roughly corresponding to successive `generations' of planned IM experiments. These are:
\begin{itemize}
 \item {\bf \StageOne} -- Small, specialised HI experiment focused on a relatively narrow redshift range, intended to provide initial detections of the BAO and other first cosmological results. \StageOne\ experiments are envisaged as either surveys on existing general-purpose arrays, or relatively cheap purpose-built telescopes using multi-feed receivers to improve sensitivity.
 
 \item {\bf \StageTwo} -- Larger interferometric experiment with enhanced sensitivity, covering a wider range of redshifts. \StageTwo\ experiments are intended to cover a substantial survey volume, with the aim of producing constraints on cosmological parameters that are competitive with contemporary (DETF\footnote{See \cite{Albrecht:2006um}.} Stage II/III) LSS surveys. They are likely to be either purpose-built HI arrays with a large number of receivers, or surveys on forthcoming `SKA-precursor' arrays such as MeerKAT and ASKAP.
 
 \item {\bf \StageThree} -- Survey on a future large array, covering a wide redshift range over most of the sky. \StageThree-type surveys will compete with other large (DETF Stage IV) experiments to produce the most precise cosmological parameter estimates, and will be able to probe novel HI-only effects for the first time. The only planned experiments of this type so far are the Phase I SKA arrays, {\corr although the full CHIME and Tianlai configurations could also fall into this class.}
\end{itemize}

\begin{table*}[t]
\hspace{-2em}
{\renewcommand{\arraystretch}{1.3} \begin{tabular}{|c|c l|c|c|c|c|c|c|c|c|c|c|c|r|}
\hline
 & & \multirow{2}{*}{\bf Experiments} & $T_\mathrm{inst}$ & \multirow{2}{*}{$N_d \times N_b$} & $D_\mathrm{dish}$ & $D_\mathrm{min}$ & $D_\mathrm{max}$ & $\nu_\mathrm{crit}$ & $\nu^\mathrm{IM}_\mathrm{max}$ & $\nu^\mathrm{IM}_\mathrm{min}$ & $\Delta\nu^\mathrm{IM}$ & \multirow{2}{*}{$z_\mathrm{min}$} & \multirow{2}{*}{$z_\mathrm{max}$} & $S_\mathrm{area}$~ \\
 & & & $[\mathrm{K}]$ &  & $[\mathrm{m}]$ & $[\mathrm{m}]$ & $[\mathrm{m}]$ & $[\mathrm{MHz}]$ & $[\mathrm{MHz}]$ & $[\mathrm{MHz}]$ & $[\mathrm{MHz}]$ & & & $[\mathrm{deg}^2]$ \\
\hline
\parbox[t]{2mm}{\multirow{3}{*}{\rotatebox[origin=c]{90}{Ref.}}} & & \StageOne\ & 50 & $1 \times 50$ & 30.0 & -- & -- & -- & 1100 & 800 & 300 & 0.29 & 0.77 & 5,000 \\
 & $\bullet$ & \StageTwo\ & 35 & $160 \times 1$ & 4.0 & 4.0 & 53.0 & 1000 & 1000 & 600 & 400 & 0.42 & 1.37 & 2,000 \\
 & & \StageThree\ & 20 & $250 \times 1$ & 15.0 & -- & -- & -- & 1100 & 400 & 700 & 0.29 & 2.55 & 25,000 \\
\hline
\parbox[t]{2mm}{\multirow{7}{*}{\rotatebox[origin=c]{90}{Existing Facility}}} & & GBT & 29 & $1 \times 1$ & 100.0 & -- & -- & -- & 920 & 680 & 240 & 0.54 & 1.09 & 100 \\
 & & GBT-HIM & 33 & $1 \times 7$ & 100.0 & -- & -- & -- & 900 & 700 & 200 & 0.58 & 1.03 & 1,000 \\
 & & GMRT & 70 & $30 \times 1$ & 45.0 & -- & -- & -- & 1420 & 1000 & 420 & 0.00 & 0.42 & 1,000 \\
 & & JVLA & 70 & $27 \times 1$ & 25.0 & -- & -- & -- & 1420 & 1000 & 420 & 0.00 & 0.42 & 1,000 \\
 & & Parkes & 23 & $1 \times 13$ & 64.0 & -- & -- & -- & 1420 & 1155 & 265 & 0.00 & 0.23 & 5,000 \\
 & & VLBA & 27 & $10 \times 1$ & 25.0 & -- & -- & -- & 1420 & 1200 & 220 & 0.00 & 0.18 & 5,000 \\
 & $\blacktriangle$ & WSRT + APERTIF & 52 & $14 \times 37$ & 25.0 & -- & -- & -- & 1300 & 1000 & 300 & 0.09 & 0.42 & 25,000 \\
\hline
\parbox[t]{2mm}{\multirow{6}{*}{\rotatebox[origin=c]{90}{Targeted IM}}} & $\bullet$ & BAOBAB-128 & 40 & $128 \times 1$ & 1.6 & 1.6 & 26.0 & -- & 900 & 600 & 300 & 0.58 & 1.37 & 1,000 \\
 & & BINGO & 50 & $1 \times 50$ & 25.0 & -- & -- & -- & 1260 & 960 & 300 & 0.13 & 0.48 & 5,000 \\
 & $\Diamond$ & CHIME & 50 & $1280 \times 1$ & 20.0 & -- & -- & -- & 800 & 400 & 400 & 0.77 & 2.55 & 25,000 \\
 & & FAST & 20 & $1 \times 20$ & 500.0 & -- & -- & -- & 1000 & 400 & 600 & 0.42 & 2.55 & 2,000 \\
 & $\bullet$ & MFAA & 50 & $3100 \times 1$ & 2.4 & 0.1 & 250.0 & 950 & 950 & 450 & 500 & 0.49 & 2.16 & 5,000 \\
 & $\Diamond$ & Tianlai & 50 & $2048 \times 1$ & 15.0 & -- & -- & -- & 950 & 550 & 400 & 0.49 & 1.58 & 25,000 \\ 
\hline
\parbox[t]{2mm}{\multirow{4}{*}{\rotatebox[origin=c]{90}{Pre-SKA}}}  & $\blacktriangle$ & ASKAP & 50 & $36 \times 36$ & 12.0 & -- & -- & -- & 1000 & 700 & 300 & 0.42 & 1.03 & 25,000 \\
 & & KAT7 & 30 & $7 \times 1$ & 13.5 & -- & -- & -- & 1420 & 1200 & 220 & 0.00 & 0.18 & 2,000 \\
 & & MeerKAT (B1) & 29 & $64 \times 1$ & 13.5 & -- & -- & -- & 1015 & 580 & 435 & 0.40 & 1.45 & 25,000 \\
 & & MeerKAT (B2) & 20 & $64 \times 1$ & 13.5 & -- & -- & -- & 1420 & 900 & 520 & 0.00 & 0.58 & 25,000 \\
\hline
\parbox[t]{2mm}{\multirow{8}{*}{\rotatebox[origin=c]{90}{SKA Phase I}}} & & SKA1-MID (B1) Autocorr. & 28 & $190 \times 1$ & 15.0 & -- & -- & -- & 1050 & 350 & 700 & 0.35 & 3.06 & 25,000 \\
 & $\circ$ & SKA1-MID (B1) Interferom. & 28 & $190 \times 1$ & 15.0 & -- & -- & -- & 1050 & 350 & 700 & 0.35 & 3.06 & 100 \\
 & & SKA1-MID (B2) Autocorr. & 20 & $190 \times 1$ & 15.0 & -- & -- & -- & 1420 & 900 & 520 & 0.00 & 0.58 & 25,000 \\
  & $\circ$ & SKA1-MID (B2) Interferom. & 20 & $190 \times 1$ & 15.0 & -- & -- & -- & 1420 & 900 & 520 & 0.00 & 0.58 & 50 \\
 & $\blacktriangle$ & SKA1-SUR (B1) & 50 & $60 \times 36$ & 15.0 & -- & -- & 710 & 900 & 400 & 500 & 0.58 & 2.55 & 25,000 \\
 & $\blacktriangle$ & SKA1-SUR (B2) & 30 & $96 \times 36$ & 15.0 & -- & -- & 1300 & 1150 & 650 & 500 & 0.23 & 1.18 & 25,000 \\
 & & SKA1-MID + MeerKAT (B1)$^\dagger$ & -- & -- & -- & -- & -- & -- & 1050 & 350 & 700 & 0.35 & 3.06 & 25,000 \\
 & & SKA1-MID + MeerKAT (B2)$^\dagger$ & -- & -- & -- & -- & -- & -- & 1420 & 900 & 520 & 0.00 & 0.58 & 25,000 \\
\hline
\end{tabular} }
\caption{{\corr Telescope configurations used in this paper. The assumed observing mode of each telescope is denoted by: (~) single-dish; ($\blacktriangle$) single-dish with phased array feed; ($\circ$) dish interferometer; ($\bullet$) dense aperture array; ($\Diamond$) cylinder interferometer. Some instruments can operate over a wider frequency range than shown here; where this is the case, our values correspond to the most appropriate $\nu_\mathrm{max}$ for IM, or we include multiple bands. $^\dagger$For combined arrays, in redshift bins where the bands overlap, we find $T_{\rm inst}$, $D_{\rm dish}$ by averaging the values for each sub-array, weighted by the no. of dishes.}}
\label{tab:experiments}
\end{table*}

We have chosen representative configurations for each of these classes (see Table \ref{tab:experiments}) that will be used to illustrate the expected progress of HI IM experiments over the next decade. We have also forecasted for the following real (existing, proposed, or plausible) experiments:
\begin{description}
\item[ASKAP] An SKA pathfinder consisting of thirty-six 12m dishes, each with 36-element PAFs, located at the eventual site of SKA1-SUR in Australia \citep{2008ExA....22..151J}.
\item[BAOBAB] {\corr Proposed compact array of 128 $1.6$m tiles with 4 dipoles per tile, co-located with GBT or SKA1-MID \citep{2013AJ....145...65P}.}
\item[BINGO] A proposed 40m (25m illuminated) multi-receiver single-dish telescope in South America \citep{2013MNRAS.434.1239B}.
\item[CHIME] A proposed array made up of $20 \times 100$m cylinders ($20 \times 80$m illuminated), based in British Columbia, Canada. There is a pathfinder with 2 half-length cylinders, and a planned full experiment with 5 \citep{chime}.
\item[FAST] {\corr A proposed multi-beam system on the 500m single-dish telescope currently under construction in south-west China \citep{Smoot:2014oia}.}
\item[GBT] A 100m single-dish telescope in West Virginia (USA). GBT has already been used for preliminary detections of the HI signal \citep{Chang:2010jp, Switzer:2013ewa, 2013ApJ...763L..20M}.
\item[GBT-HIM] {\corr A planned seven-beam receiver system on GBT \citep{2014era..conf50102C}.}
\item[GMRT] Array of thirty 45m dishes in Pune, India \citep{swarup1991giant}.
\item[JVLA] An array of twenty-seven 25m dishes, based in New Mexico, USA \citep{jvla}.
\item[KAT7] An SKA pathfinder made up of seven 12m dishes, on the planned site of SKA1-MID \citep{kat7}.
\item[MeerKAT] An SKA pathfinder with sixty-four 13.5m dishes, on the site of SKA1-MID \citep{2009IEEEP..97.1522J}. Has a choice of two frequency bands.
\item[MFAA] {\corr A proposal for a mid-frequency aperture array component of Phase II of the SKA \citep{mfaa}.}
\item[Parkes] A single 64m dish (with 13 beams) in NSW, Australia \citep{parkes}.
\item[SKA1-MID] A planned SKA Phase I configuration with one hundred and ninety 15m dishes, based in the Northern Cape, South Africa \citep{dewdney2013ska1}. Can be extended to incorporate the 64 MeerKAT dishes.
\item[SKA1-SUR] A planned SKA Phase I configuration with sixty 15m dishes, each with 36-element PAFs, based in Western Australia \citep{dewdney2013ska1}. Can be extended to incorporate the ASKAP dishes.
\item[Tianlai] {\corr A proposed array of eight $15 \times 120$m cylinders to be built in north-west China \citep{2012IJMPS..12..256C}.}
\item[VLBA] An array of ten 25m dishes distributed across North America \citep{napier1994}.
\item[WSRT + APERTIF] A proposed upgrade to WSRT that uses a phased array feed (PAF) in the focal plane to produce multiple beams on the sky \citep{Oosterloo:2010wz}.
\end{description}

This is intended to be a relatively exhaustive list of current and {\corr planned HI intensity mapping experiments at $z \lesssim 3$}, but inevitably some have been omitted due to a lack of publicly-available specifications. We have made our code publicly available,\footnote{\url{https://gitlab.com/radio-fisher/bao21cm}} so forecasts can be performed when specifications become available.

The instrumental parameters used for each experiment are listed in Table \ref{tab:experiments}. The survey area is chosen (between $10$ and 25,000 $\mathrm{deg}^2$) to maximise the DE figure of merit (see Sect. \ref{sect-parameters}), and the survey time is assumed to be $t_\mathrm{tot} = 10^4$ hours for all experiments. For interferometers, we use either the baseline density calculated from the actual array layout (see App. \ref{app-baselines}), or the $n(u) \sim \mathrm{const.}$ approximation of Eq. (\ref{eqn-nu-const}). For simplicity, we consider only single-dish mode for some telescope arrays, even though they are capable of interferometric measurements.

It is useful to be able to compare the performance of IM experiments with competing probes, such as galaxy redshift surveys. To this end, we also produce forecasts for a fiducial DETF Stage IV spectroscopic galaxy redshift survey, similar to {\corr DESI, Euclid or WFIRST,} with an expected yield of $\sim 6 \times 10^7$ spectroscopic redshifts. We take the survey to cover roughly a third of the sky ($f_\mathrm{sky} = 0.35$) over a redshift range of $0.65 \leq z \leq 2.05$, with redshift distribution taken from \citet{2013LRR....16....6A} (Euclid reference case, Table 1.3). The bias is taken to evolve as $b(z) = \sqrt{1+z}$. We forecast for {\corr the same set of cosmological parameters as IM experiments, with the same fiducial values,} including relevant nuisance parameters like $\sigma_\mathrm{NL}$. We use the redshift scaling from \citet{Smith:2002dz} to set $k_\mathrm{max} = k_\mathrm{NL} (1 + z)^{2 / (2 + n_s)}$, and choose $k_\mathrm{min} = 2\pi/(V_\mathrm{bin})^\frac{1}{3}$. Constraints are quite sensitive to the choice of $k_\mathrm{max}$ \citep{White:2008jy}, but mostly insensitive to $k_\mathrm{min}$ (if chosen sufficiently small).

\subsection{Prior information}

Intensity mapping experiments cannot constrain all cosmological quantities of interest on their own; information from other probes must be added in order to break degeneracies and improve precision. High quality data are already available from a range of other sources, including CMB temperature and polarisation anisotropies at high and low-$\ell$, galaxy redshift surveys at high and low redshift, weak gravitational lensing, and supernova distance measurements. By the time of the first intensity mapping surveys with the SKA, however, the number of experiments for each of these probes, and their precision, will have risen sharply.

\begin{figure}[t]
\includegraphics[width=\columnwidth]{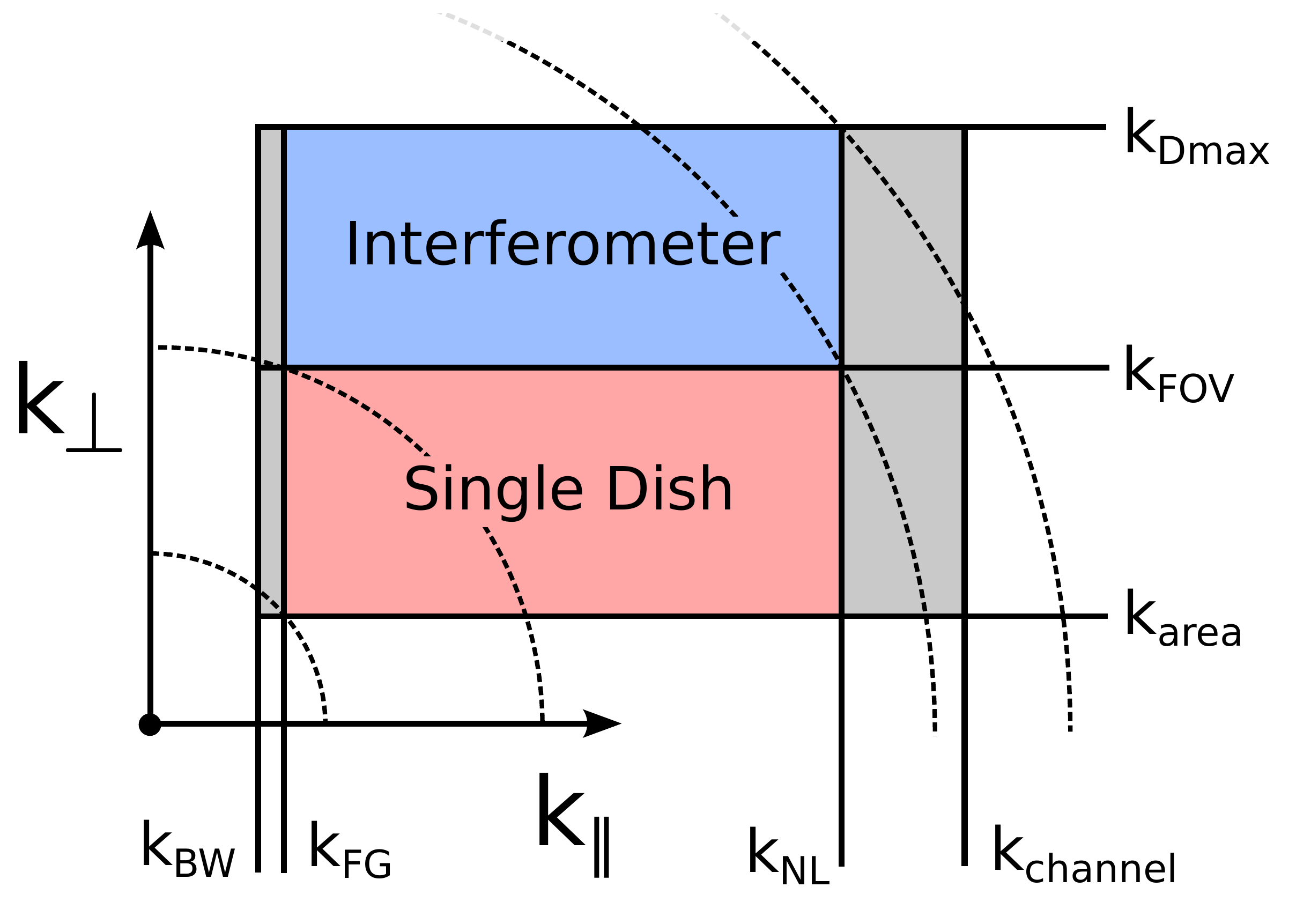}
\caption{Schematic illustration of the ranges of radial and transverse wavenumbers that the two types of experiment are sensitive to. The dotted lines show the ranges in absolute wavenumber $|k|$; single-dish experiments are sensitive to smaller $|k|$ due to their lower $k_\perp^\mathrm{min}$, while interferometers can see larger $|k|$ on account of their high angular resolution. The two types of experiment are complementary in terms of their angular sensitivity, but are subject to the same constraints in frequency space.}
\label{fig-pk-direction}
\end{figure}

While the best constraints will ultimately be obtained by combining information from all relevant experiments, our intention here is not to provide an exhaustive account of the expected state of observational cosmology in ten years' time. Instead, we will (conservatively) focus only on the CMB as an external probe in this paper.


CMB data provides a high-redshift distance measurement that is vital for anchoring the low-$z$ distance measures that most effectively probe dark energy. It also yields information about the shape and normalisation of the matter power spectrum, the matter content at $z \simeq 1090$, and the physical scale of the BAO.
 
We use the DETF Planck prior from \citet{Albrecht:2009ct}, which assumes temperature and E-mode polarisation data over 70\% of the sky for the 70, 100, and 143 GHz channels out to $\ell=2000$. We rescale the Fisher matrix to reflect our different fiducial cosmology, and then project it out to the following variables (fixing all others): 
\be
\{ h, \Omega_b h^2, \Omega_\mathrm{DE}, \Omega_K, w_0, w_a, n_s, \sigma_8 \}. \nonumber
\ee
Note that the optical depth to last scattering, $\tau$, which depends on the cosmic reionisation history, has been marginalised over in the original DETF Fisher matrix. We ignore constraints from CMB lensing, high-$\ell$ CMB experiments and B-mode polarisation, although in principle these would provide additional information on $h$, $\sigma_8$, $n_s$, and the effective number of neutrino species.

\section{Comparison with galaxy redshift surveys} \label{sect-zsurvey}

It is instructive to compare IM surveys of large scale structure with `conventional' redshift surveys, by which we mean surveys that catalogue individual galaxies in angle and redshift. These offer some of the most stringent constraints on cosmological parameters to date, and are likely to do so for some time as experiments like the Dark Energy Survey (DES) \citep{2013AAS...22133501F} and Euclid \citep{2013LRR....16....6A} come online.

In this section, we will compare IM and galaxy redshift surveys by looking directly at constraints on the power spectrum, $P(k)$. To do this we divide up a range of wavenumbers into bins, $\Delta_a = [k_a,k_{a+1}]$, and assign a constant value, $P_a$ to the power spectrum in each bin. The exercise is then to forecast errors for each $P_a$.

\begin{figure}[t]
\includegraphics[width=\columnwidth]{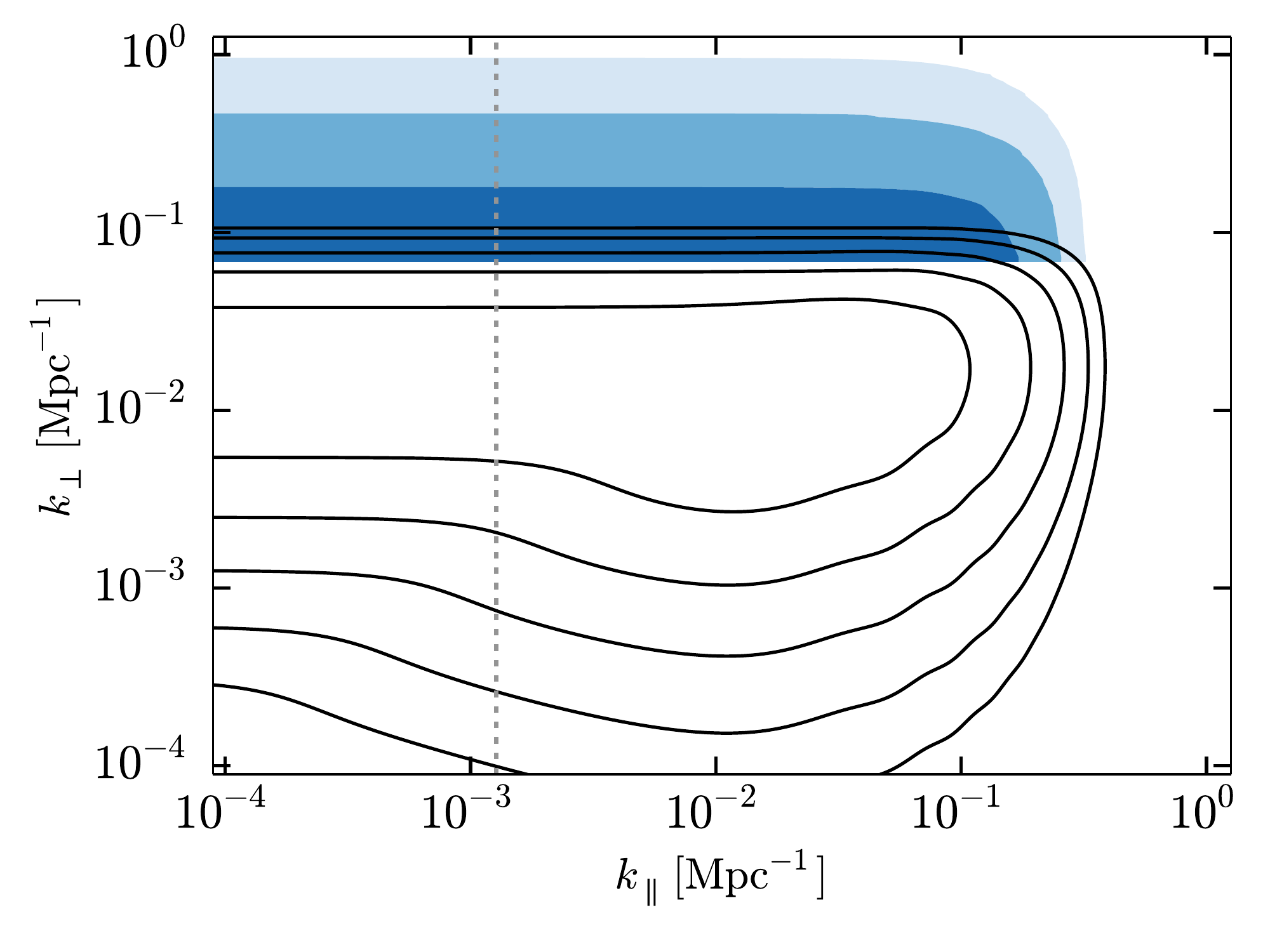}
\caption{{\corr The normalised effective volume $V_\mathrm{eff}(k_\perp, k_\parallel)/V_\mathrm{bin}$ at $z \approx 1$, for SKA1-MID Band 1 in single-dish mode (black contours) and interferometer mode (shaded blue contours). Foregrounds have not been included, but the effective minimum $k_\parallel$, given by $k_{\rm FG} \simeq k_{\rm BW}$, is shown as a dashed grey line. The contours are for values $[0.9, 0.5, 0.1, 0.01, 0.001]$, where $1.0$ is the maximum (implying a cosmic variance-limited measurement). Only the last three contours (i.e. $<0.5$) appear for the interferometer mode due to its lower sensitivity.}}
\label{fig-Veff}
\end{figure}

The main quantities that describe a redshift survey are the survey area, $S_{\rm area}$, and the number density of sources as a function of redshift, ${\bar n}(z)$, which in turn allows us to define a survey depth. As with the IM survey, we can define the survey volume, $V_{\rm bin}\simeq r^2({\bar z}) (dr/dz) \Delta z \,S_{\rm area}$ (where ${\bar z}$ is the mean redshift of the survey). The Fisher matrix is then \citep{1994ApJ...426...23F,1998ApJ...499..555T}
\bea
F^{\rm gal}_{ij} &=& \frac{1}{2}\int_{k_\mathrm{min}}^{k_\mathrm{max}}\frac{ d^3k}{(2\pi)^3}[\partial_i \ln P_{\rm tot}({\bf k})\partial_j \ln P_{\rm tot}({\bf k})]V_{\rm eff}({\bf k}) \nonumber\\
V_{\rm eff} &=& V_{\rm bin}\left[\frac{{\bar n}(z)P_{\rm tot}({\bf k})}{1+{\bar n}(z)
P_{\rm tot}({\bf k})}\right]^2. \label{fisherz}\label{eqn-Veff}
\eea
Shot noise plays a crucial role, dominating if ${\bar n}$ is too small; as does cosmic variance, via the $V_{\rm bin}$ term. The effective volume tells us how well different parts of Fourier space are sampled. Applying (\ref{fisherz}) to the binned power spectrum, we recover the well-known result
\bea
\left \langle\left(\frac{\Delta P_a}{P_a}\right)^2\right \rangle=\left[
\frac{1}{2} \int_{V_n} \frac{d^3k}{(2\pi)^3}V_{\rm eff}({\bf k})\right]^{-1}, \label{dPnz}
\eea
which has $V_{\rm eff}$ at the heart of the expression. It is only in regions where ${\bar n}P\gg 1$ that the power spectrum can be measured well. In this regime the fundamental limitation becomes cosmic variance, which is set by the number of modes sampled in each bin, $N_a\sim V_{\rm bin} k_a^2\Delta k_a / 2\pi^2$, where $\Delta k_a$ is the width of the corresponding bin.

The equivalent expression for an IM survey is
\bea
\left(\frac{\Delta P_a}{P_a}\right)^2=\left[
\frac{1}{2}U_{\rm bin}\int_{V_n} \frac{d^2q \, dy}{(2\pi)^3}\left(\frac{C^S({\bf q},y)}
{C^T({\bf q},y)}\right)^2\right]^{-1} \label{deltap}
\eea
By analogy with (\ref{eqn-Veff}) and (\ref{dPnz}), we can define an effective volume, $V^{\rm IM}_{\rm eff}$, and a pseudo number density, ${\bar n}^{\rm IM}P(\mathbf{k})\equiv C^S/C^N$, such that
\bea
{\bar n}^{\rm IM}(z,{\bf k}) = \left(\frac{T_b}{T_{\rm sys}}\right)^2 \frac{ t_{\rm tot} \Delta \nu}{U_{\rm bin}} B^2_\perp B_\parallel {\cal I}^{-1}. \label{eqn-nim}
\eea
The foreground term has been left out for the time being. With (\ref{eqn-nim}) in hand, we are now in a position to exploit the analogy with conventional redshift surveys to better understand the properties of IM experiments. An illustration of the various scales relevant to single-dish and interferometric experiments is shown in Fig. \ref{fig-pk-direction}, and an example of $V^\mathrm{IM}_\mathrm{eff}$ is shown in Fig. \ref{fig-Veff}.

The first thing to notice is that $V^{\rm IM}_{\rm eff}$ is highly anisotropic. In the case of a single dish, information on angular scales smaller than the instrumental beam is washed out; for dishes of diameter $D_\mathrm{dish}$, scales $k_\perp \gtrsim D_\mathrm{dish} / r \lambda$ are suppressed. For an interferometer, it is possible to probe much smaller angular scales (up to $k_\perp \sim 2 \pi u_\mathrm{max}/r$), although the transverse Fourier plane will be sampled much less homogeneously than for a single dish, depending on the instrument's $uv$ coverage.

Along the radial direction, we expect foreground subtraction to throw away information on scales of order the total bandwidth, $k_\parallel \lesssim k_\mathrm{FG}$, and on smaller scales non-linear velocities smear out information for $k \gtrsim 0.15$ Mpc$^{-1}$. As discussed in Section \ref{sect-cn}, the channel bandwidth also imposes an effective radial beam, although this is generally at higher $k_\parallel$ than the non-linear cutoff.

Using Eq. (\ref{deltap}), we can now piece together how well the power spectrum can be measured by various experiments. On large scales, cosmic variance and the survey size are the limiting factor. For a single-dish experiment with a relatively isotropic survey volume, we can sample the longest wavelength modes in the radial and transverse directions equally well, so that $({\Delta P}/{P})^2 \propto k^{-3}$. On smaller scales, the beam size will severely limit how small a transverse scale we can probe, so only radial modes will be properly sampled, implying $({\Delta P}/{P})^2 \propto k^{-1}$. In the radial direction, we will eventually come up against the non-linear velocity scale, preventing us from extracting information on scales smaller than $\sigma_\mathrm{NL}$.

For an interferometric experiment, the situation is reversed, and is in some sense complementary, as shown in Fig. \ref{fig-pk-direction}. With sufficiently long baselines, it is possible to probe very small angular scales. We then expect to have $({\Delta P}/{P})^2 \propto k^{-3}$ up until the non-linear scale is reached in the radial direction, beyond which only transverse scales contribute, such that $({\Delta P}/{P})^2 \propto k^{-2}$ until the maximum transverse resolution is hit. Conversely, interferometers are fundamentally unable to probe any scale larger than that corresponding to their minimum baseline (which for a dense array is roughly the dish diameter, which gives the beam size in single-dish mode).

\begin{figure}[t]
\includegraphics[width=\columnwidth]{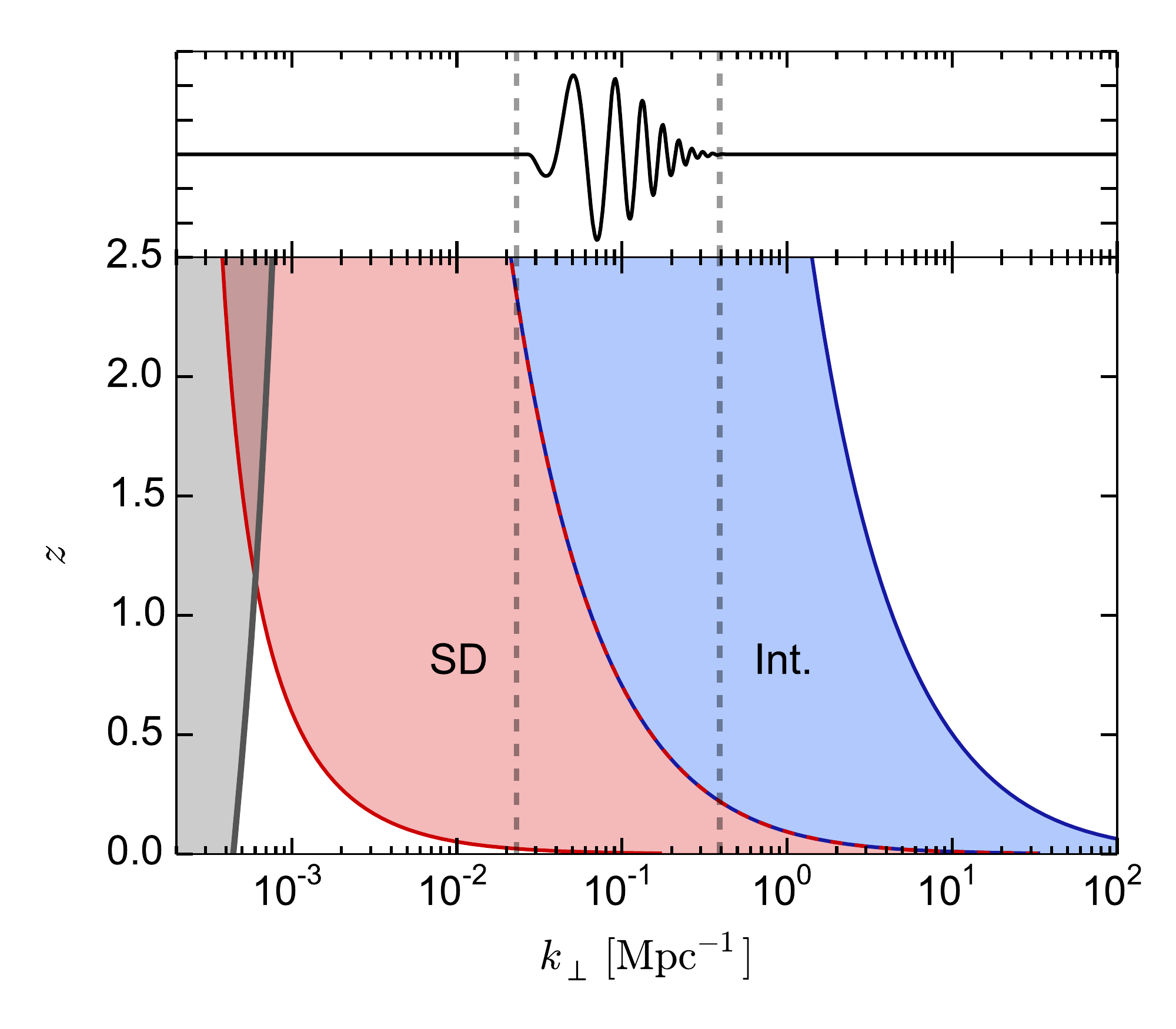}
\vspace{-2em}\caption{Redshift evolution of the minimum/maximum transverse scales (filled regions) for illustrative interferometer (blue) and single-dish (red) experiments. The BAO are plotted for comparison. The dishes have diameter $D_\mathrm{dish} = 15$m, the min./max. interferometer baselines are $D_\mathrm{min} = 15$m and $D_\mathrm{max} = 1000$m, and the survey has bandwidth $\Delta \nu = 600$ MHz and area $S_\mathrm{area} = 25,000$ sq. deg. The shaded grey region denotes superhorizon scales, $k < k_H = 2 \pi / r_H$.}
\vspace{-1em}
\label{fig-resolution-z}
\end{figure}

In summary, the important scales for IM are:
\vspace{-0.7em}\[\arraycolsep=2pt\def\arraystretch{1.8} \begin{array}{ccll}
k^\mathrm{min}_\parallel &\sim& k_\mathrm{FG} = 2 \pi / (r_\nu \Delta{\tilde \nu}_{\rm tot}) & \\
k^\mathrm{max}_\parallel &\sim& k_\mathrm{NL} = 1 / \sigma_\mathrm{NL} & \\
k^\mathrm{min}_\perp &\sim& k_\mathrm{area} = 2\pi / \sqrt{r^2 S_{\rm area}} & \;\;\mathrm{(single ~dish)} \\
 &\sim& k_{D_\mathrm{min}} = 2 \pi D_\mathrm{min} / r \lambda & \;\;\mathrm{(interferom.)} \\
k^\mathrm{max}_\perp &\sim& k_\mathrm{FOV} = 2 \pi D_{\rm dish} / r \lambda & \;\;\mathrm{(single ~dish)} \\
 &\sim& k_{D_\mathrm{max}} = 2 \pi D_\mathrm{max} / r \lambda & \;\;\mathrm{(interferom.)} \nonumber
\end{array} \]
The redshift dependence of the transverse scales for an example setup is shown in Fig. \ref{fig-resolution-z}. Note that these are only the minimum/maximum scales that can be probed {\it in principle} by a given instrument -- the sensitivity to scales within these ranges will vary, so it may not be possible to constrain the more extreme scales in practise.

\section{Expansion, growth, and the acoustic peak}

We begin our exploration of the capabilities of IM experiments by focusing on a few key observables. These variously constrain the growth of large-scale structure and the expansion and geometry of the Universe: The positions of the acoustic peaks and the overall shape of the power spectrum in both the radial and transverse directions can be used as distance indicators to place constraints on $D_A(z)$ and $H(z)$, and redshift space distortions make it possible to measure the growth rate, $f(z)$.

In addition to being of intrinsic cosmological interest, these also serve as useful models for other observables. For example, the detection of the acoustic peaks is a comparable problem to measuring other `shape' features of the power spectrum, such as scale-dependent bias. We will therefore devote this section to understanding the detailed characteristics of the measurements on these observables that can be made with IM experiments. Throughout, we will forecast for the following parameter set (without any external priors):
\be
\{ A(z), [b_\mathrm{HI} \sigma_8](z), [f \sigma_8](z), D_A(z), H(z), \sigma_\mathrm{NL} \} \nonumber .
\ee

\begin{figure}[t]
\includegraphics[width=\columnwidth]{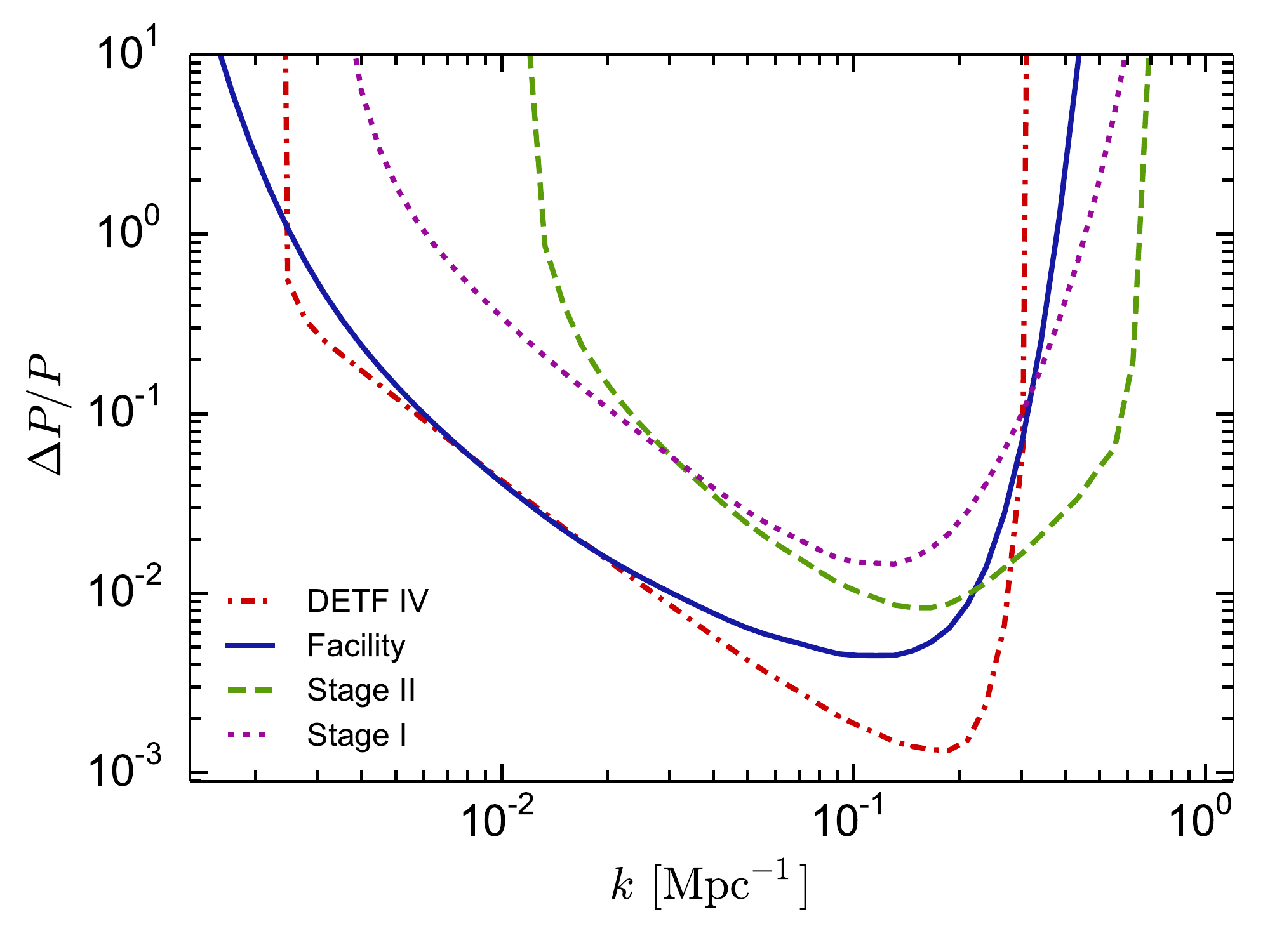}
\caption{Fractional constraints on $P(k)$ for the set of reference experiments, combined over the whole redshift range of each experiment, with 20 bins per decade in $k$.}
\label{fig-pk-constraints}
\end{figure}

\subsection{Detectability of Baryon Acoustic Oscillations}

\begin{figure}[t]
\vspace{-3em}\includegraphics[width=\columnwidth]{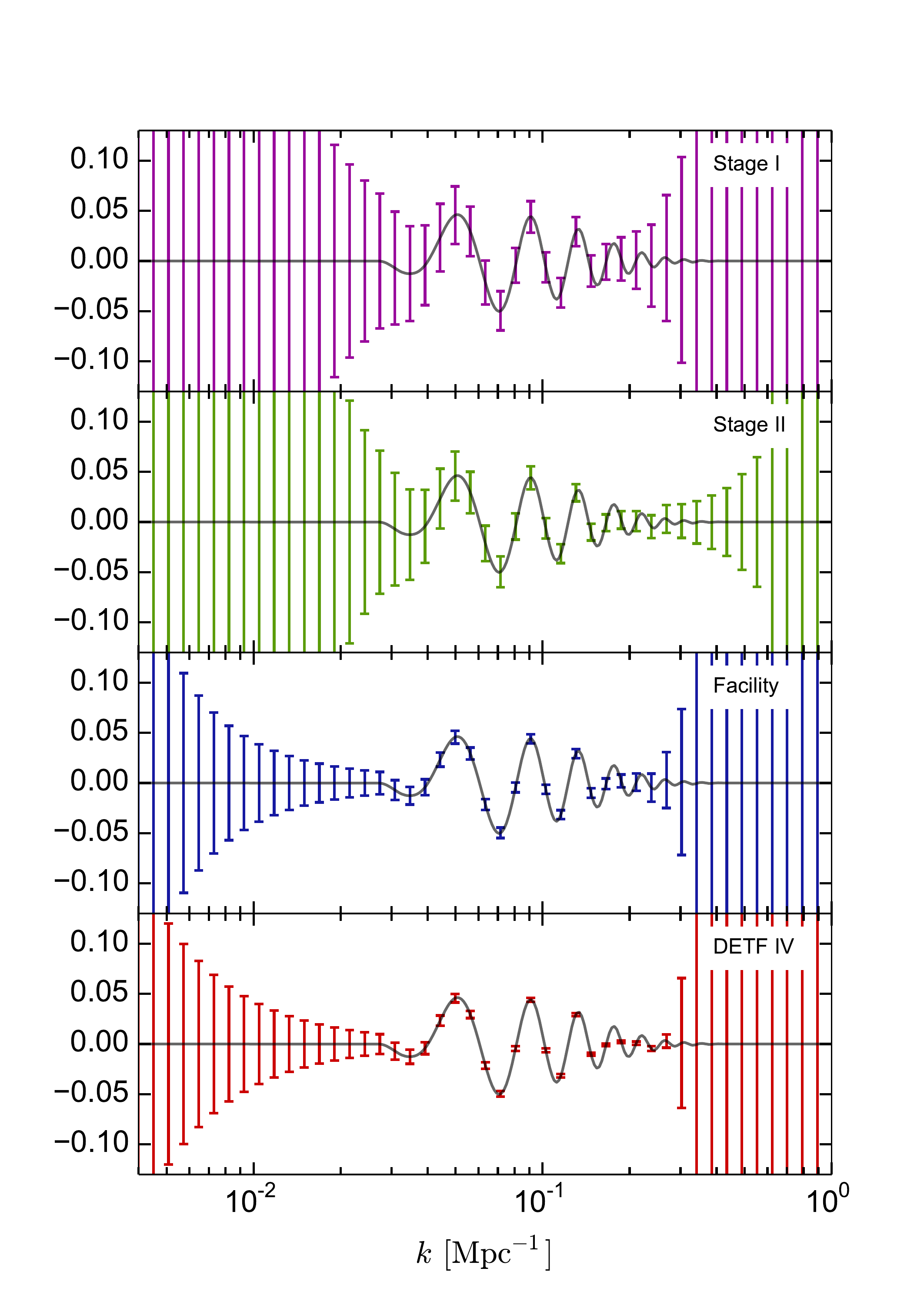}
\caption{Forecast constraints on the BAO wiggles, combined over the whole redshift range for each of the reference surveys.}
\label{fig-fbao}
\end{figure}

The Baryon Acoustic Oscillations are a `statistical standard ruler' that forms the primary distance measure in surveys of large-scale structure. We can get an idea of the detectability of the BAOs by looking at the fractional errors on $P(k)$, using Eq. (\ref{deltap}). These are shown for the reference surveys in Fig. \ref{fig-pk-constraints}, and are overplotted on the BAO wiggle function (to be defined shortly) in Fig. \ref{fig-fbao}. All three IM surveys are capable of strongly detecting the BAO feature when the constraints are combined over their full redshift ranges. \StageThree\ approaches the cosmic variance limit (represented by the DETF Stage IV survey out to $k \sim 0.1 \,\mathrm{Mpc}^{-1}$) over a substantial fraction of the scales relevant to the BAO, mostly due to the sensitivity of its single-dish component. This also helps to put sub-10\% level constraints on the power spectrum on scales slightly larger than the matter-radiation equality peak, $k_\mathrm{eq}\approx 10^{-2}$ Mpc$^{-1}$. Its interferometric component provides constraints on smaller scales, achieving $\sim 10\%$ errors on $P(k)$ out to $k \approx 1 \,\mathrm{Mpc}^{-1}$.

\begin{figure*}[t]
\centering{
\vspace{-1em}\includegraphics[width=1.9\columnwidth]{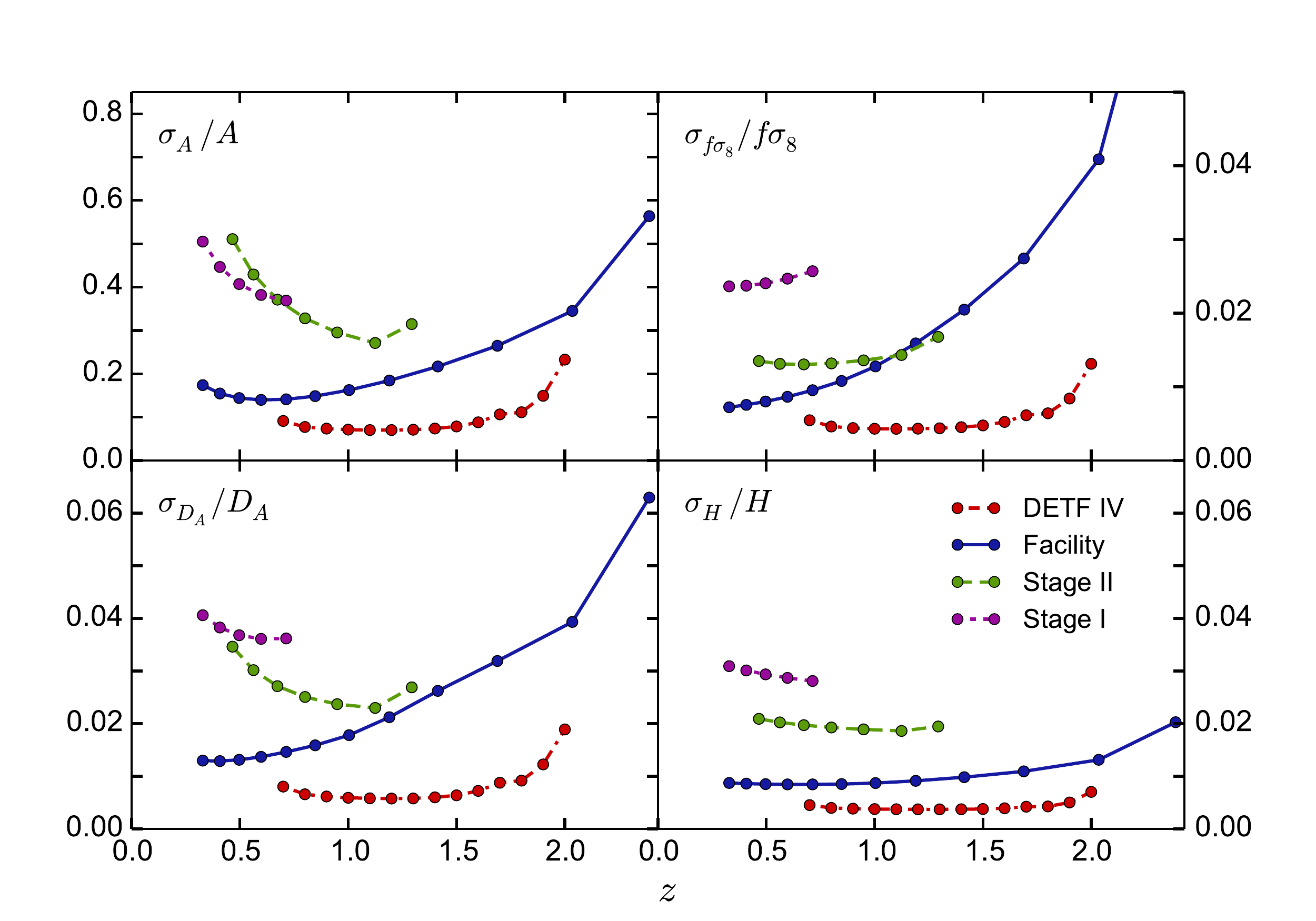} }
\caption{Fractional errors on $A(z)$, $f \sigma_8(z)$, $D_A(z)$, and $H(z)$, as a function of redshift.}
\label{fig-redshift-fns}
\end{figure*}

The interferometric \StageTwo\ survey is sensitive to generally smaller scales, but still achieves good constraints on the BAO thanks to its coverage out to intermediate redshifts ($z \sim 1.4$). The \StageOne\ survey can comfortably detect the BAO despite its significantly lower sensitivity than \StageThree, but leaves smaller scales unconstrained.

Alternatively, one can look at the detectability of the BAO feature as a whole. We follow a similar approach to \citep{2003ApJ...594..665B} and split the matter power spectrum, $P(k)$, into a `smooth' part, $P_{\rm smooth}(k)$, and an oscillatory part,
\bea
f_{\rm bao}(k) = \frac{P(k) - P_{\rm smooth}(k)}{P_{\rm smooth}(k)}.
\eea
We then introduce an amplitude parameter, $A$, such that
\be
P(k) = [1 + A f_{\rm bao}(k)] P_{\rm smooth}(k). \label{eqn-Abao}
\ee
Constraints on $A$ therefore give a measure of the detectability of the BAO feature.

The splitting of $P(k)$ between smooth and oscillatory parts is somewhat arbitrary. We attempt to construct a `purely oscillatory' $f_\mathrm{bao}(k)$ -- i.e. one that lacks a smooth overall trend in $k$ -- as follows. First, we use CAMB to calculate $P(k)$ for the fiducial cosmological model over a range of sample points in $k$. We then choose two reference values of $k$ that bound the region in which the oscillations are significant ($k \approx 0.02$ and 0.45 Mpc${}^{-1}$ for our fiducial cosmology), and construct a cubic spline for $\log P(k)$ as a function of $\log k$ using all points {\it outside} that region. Next, we construct a preliminary oscillatory function by dividing the sampled $P(k)$ by the splined function (not its logarithm), then fit another cubic spline to the result and find the zeros of its second derivative with respect to $k$. These are the points at which the first derivatives of the oscillatory function are maximal/minimal, and in some sense define `mid-points' of the function -- its overall trend. We construct a cubic spline through these too, and then divide the preliminary oscillatory function by it to `de-trend'. This leaves $f_\mathrm{bao}(k)$ as the final result (Fig. \ref{fig-fbao}). Unlike other methods, which look at ratios of the form $P(k, \Omega_b \!\neq\! 0) / P(k, \Omega_b\!=\!0)$ to pick out oscillations \citep{Rassat:2008ja}, this method is essentially model-independent for a given fiducial $P(k)$.

The constraint on the overall amplitude of the BAO feature, $A$, is plotted as a function of redshift for the reference surveys in Fig. \ref{fig-redshift-fns}. \StageThree\ is capable of $> 3 \sigma$ detections of the BAO feature out to $z \approx 1.5$, but makes progressively weaker detections at higher redshift, predominantly due to its limited angular resolution in single-dish mode. In comparison, the \StageTwo\ survey's constraints degrade much less rapidly with redshift, owing to its greater sensitivity to smaller angular scales (which translate to intermediate physical scales at higher $z$).

\begin{figure}[t]
\includegraphics[width=\columnwidth]{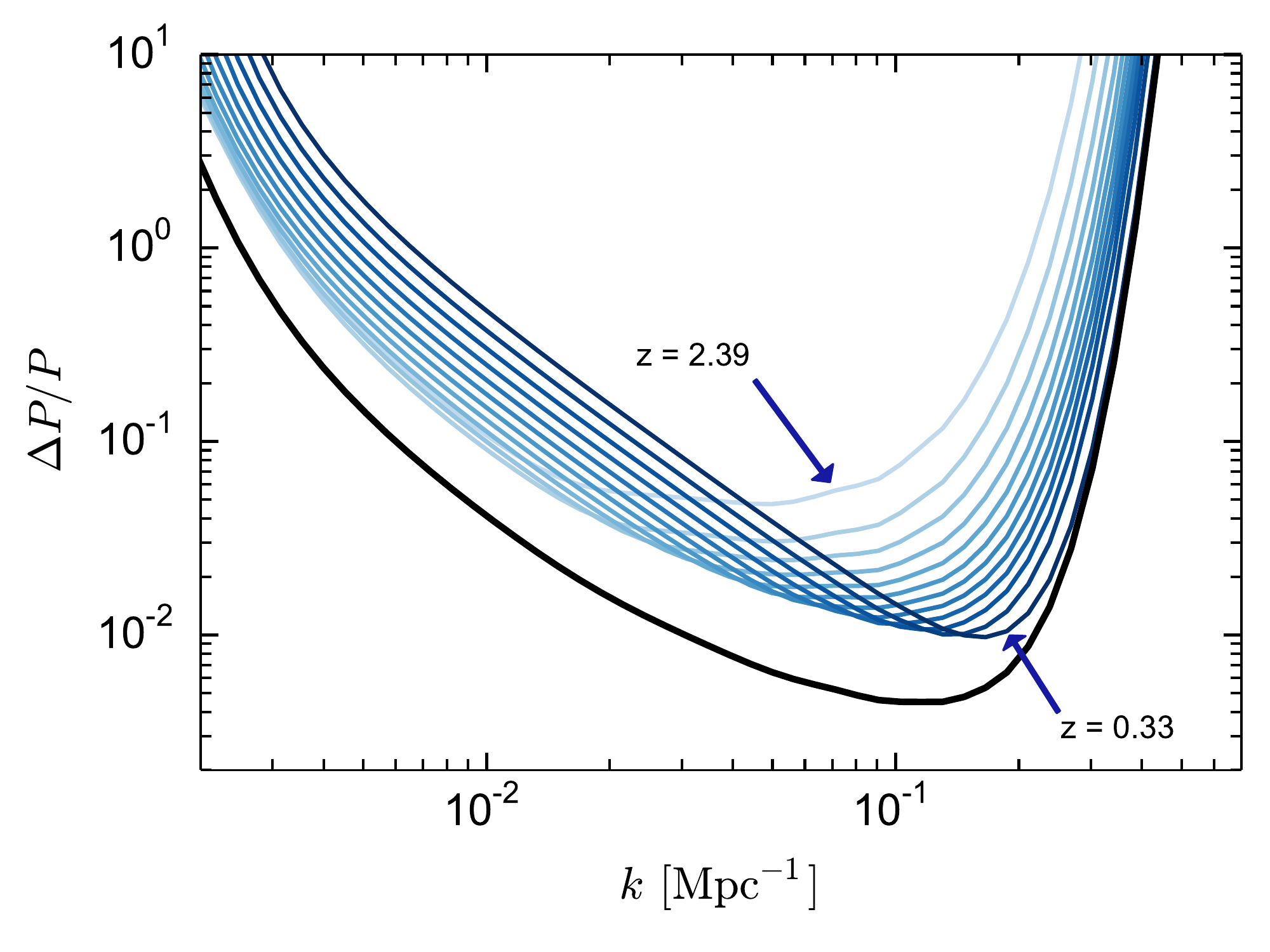}
\caption{Fractional constraints on $P(k)$ in each redshift bin, for the \StageThree\ experiment. The thick black line is the total constraint, summed over all redshift bins.}
\label{fig-pk-zbins}
\end{figure}

Fig. \ref{fig-pk-zbins} plots the errors on $P(k)$ for \StageThree\ as a function of both scale and redshift. For $k \gtrsim 0.1 \, \mathrm{Mpc}^{-1}$, most of the information comes from low redshifts, where the amplitude of the power spectrum is largest. At smaller $k$, however, the volume of the redshift bin begins to matter, as the increase in bin volume with $z$ allows progressively larger scales to be probed. For \StageThree, the constraints on intermediate BAO scales ($k \sim 0.07\, \mathrm{Mpc}^{-1}$) come from a mixture of low and intermediate redshift bins, with the high redshift bins taking over on larger scales.

\subsection{Constraints on $D_A(z)$ and $H(z)$} \label{sect-dahz}

The angular diameter distance, $D_A(z)$, and the expansion rate, $H(z)$, are measures of distance in the transverse and radial directions respectively. To include them in our forecasts, we introduce \citep{2003ApJ...594..665B}
\bea
\alpha_\perp &\equiv& \frac{r^\mathrm{fid}}{r} = \frac{D^\mathrm{fid}_A(z)}{D_A(z)} \\  
\alpha_\parallel &\equiv& \frac{r^\mathrm{fid}_\nu}{r_\nu} = \frac{H(z)}{H^\mathrm{fid}(z)}
\eea
where $r^\mathrm{fid}$ and $r^\mathrm{fid}_\nu$ are the fiducial $\Lambda$CDM values of $r$ and $r_{\nu}$ at a given redshift. We then replace $q\rightarrow \alpha_\perp q$ and $y\rightarrow \alpha_\parallel y$ in Eq. (\ref{sigcov}) to get
\bea
C^S({\bf q},y) &=& T^2_b \frac{\alpha_\perp^2\alpha_\parallel}{r^2 r_\nu}  F_\mathrm{RSD}\left ( \frac{\alpha_\perp \bf q}{r},\frac{\alpha_\parallel y}{ r_\nu} \right ) D^2(z) \nonumber \\
 &\times& P \left ( k = \sqrt{ \left ( \frac{\alpha_\perp {\bf q}}{r}\right )^2 + \left ( \frac{\alpha_\parallel y}{r_\nu} \right )^2 } \right ). \label{eqn:cs}
\eea

The distance measures enter this expression in four places: (a) an overall factor of $\alpha_\perp^2\alpha_\parallel$, related to the physical volume of the survey; (b) a distortion of the angular ($\mu$) dependence of the RSDs; (c) a shift in the non-linear cutoff scale of the RSDs; and (d) a shift of the whole isotropic power spectrum, $P(k)$, that can be further subdivided into corresponding shifts in the BAO feature and smooth power spectrum. The latter two, (c) and (d), are both due to the remapping of $k$, shown explicitly in the argument of $P(k)$ in Eq. (\ref{eqn:cs}) \citep{2011MNRAS.410.1993S}.

Due to modelling uncertainties and degeneracies with other parameters, it is not necessarily desirable to use all of these terms to measure distances from real data. The BAO feature is the standard choice of distance measure, owing to its robustness to systematic error; the acoustic scale shifts only slightly when non-linearities are introduced \citep{Smith:2007gi, Crocce:2007dt}, and smooth variations such as scale-dependent bias also have a relatively minor impact \citep{Zhang:2008ia} (although corrections must still be made for precision measurements). Anisotropies of the correlation function can also be used to measure distances \citep{Alcock:1979mp, Kaiser:1987qv}, through redshift space distortions (b) and the Alcock-Paczynski effect (a, c, and d), although these are more sensitive to the detailed modelling of the power spectrum \citep{Reid:2012sw}. Thus, a particularly conservative analysis might only derive distances from the BAO, and discard information from the other terms.

\begin{figure}[t]
\vspace{-0.5em}
\includegraphics[width=\columnwidth]{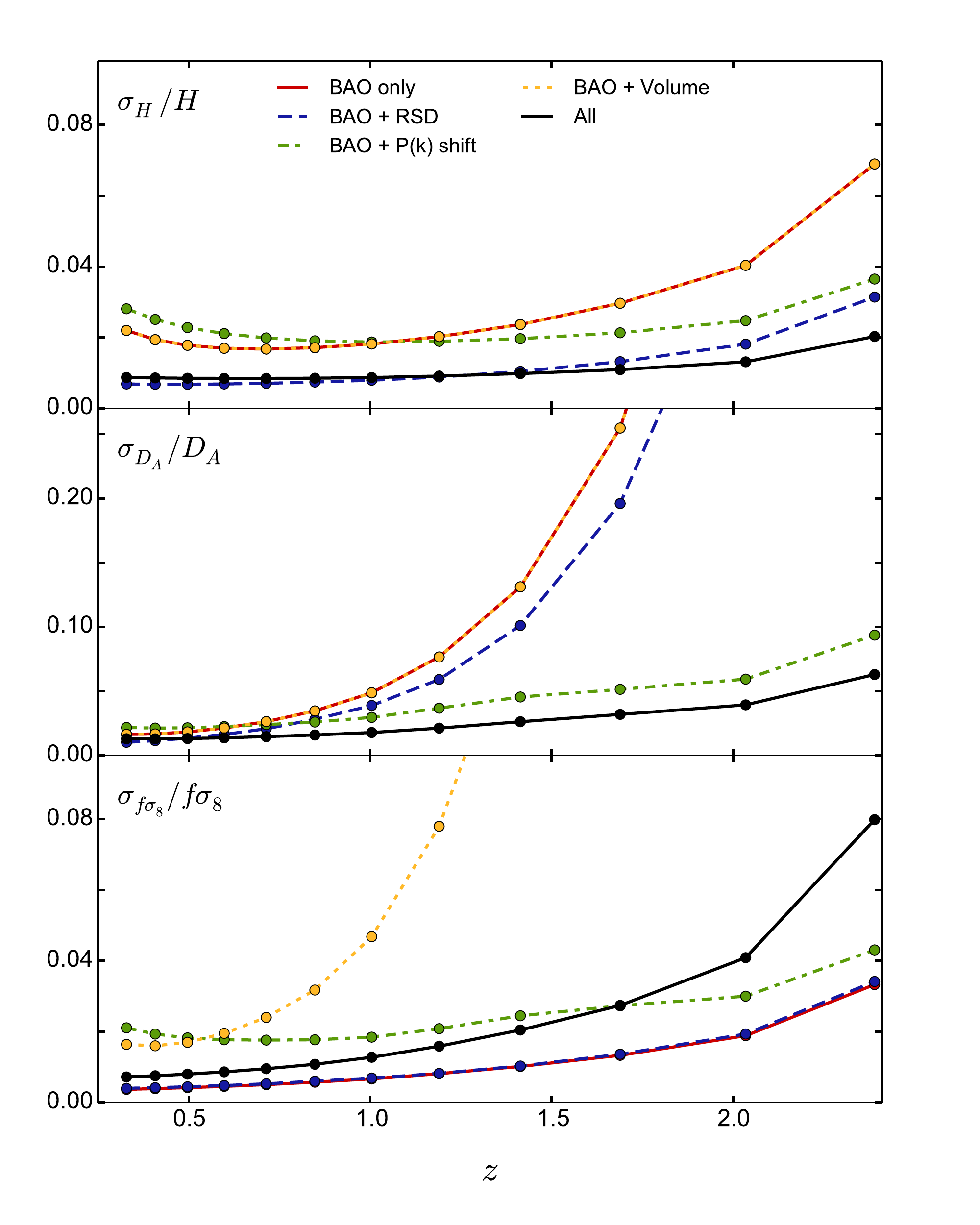}
\caption{Fractional errors on $H(z)$, $D_A(z)$, and $f\sigma_8(z)$ for the \StageThree\ experiment, for various combinations of distance terms being switched on and off in the Fisher matrix calculation. The RSDs are taken to include both the angle-dependence and non-linear cut-off terms (see text).}
\label{fig-distance-terms}
\end{figure}

Fig. \ref{fig-distance-terms} shows the effect of neglecting some of the distance terms for the \StageThree\ experiment. By using only the BAO, one is discarding a substantial amount of useful information, as shown by the comparatively poor constraints on $D_A$ and $H$. This is partially compensated by the reduced risk of systematic error, and the improved growth rate constraints that are due to weaker degeneracies with other parameters when compared to the other distance terms. Including broadband information -- i.e. the shift in (smooth) $P(k)$ -- reduces the error on $D_A$ by a factor of $2-5$ over the entire redshift range, and is especially beneficial at higher $z$, where the BAO-only constraints degrade rapidly due to the limited angular sensitivity of the telescope. Adding the RSD terms helps to distinguish between the radial and transverse directions, which also reduces the error on $H(z)$.

Disregarding any of the distance terms can also substantially alter the correlation structure of the Fisher matrix, which has a knock-on effect on constraints for other parameters. This can be seen clearly for $f(z)$ in Fig. \ref{fig-distance-terms}, which has a substantial scatter in fractional error depending on which distance measures are used. To most accurately reflect the interdependencies of the various cosmological parameters, we will use all of the distance measure terms in what follows. The increased uncertainty that comes from marginalising over nuisance parameters such as $\sigma_\mathrm{NL}$ helps compensate for the increased risk of systematic error that attends some of the measures, which means that this is not too optimistic of us.

Fig. \ref{fig-redshift-fns} shows the fractional constraints that can be achieved on $D_A(z)$ and $H(z)$ for the full set of reference surveys. \StageThree\ measures the expansion rate to better than {\corr 1\% out to $z\approx 1.7$, and stays within 2\% as far out as $z=2.5$.} This is roughly a factor of two worse than the DETF Stage IV reference survey, although the redshift range covered by \StageThree\ is significantly larger. \StageTwo\ also {\corr obtains $\sim \! 2\%$} constraints across its full redshift range, {\corr and \StageOne\ hovers around 3\%.}

The picture is somewhat different for $D_A(z)$. The fractional errors for \StageThree\ increase from $\sim 1.5\%$ at $z \approx 0.4$ up to $4\%$ at $z=2$, compared with the relatively flat errors for the galaxy survey that {\corr remain below $1\%$} for most of the redshift range. \StageTwo's errors are also {\corr relatively flat as a function of $z$, staying at around the $2.5\%$ mark.}

The limited angular resolution of the single-dish experiments is the cause of this behaviour. $H(z)$ is most sensitive to the resolution in the radial (frequency) direction, which is essentially the same for all experiments and does not evolve appreciably with redshift (being set by the non-linear scale rather than the channel bandwidth, as discussed in Sect. \ref{sect-zsurvey}). The $D_A(z)$ constraints depend more on the sensitivity to transverse physical scales, however, which differs between interferometers and single-dish experiments. For single-dish, $k^\mathrm{max}_\perp$ tends to be relatively small even at $z=0$ for moderately-sized dishes, and continues to decrease (shift to larger scales) as $z$ increases, as shown in Fig. \ref{fig-resolution-z}. As this happens, useful distance information from smaller scales is lost. The same happens for interferometers, but $k^\mathrm{max}_\perp$ is typically much larger, so the most useful transverse scales remain resolved. In fact, since $k^\mathrm{min}_\perp$ is also decreasing, additional distance information becomes available from larger scales.

It is worth noting that the effect would be different if only the BAO were being used as distance measures -- both the $D_A(z)$ {\it and} $H(z)$ constraints would be affected by the loss of resolution in the transverse direction \citep{2013MNRAS.434.2008S}. To see this, consider a simplified model of the correlation function consisting of the sum of a smooth component and a feature, $\xi_{\rm BAO}(r)=A\exp[-(r-r_{\rm BAO})^2/2\sigma^2]$. The response to the loss of resolution along the transverse direction corresponds to convolving the correlation function with a window function such that
\bea
{\tilde \xi}_{\rm BAO}(r_\parallel,{\bm r_\perp})=\int d^2r'_\perp W({\bm r}_\perp-{\bm r}'_\perp){\xi}_{\rm BAO}\left(\sqrt{r^2_\parallel+r^{'2}_\perp}\right) \nonumber
\eea
which can be rearranged to have the form
\bea
{\tilde \xi}_{\rm BAO}(r_\parallel, {\bm r}_\perp)=\int d^2\Delta W(\Delta){\xi}_{\rm BAO}\left(\sqrt{r^2+2{\bm \Delta}\cdot {\bm r}_\perp+\Delta^2}\right). \nonumber
\eea
The smoothed correlation function remains a function of $r$ and $r_\perp$, which means that the degradation of signal due to the smoothing is almost democratically taken up by both the transverse and parallel directions. This can be seen explicitly in Fig. \ref{fig-distance-terms} for the BAO-only curves for $D_A(z)$ and $H(z)$, which both show a much stronger evolution with redshift than any other combination of distance measures. The inclusion of additional distance measures is therefore necessary to help limit the effect of the poor angular resolution of single-dish experiments.

\begin{figure}[t]
\includegraphics[width=\columnwidth]{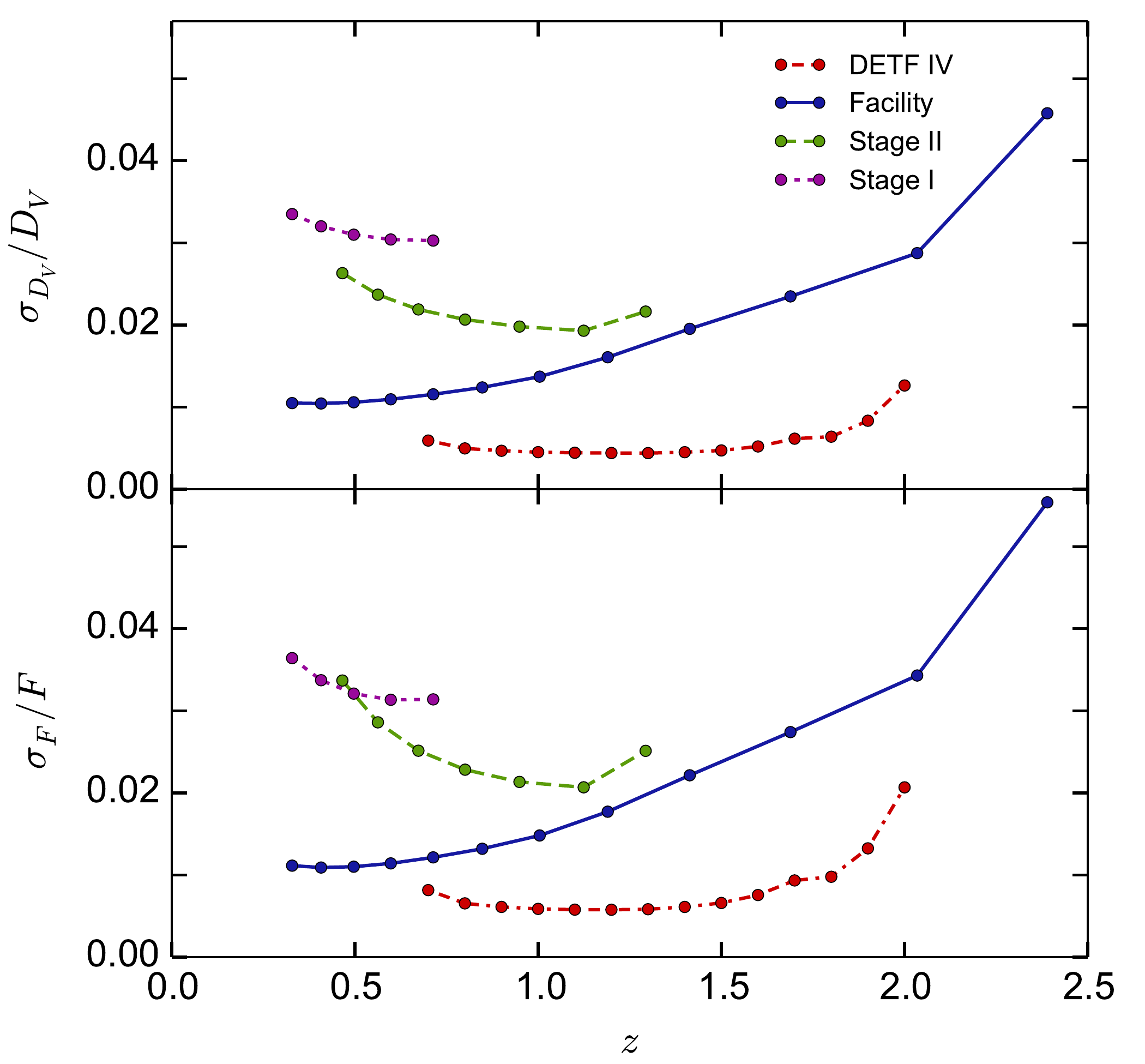}
\caption{Fractional errors on the volume distance, $D_V(z)$, and Alcock-Paczynski distortion, $F(z)$, for our reference surveys.}
\label{fig-lss-distances}
\end{figure}

The various distance measures for LSS surveys depend on combinations of $D_A$ and $H$ rather than constraining them individually. For example, moments of the correlation function \citep{Reid:2012sw} give the volume distance and Alcock-Paczynski terms,
\bea
D_V(z) &=& \left [ (1+z)^2 D_A^2 \frac{c z}{H(z)} \right ]^\frac{1}{3} \\
F(z) &=& (1+z) D_A(z) H(z) / c.
\eea
In some sense, these quantities define redshift-dependent figures of merit -- many other studies forecast directly in terms of $D_V$, for example, and surveys are compared in terms of the errors that they can achieve on this parameter at a given redshift. We have therefore presented results for both $D_A$ and $H$ (Fig. \ref{fig-redshift-fns}), and $D_V$ and $F$ (Fig. \ref{fig-lss-distances}) to facilitate comparison with previous studies.

\subsection{Constraints on the growth rate, $f(z)$} \label{sect-f}

The other key observable provided by LSS surveys is the growth rate, which can be measured from the anisotropy of the correlation function in redshift space.

The growth rate has two major roles: first, as another measure of distance, since it can be related to the expansion rate (albeit in a model-dependent way); and second, as a non-geometric probe of gravity over cosmic time. The former is useful for helping to break parameter degeneracies that can crop up when only geometric distance measures are used. The latter is of importance in distinguishing theories of dark energy and modified gravity \citep{Hall:2012wd, Masui:2009cj}; it is often the case that theories can be made to have the same expansion history, for example, but differ in growth rate. (We will return to this in Sect. \ref{sect-parametrised-growth}.)

\begin{figure}[t]
\includegraphics[width=\columnwidth]{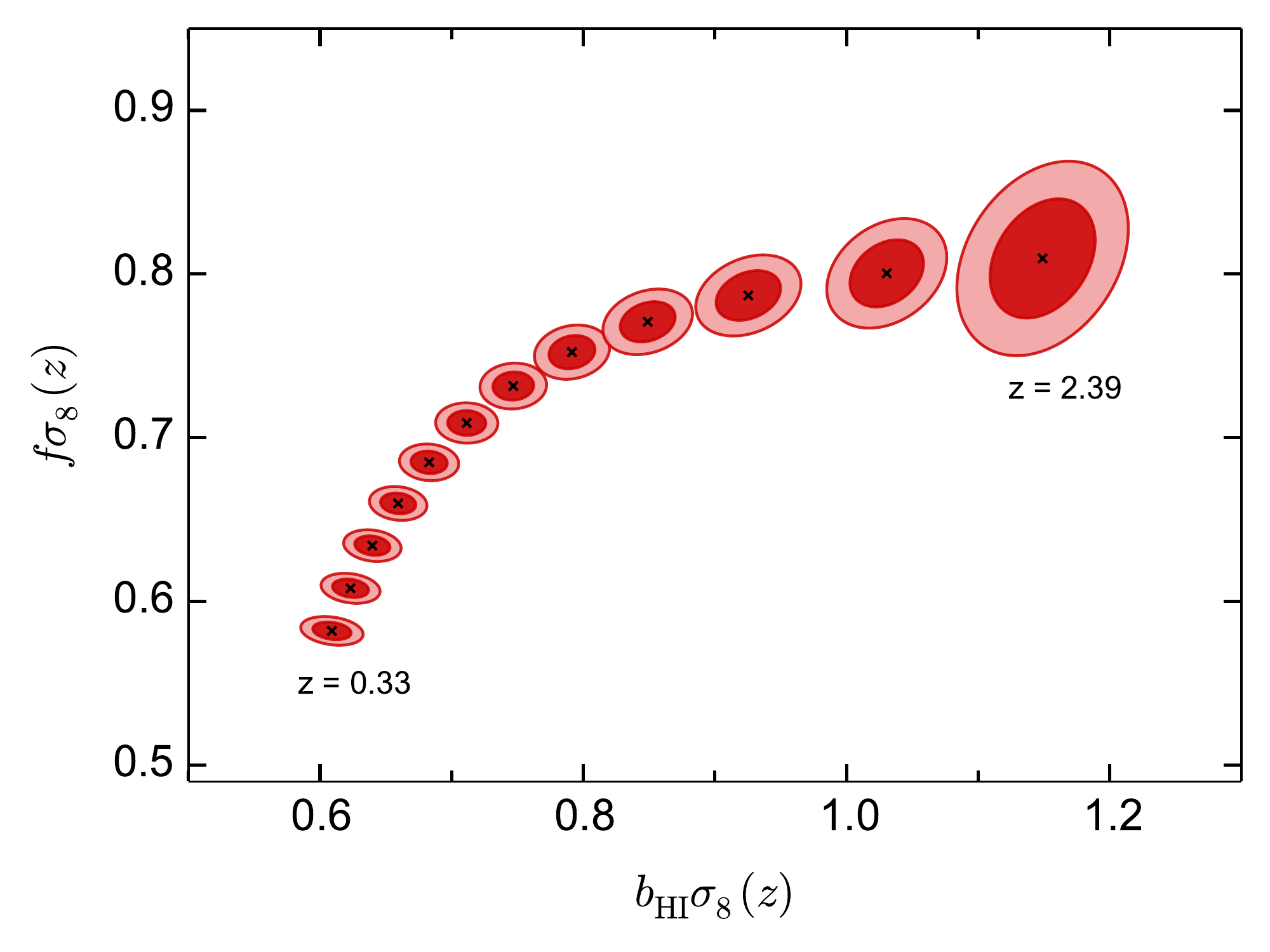}
\caption{Constraints on $b_\mathrm{HI}\sigma_8$ and $f \sigma_8$ as a function of redshift, for the \StageThree\ survey.}
\label{fig-fz-bz}
\vspace{-0.6em}
\end{figure}

In what follows, we will concentrate on the linear growth rate, $f(z)$. HI intensity mapping experiments are also capable of probing non-linear growth, but modelling uncertainties on small scales reduces their usefulness for constraining cosmological parameters. Non-linear effects are discussed in more detail in Sects. \ref{sect-parametrised-growth} and \ref{sect-nonlinear}.

The linear growth rate appears in two places in Eq. \ref{eqn:cs}; explicitly, in the angle-dependent RSD factor, and implicitly, through the linear growth factor, $D(z)$, that gives the redshift evolution of the power spectrum. The two are related by $f(z) = d \log D/d \log a$. Because $D(z)$ appears as an overall factor of the signal covariance, it is degenerate with other parameters, making it hard to measure in an unbiased way. We will assume that $D(z)$ provides no new information on the growth rate here (i.e. we neglect its derivative w.r.t. $f$ in the Fisher matrix).

Various other degeneracies crop up when measuring the growth rate from RSDs. The signal covariance is proportional to $C^S \propto \left [ b_\mathrm{HI}(z) + f(z) \mu^2 \right ]^2 D^2(z) T_b^2(z) \sigma_8^2$.
If no functional form is assumed for any of these terms (i.e. they are left free in each redshift bin), there are only two quantities that can be uniquely distinguished from this expression: an overall amplitude, and a factor of the angle dependence, e.g. $C^S \propto \kappa^2(z) \left [ 1 + \beta(z) \mu^2 \right ]^2$, where $\beta = f(z) / b_\mathrm{HI}(z)$ and $\kappa = \sigma_8 D(z) T_b(z) b_\mathrm{HI}(z)$. Alternatively, one can merge the linear growth factor with the overall normalisation to give a redshift-dependent normalisation, $\sigma_8(z) = \sigma_8 D(z)$, and write the two RSD functions as $T_b f \sigma_8$ and $T_b b_\mathrm{HI} \sigma_8$.

Either way, there are at least three unknowns to be determined from two functions, so it is clear that more information is needed to unpick the degeneracy. The CMB gives a prior on $\sigma_8(z\simeq 1090)$, $D(z)$ can in principle be determined from $f(z)$, and several models for the bias and brightness temperature exist, although there is significant disagreement between them (see Sect. \ref{sect-HI-evol}). The brightness temperature will be measurable from the non-fluctuating part of the HI signal when future IM experiments come online, though, so this can reasonably be taken as a given quantity -- $T_b(z)$ is fixed to its fiducial form throughout this paper. We resolve the remaining degeneracy by treating the combinations $f(z)\sigma_8(z)$ and $b_\mathrm{HI}(z)\sigma_8(z)$ as independent parameters, with both being free functions of redshift. Fig. \ref{fig-fz-bz} shows the joint constraints on them as a function of $z$ for the \StageThree\ survey.

Fig. \ref{fig-redshift-fns} shows the constraints on $f \sigma_8(z)$ that can be achieved by our set of reference surveys. Sub-2\% errors are possible for the \StageThree\ and \StageTwo\ experiments out to a redshift of $z \sim 1.2$, despite (pessimistically) taking the bias to be a completely free function of redshift. As shown in Fig. \ref{fig-distance-terms}, constraints on the growth rate are sensitive to the choice of other distance measures; for \StageThree\ at least, using only BAO to measure distances would result in a $\sim50\%$ reduction in the error on $f\sigma_8(z)$ across the whole redshift range, albeit at the cost of significantly degraded $H(z)$ and $D_A(z)$ measurements.

As shown in Fig. \ref{fig-redshift-fns}, the errors on $f\sigma_8(z)$ for the \StageOne\ and \StageThree\ surveys increase significantly with redshift, while the evolution is less severe for \StageTwo. As with the angular diameter distance (see previous section), this is mostly due to the limited angular resolution of the dish-based surveys.

\section{Cosmological parameters} \label{sect-parameters}

In the previous section, we assessed how well HI intensity mapping experiments will be able to measure the geometry, expansion and growth rate of the Universe. We will now discuss how these map to constraints on the cosmological parameters that characterise the standard $\Lambda$CDM model, including extensions such as a time-varying equation of state of dark energy, non-zero spatial curvature, and a modified growth index.

\begin{figure*}[t]
\includegraphics[width=2.1\columnwidth]{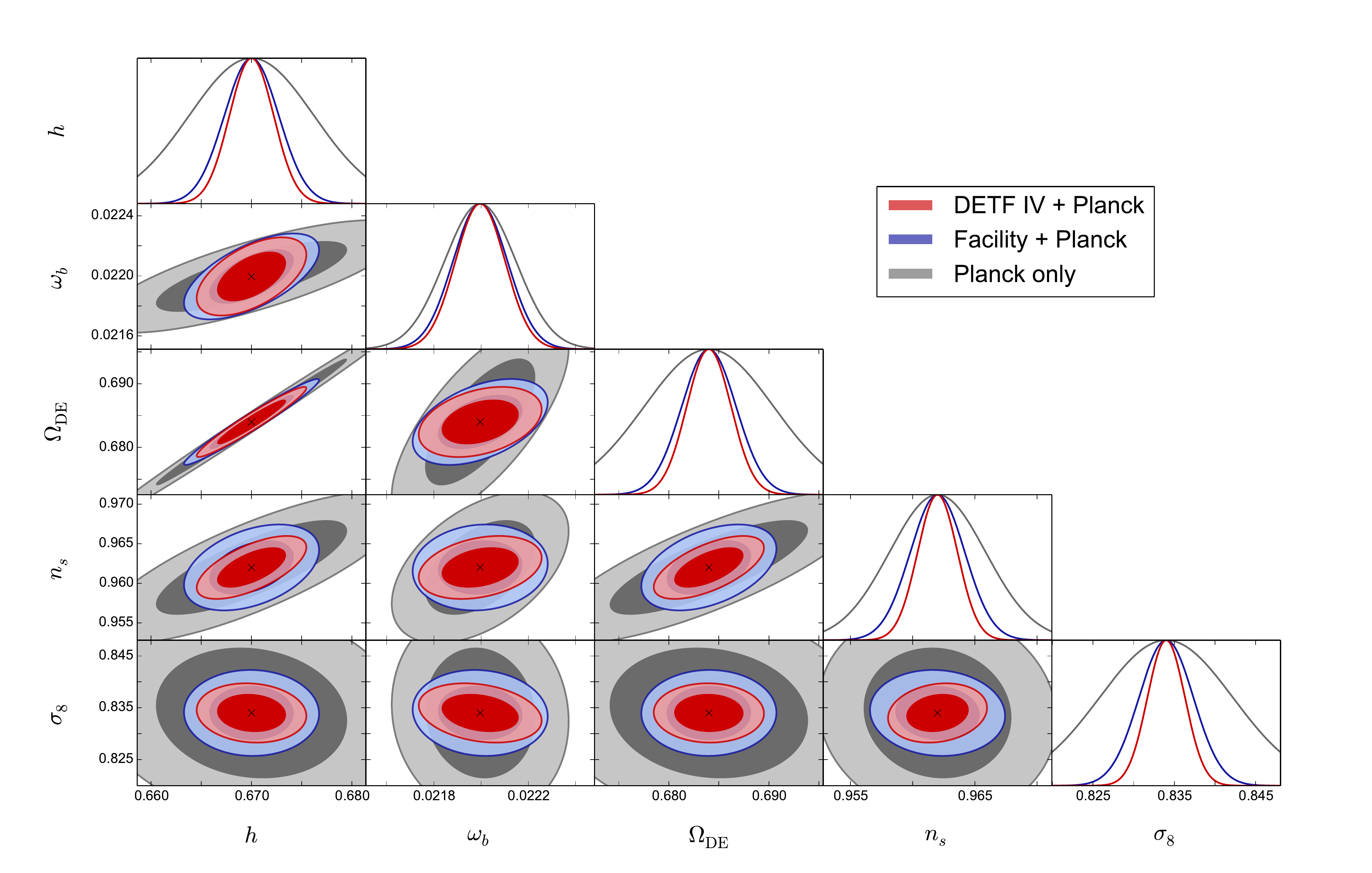}
\caption{Forecasts for a six-parameter $\Lambda$CDM model (the sixth parameter, $\tau$, was marginalised over in advance in the Planck Fisher matrix). This model has fixed $\Omega_K = 0$, $w_0 = -1$, and $w_a = 0$.}
\label{fig-6param}
\end{figure*}

One can map the functions of redshift into the set of cosmological parameters using a simple linear transformation of the Fisher matrix,
\be \label{eqn-project-params}
F(\bm{\beta}) = \sum_i M_i^T F_i(\bm{\alpha}) M_i,
\ee
where $\bm{\alpha} = \{ f(z), D_A(z), H(z) \}$ are the old parameters, $\bm{\beta} = \{ h, \Omega_\mathrm{DE}, \Omega_K, w_0, w_a, \gamma \}$ are the new parameters, $F_i$ is the Fisher matrix in a bin with redshift $z_i$, and the transformation matrix is given by $M_{j k}(z_i) = \partial \alpha_j (z_i) / \partial \beta_k$ \citep{Albrecht:2009ct}. The derivatives required for the transformation matrix are all analytical; for completeness, we present them in Appendix \ref{app-derivs}.

To complete the set of cosmological parameters, we must also include information on the shape and normalisation of the initial power spectrum, $\{ n_s, \sigma_8 \}$. These parameters are derived directly from the signal model of Eq. (\ref{sigcov}), and do not depend on the functions of redshift from the previous section (i.e. we have separated $f(z)$ and $\sigma_8$). Note that we do not use the shape of the power spectrum to constrain any other parameters, such as $\Omega_M$ or $\Omega_b$, even though in principle it does depend on them.

Carrying over the remaining parameters from the previous section, the full set is now
\be
\{ h, \Omega_\mathrm{DE}, \Omega_K, w_0, w_a, n_s, \sigma_8, \gamma, A(z), b_\mathrm{HI}(z), \sigma_\mathrm{NL}, \omega_b \}. \nonumber
\ee
The baryon density, $\omega_b = \Omega_b h^2$, is not constrained directly by HI experiments, but is included in the Planck prior. The total matter density (CDM + baryons) is fixed by $\Omega_M = 1 - \Omega_K - \Omega_\mathrm{DE}$, so we do not include it separately. The HI bias is free in each redshift bin, and we have taken $\sigma_8$ to be constant in redshift.

In what follows, we focus on the higher-end reference experiments, \StageTwo\ and \StageThree, although marginal (1D) constraints are provided for all of the experiments listed in Sect. \ref{sect-experiments}. We will also consider the effect of adding prior information from the CMB.

\begin{table}[b]
\hspace{-0.5em}
{\renewcommand{\arraystretch}{1.2}\begin{tabular}{|l|c|c|c|c|c|}
\hline
\parbox[t]{4mm}{\multirow{2}{*}{\bf Experiments}} & $h$ & $\omega_b$ & $\Omega_\mathrm{DE}$ & $n_s$ & $\sigma_8$ \\
 & $/ 10^{-3}$ & $/ 10^{-4}$ & $/ 10^{-3}$ & $/ 10^{-4}$ & $/ 10^{-3}$ \\
\hline
Planck + \StageOne\ & 5.8 & 1.5 & 6.1 & 36.9 & 7.0 \\
Planck + \StageTwo\ & 5.1 & 1.4 & 5.3 & 32.8 & 6.0 \\
Planck + \StageThree\ & 2.7 & 1.2 & 2.7 & 21.9 & 3.3 \\
Planck + DETF IV & 2.2 & 1.0 & 2.2 & 16.0 & 2.3 \\
\hline
Planck + WMAP & 12 & 2.8 & 17 & 73.0 & 12 \\ 
Planck+WP+BAO & 7.8 & 2.5 & 10 & 57.0 & 11 \\ 
\hline
\end{tabular} }
\caption{{\corr Forecast 1$\sigma$ marginal errors on vanilla $\Lambda$CDM model parameters for the set of reference surveys, compared with current constraints from Planck (temperature-only) and WMAP \citep{2013arXiv1303.5076P}.}}
\label{tbl:cosmo-6param-w0wafixed}
\end{table}

\subsection{`Vanilla' $\Lambda$CDM}

The current consensus is that cosmological data are well-described by the a flat $\Lambda$CDM model of structure formation that can be characterised in terms of six parameters: the Hubble parameter, $H_0=100h$ km\,s$^{-1}$\,Mpc$^{-1}$, the density of dark energy (or cosmological constant), $\Omega_{\rm DE}$, the physical density of baryons, $\omega_b$, the linear amplitude of density fluctuations, parametrised by $\sigma_8$, the spectral index of primordial density perturbations, $n_s$, and the optical depth to last scattering, $\tau$. In this section, we examine the constraints that intensity mapping experiments will be able to put on this model when combined with CMB data from Planck. Parameters that extend the `vanilla' $\Lambda$CDM model ($w_0, w_a, \Omega_K, \gamma$) are fixed to their fiducial values in this section, and $\Omega_\mathrm{DE} = \Omega_\Lambda$.

Fig. \ref{fig-6param} presents forecasts for five of the six parameters for the \StageThree\ experiment, compared with Planck-only and the DETF Stage IV galaxy redshift survey. Although the reionisation history will have a significant role in the evolution of the HI density and bias, we are focusing on sufficiently late times that our constraints will essentially be insensitive to variations of $\tau$, within current constraints, and so we leave it out of the plot. (In fact, it has already been marginalised over in the Planck prior Fisher matrix.) We do not directly constrain $\omega_b$ with IM experiments either, but as it is strongly correlated with other parameters in the Planck prior, we leave it in.

As expected, there is a modest improvement over Planck alone by a factor of a few. The Planck-only constraints are mostly limited by strong correlations between parameters, so the role of the IM survey is primarily to break degeneracies. Future high resolution experiments such as ACTPol \citep{Niemack:2010wz} and SPTpol \citep{Austermann:2012ga} will be able to measure weak lensing of the CMB to sufficient accuracy that constraints from the CMB {\it alone} should be competitive with IM (and redshift surveys). This is contingent on the assumption of a fixed $w = -1$, however.

The biggest effect of adding IM data (or indeed any LSS data) is to substantially improve the constraints on $h$ and $\Omega_\mathrm{DE}$. As shown in Fig. \ref{fig-6param}, the two are strongly correlated for Planck alone, as the CMB only measures the combination $\Omega_M h^3 = (1 - \Omega_\mathrm{DE}) h^3$ \citep{2013arXiv1303.5076P}. The distance measures at late times depend on different combinations of these parameters and so help to break this degeneracy. This has a knock-on effect on other parameters that are correlated with them, especially $n_s$.

Table \ref{tbl:cosmo-6param-w0wafixed} summarises the vanilla $\Lambda$CDM constraints for the full set of reference experiments. The majority of future HI surveys are capable of improving on constraints from Planck plus existing LSS datasets; notably, $H_0$ and $\Omega_\mathrm{DE}$ are determined at the sub-1\% level by all but the smallest IM experiments.

High-end \StageThree-class experiments should even be competitive with a DETF Stage IV galaxy survey (c.f. Euclid or LSST), although this is only likely to be the case for parameters constrained by the distance measures, such as $\Omega_\mathrm{DE}$. Those that depend more on the power spectrum at smaller scales will not be quite as close, because the galaxy surveys measure $P(k)$ significantly better at relatively high wavenumbers of $k \sim 0.1$ Mpc$^{-1}$ (Fig. \ref{fig-pk-constraints}); our IM experiments fall behind on these scales due to limited (single-dish) angular resolution.

\subsection{Dark energy equation of state}

The driver for the majority of the cosmological surveys currently under development is to find precision constraints on the dark energy equation of state, $w(z)$, and in doing so to infer the physical nature of the substance that appears to be driving the accelerated expansion of the Universe. HI intensity mapping provides a way of constraining $w(z)$ with considerable precision, using the full combination of $D_A(z)$, $H(z)$, and $f(z)$ reconstructed over a broad range of redshifts.

While the evolution of the equation of state parameter depends on the underlying dark energy theory, and as such could take any number of functional forms, it is nevertheless useful to work with a simple expansion about $z=0$,
\be
w(a) \approx w_0 + \frac{z}{1 + z} w_a.
\ee
This commonly-used parametrisation should be reasonably accurate at late times, but will not capture more exotic behaviour at $z \gg 1$. The corresponding dark energy density evolves with redshift as
\be
\Omega_\mathrm{DE}(z) = \Omega_{\mathrm{DE},0} \exp [3 w_a z / (1+z)] \, (1+z)^{3(1 + w_0 + w_a)}.
\ee
The overall sensitivity of an experiment to a varying equation of state can be summarised (to some extent) by the dark energy {\it figure of merit}, defined by the Dark Energy Task Force as \citep{Albrecht:2009ct}
\be
\mathrm{FOM} = 1 \big/ \sqrt{\mathrm{det}(F^{-1}|_{w_0, w_a})},
\ee
which is proportional to the reciprocal of the area enclosed by the 68\% contour of the $(w_0, w_a)$ error ellipse for Fisher matrix $F$.

\begin{figure}[t]
\includegraphics[width=\columnwidth]{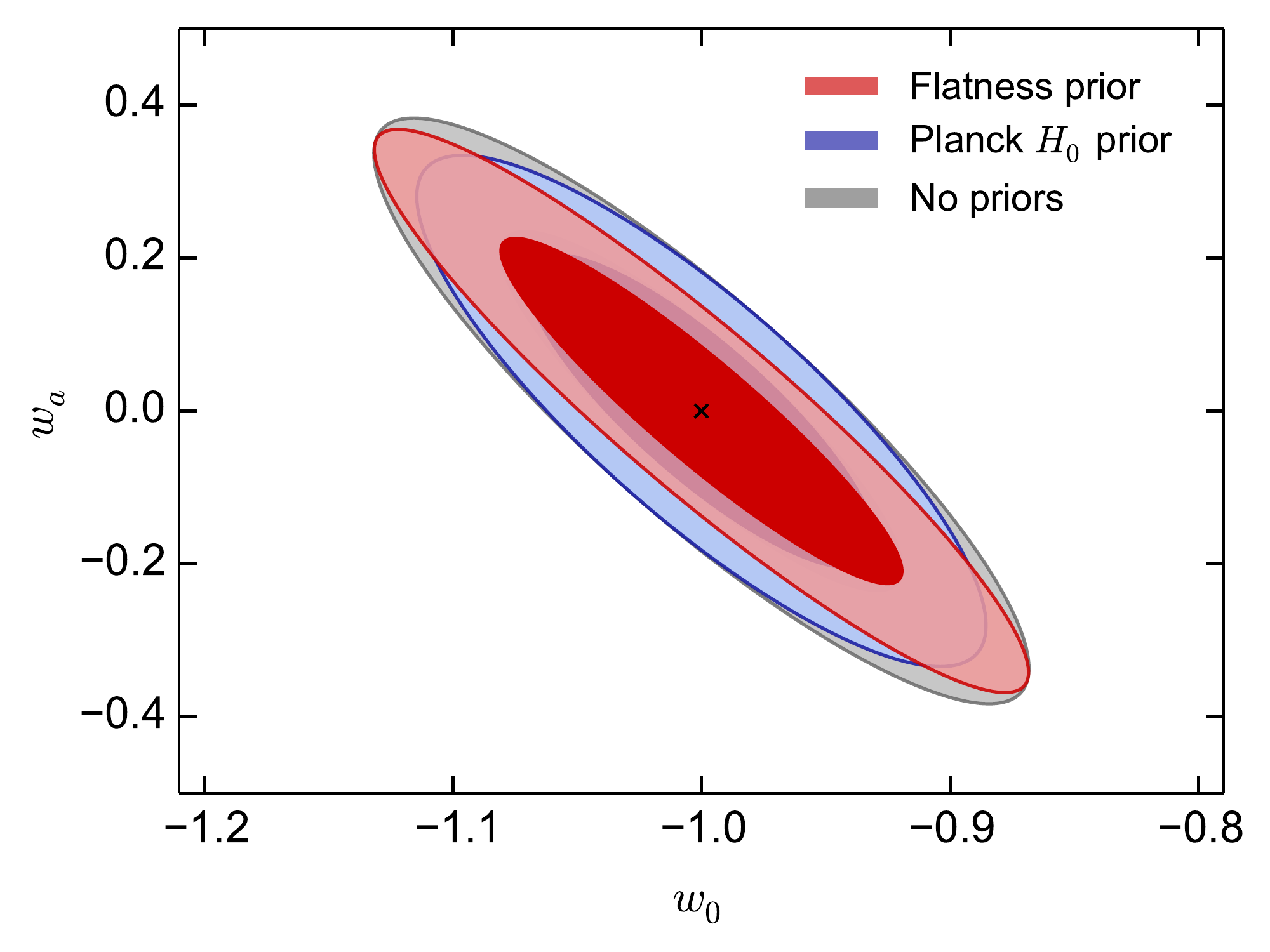}
\vspace{-2em}\caption{Effect of various priors on $w_0-w_a$ constraints, for \StageThree\ + Planck. $\Omega_K$ is already well-constrained by the combination of CMB and HI data, so the flatness prior has only a small effect. Additional $H_0$ information has a larger effect in breaking the degeneracy.} 
\label{pub-w0wa-priors}
\end{figure}

\begin{figure}[t]
\includegraphics[width=\columnwidth]{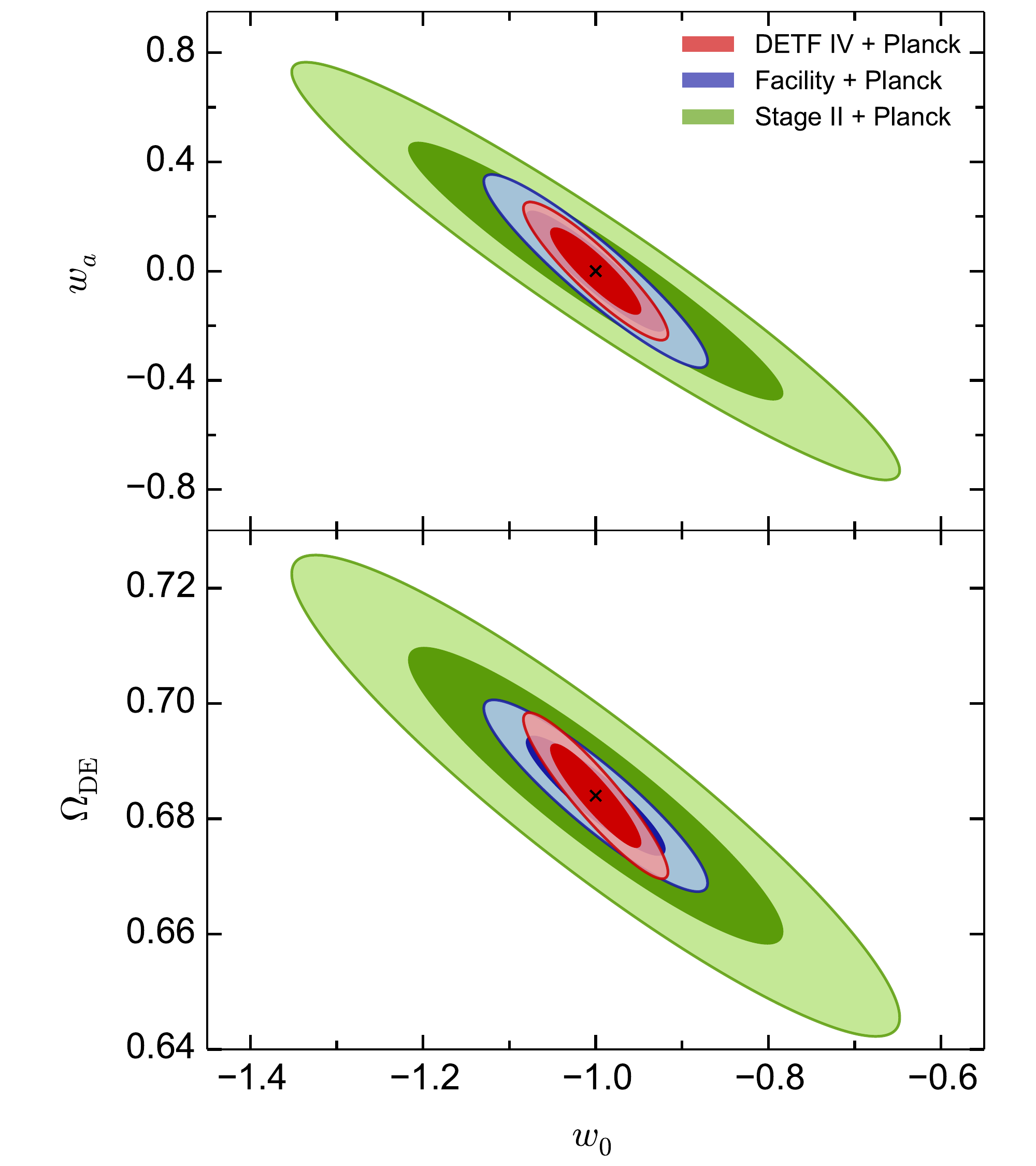}
\vspace{-2em}\caption{Top Panel: Forecast constraints on $w_0$ and $w_a$, including the Planck prior. We have assumed flatness ($\Omega_K=0$), and fixed $\gamma$ to its fiducial value. The DETF figures of merit for the \StageTwo, \StageThree, and DETF Stage IV surveys are 95, 358 and 712 respectively. Bottom Panel: Forecast constraints on $w_0$ and $\Omega_{\rm DE}$ for the same setup.} 
\label{fig-w0waokfix}
\end{figure}

The foremost task in understanding the nature of dark energy is to determine whether the equation of state differs from that of a cosmological constant, $w=-1$. Current constraints on $w_0$ and $w_a$ are relatively weak; the combination of Planck with SNLS supernova data does give values that are slightly in tension with a pure cosmological constant \citep{2013arXiv1303.5076P}, but the significance fades when other datasets are used instead. Fig. \ref{fig-w0waokfix} shows the improved constraints that can be expected on $w_0$, $w_a$, and $\Omega_\mathrm{DE}$ for the combination of our reference experiments with Planck, assuming flatness. Despite the addition of IM data, the parameters remain strongly correlated, so even substantial deviations from $w=-1$ will not necessarily be picked up. Nevertheless, a substantial fraction of the $w_0-w_a$ plane can be excluded by IM + Planck, so a successful detection is still possible if the real values lie orthogonal to the degeneracy direction. 1D marginal constraints for the full set of extensions to $\Lambda$CDM that we are considering here (including $w_0$ and $w_a$) are given in Table \ref{tbl:eosparams} for all of the experiments from Sect. \ref{sect-experiments}.

If one takes the possibility of a varying equation of state seriously, $w_0$ and $w_a$ should be left free when deriving constraints on other cosmological parameters. Table \ref{tbl:eosparams} shows the effect of marginalising over the equation of state on the vanilla $\Lambda$CDM model parameters. The parameters derived from the various distance measures are strongly affected -- their 1D marginal uncertainty is typically increased by around an order of magnitude compared to the unmarginalised case shown in Table \ref{tbl:cosmo-6param-w0wafixed}. This can be understood in terms of the degeneracies shown in Fig. \ref{fig-w0waokfix}; adding new parameters always increases the overall uncertainty, but because $\Omega_\mathrm{DE}$ (and $h$) are highly correlated with $w_0$ and $w_a$, they are particularly strongly affected. Parameters that do not depend on distance measures, i.e. $n_s$ and $\sigma_8$, are less affected by the equation of state parameters, and so their marginal uncertainties increase by only a modest amount.

\begin{table*}[!tbp]
\footnotesize
\vspace{-0.5em}\begin{center}
{\renewcommand{\arraystretch}{1.3} \begin{tabular}{|l||r|r|r|r|r|r|r|r|r||r|}
\hline
\multirow{2}{*}{\bf Experiments} & \multicolumn{1}{c|}{$A$} & \multicolumn{1}{c|}{$h$} & \multicolumn{1}{c|}{$\Omega_K$} & \multicolumn{1}{c|}{$\Omega_\mathrm{DE}$} & \multicolumn{1}{c|}{$n_s$} & \multicolumn{1}{c|}{$\sigma_8$} & \multicolumn{1}{c|}{$\gamma$} & \multicolumn{1}{c|}{$w_0$} & \multicolumn{1}{c||}{$w_a$} & \multirow{2}{*}{FOM} \\
 & $/ 10^{-2}$ & $/ 10^{-3}$ & $/ 10^{-4}$ & $/ 10^{-3}$ & $/ 10^{-4}$ & $/ 10^{-3}$ & $/ 10^{-2}$ & $/ 10^{-2}$ & $/ 10^{-2}$ &  \\
\hline
Stage I & 18.9 & 32.3 & 47.9 & 22.3 & 38.5 & 8.1 & 3.6 & 31.3 & 85.2 & 13.8 \\
Stage II & 13.2 & 23.7 & 33.1 & 17.2 & 38.0 & 7.8 & 4.4 & 15.2 & 33.1 & 39.9 \\
Facility & 5.2 & 8.7 & 13.6 & 6.9 & 35.0 & 6.0 & 1.8 & 5.4 & 14.9 & 265.4 \\
\hline
GBT & 73.9 & 131.9 & 178.4 & 93.4 & 38.6 & 8.2 & 20.1 & 95.0 & 221.8 & 1.1 \\
GBT-HIM & 31.2 & 64.3 & 78.9 & 45.4 & 38.6 & 8.2 & 9.8 & 50.7 & 126.9 & 4.2 \\
GMRT & 54.3 & 37.1 & 153.0 & 19.3 & 38.5 & 8.2 & 4.1 & 35.8 & 184.2 & 7.0 \\
JVLA & 57.7 & 43.0 & 175.3 & 22.6 & 38.6 & 8.2 & 4.5 & 40.1 & 209.2 & 5.5 \\
Parkes & 51.2 & 28.4 & 322.6 & 32.6 & 38.4 & 8.2 & 2.7 & 44.7 & 335.5 & 3.6 \\
VLBA & 74.8 & 47.8 & 826.9 & 86.2 & 38.6 & 8.2 & 3.8 & 91.7 & 799.8 & 0.8 \\
WSRT + APERTIF & 11.2 & 11.1 & 41.2 & 6.5 & 37.7 & 8.0 & 1.5 & 15.1 & 66.4 & 57.6 \\
\hline
BAOBAB-128 & 24.3 & 50.2 & 71.3 & 36.6 & 38.5 & 8.1 & 9.0 & 33.3 & 71.4 & 8.0 \\
BINGO & 25.8 & 30.8 & 90.0 & 16.1 & 38.5 & 8.2 & 2.8 & 44.1 & 172.5 & 7.8 \\
CHIME & 3.0 & 8.7 & 9.7 & 7.1 & 30.2 & 5.2 & 3.4 & 5.0 & 15.1 & 288.1 \\ 
FAST & 7.5 & 13.5 & 16.0 & 10.1 & 33.5 & 6.4 & 3.2 & 7.1 & 18.5 & 144.7 \\
MFAA & 5.7 & 11.9 & 14.1 & 9.1 & 32.2 & 6.0 & 3.1 & 6.3 & 17.2 & 165.7 \\
Tianlai & 3.6 & 8.0 & 11.9 & 6.3 & 28.7 & 4.9 & 2.4 & 4.0 & 12.0 & 383.3 \\ 
\hline
ASKAP & 7.7 & 16.2 & 21.1 & 11.9 & 37.8 & 7.7 & 2.9 & 11.8 & 26.8 & 80.3 \\
KAT7 & 114.0 & 76.4 & 1182.5 & 124.1 & 38.6 & 8.2 & 5.8 & 130.1 & 1138.6 & 0.4 \\
MeerKAT (B1) & 12.2 & 24.4 & 29.4 & 17.9 & 38.1 & 7.9 & 3.6 & 17.4 & 38.4 & 35.9 \\
MeerKAT (B2) & 10.2 & 9.4 & 26.8 & 6.1 & 37.5 & 7.7 & 1.5 & 6.4 & 29.5 & 171.4 \\
\hline
SKA1-MID (B1) Autocorr. & 6.2 & 11.2 & 16.1 & 8.7 & 35.9 & 6.6 & 2.3 & 7.1 & 17.6 & 162.5 \\
SKA1-MID (B1) Interferom. & 22.3 & 29.1 & 34.3 & 19.9 & 37.2 & 7.8 & 8.7 & 13.6 & 33.8 & 45.1 \\
SKA1-MID (B2) Autocorr. & 7.6 & 7.1 & 18.6 & 5.1 & 35.9 & 7.2 & 1.3 & 3.6 & 16.4 & 410.9 \\
SKA1-MID (B2) Interferom. & 368.2 & 37.3 & 94.3 & 19.0 & 38.0 & 8.2 & 10.6 & 22.8 & 86.5 & 18.5 \\
SKA1-SUR (B1) & 5.3 & 11.9 & 15.2 & 9.3 & 35.4 & 6.7 & 3.3 & 6.5 & 16.0 & 159.5 \\
SKA1-SUR (B2) & 4.5 & 6.5 & 12.2 & 5.3 & 35.3 & 5.7 & 1.4 & 3.8 & 12.2 & 444.2 \\
SKA1-MID + MeerKAT (B1) & 6.4 & 11.6 & 16.7 & 9.0 & 36.1 & 6.8 & 2.4 & 7.5 & 18.2 & 148.9 \\
SKA1-MID + MeerKAT (B2) & 7.7 & 7.1 & 18.6 & 5.1 & 35.9 & 7.2 & 1.3 & 3.5 & 16.3 & 414.7 \\
\hline
DETF Stage IV (gal. survey) & 2.4 & 7.5 & 8.6 & 6.2 & 27.1 & 5.3 & 3.2 & 4.1 & 12.8 & 405.5 \\
\hline
\hline
{\bf Fiducial values} & {\bf 1.0} & {\bf 0.67} & {\bf 0.0} & {\bf 0.684} & {\bf 0.962} & {\bf 0.834} & {\bf 0.55} & {\bf -1.0} & {\bf 0.0} & \multicolumn{1}{|c|}{--} \\
\hline
\end{tabular} }
\end{center}
\caption{1D marginal constraints (68\% CL) on the extended $\Lambda$CDM model, including the Planck prior. The constraint on $A$ (which has been summed over all redshift bins) gives a measure of the detectability of the BAO.} 
\label{tbl:eosparams}
\end{table*}

As we have seen, even the addition of intensity mapping or other intermediate-redshift LSS data to the CMB constraints is insufficient to break all of the parameter degeneracies once $w_0$ and $w_a$ are allowed to vary. In order to precisely determine these parameters, it is therefore necessary to add more data. Distance measurements from Type Ia supernovae are the obvious candidate, since they offer orthogonal constraints on $\Omega_\mathrm{DE} - \Omega_M$ \citep{Efstathiou:1998xx}. A local measurement of $H_0$ is also useful; as shown in Fig. \ref{fig-6param}, $h$ is strongly correlated with the dark energy density, so additional information about either parameter can substantially improve the constraints on both. Fig. \ref{pub-w0wa-priors} shows the effect of adding $H_0$ data to Planck + \StageThree. We also consider the effect of allowing departures from spatial flatness; as we will see in the next section, the combination of CMB and intensity mapping data measure $\Omega_K$ well, mostly independent of dark energy, so marginalising over curvature has a relatively minor effect on the $w_0-w_a$ ellipse.

\begin{figure}[!thb]
\includegraphics[width=\columnwidth]{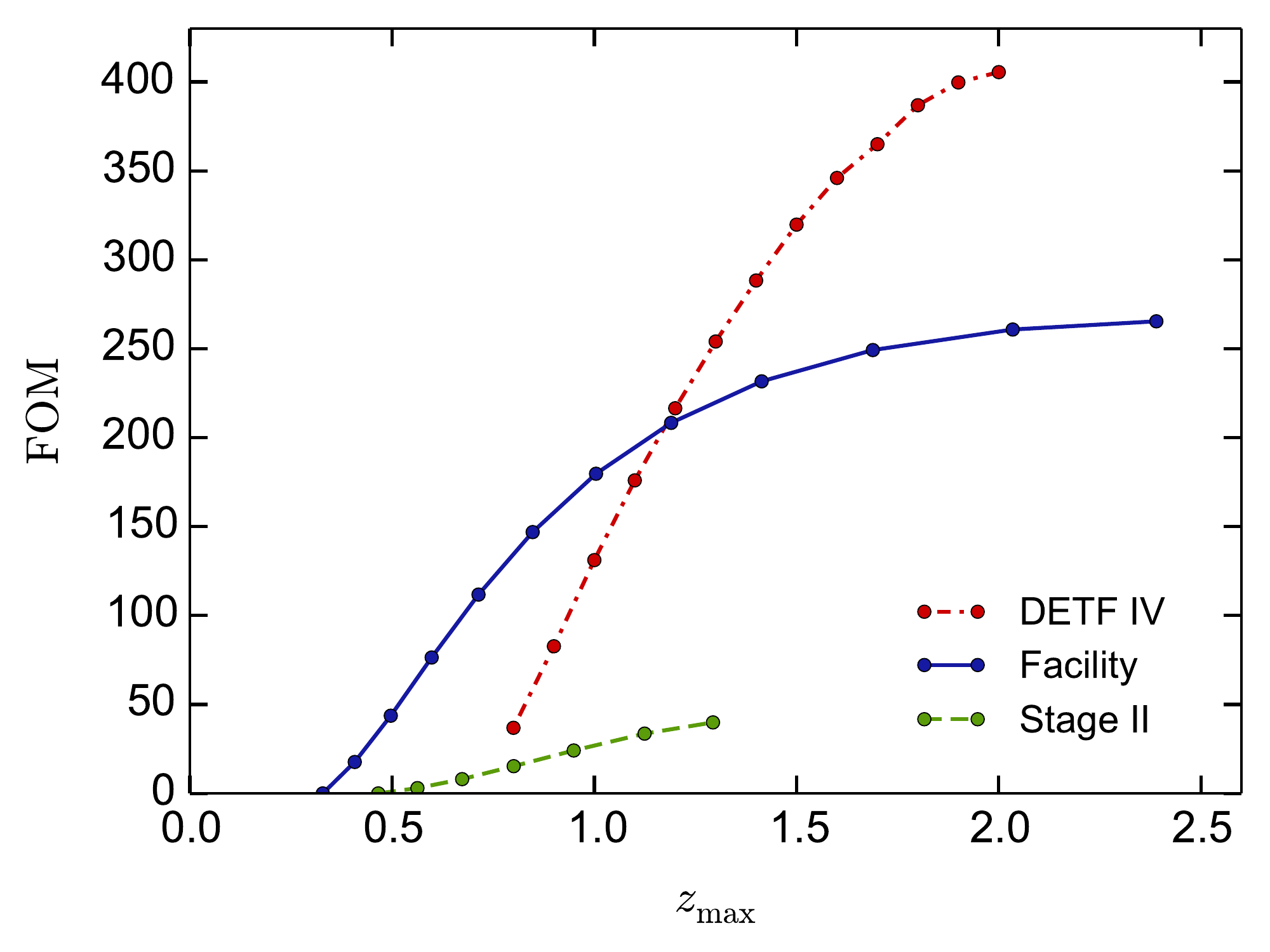}
\vspace{-1em}\caption{Improvement in dark energy FOM as a function of the maximum redshift of the survey ($\Omega_K$ and $\gamma$ marginalised).}
\label{pub-fom-fnz}
\end{figure}

Fig. \ref{pub-fom-fnz} shows the contribution to the dark energy figure of merit from each redshift bin. For our reference IM experiments, it is clear that the redshift range $z \lesssim 1.2$ is most critical; little improvement in FOM is seen above this redshift. The same cannot be said for the galaxy survey, however, which sees a roughly equal increase in FOM with each additional redshift bin across its whole $z$ range. One way of understanding this behaviour is to compare Fig. \ref{pub-fom-fnz} with the plots for $D_A(z)$, $H(z)$, and $f \sigma_8(z)$ in Fig. \ref{fig-redshift-fns}. Above $z \sim 1.2$, the angular diameter distance and growth rate constraints begin to worsen for \StageThree, but remain relatively flat for the galaxy survey. Since $w_0$ and $w_a$ are obtained from projections of these functions, it is no surprise that little is gained on the FOM at redshifts where they are poorly constrained.

\subsection{Curvature}

The potential for HI intensity mapping experiments to span extremely wide redshift ranges -- from $z \approx 0.1$ out to $z \gtrsim 2.5$ without too much difficulty -- makes them an interesting prospect for unravelling the geometric degeneracy, i.e. the interplay between dark energy and curvature. Without strong assumptions on one or the other, it is difficult to separate the effects of $\Omega_K$ and $w(z)$ using only a single type of distance measure \citep{Mortonson:2009nw, Shafieloo:2011zv}, and for the CMB power spectrum alone they are completely degenerate. As discussed in Sect. \ref{sect-dahz}, intensity mapping provides a suite of distance measures. The combination of IM and (e.g.) CMB data should therefore be very useful in separating curvature from the evolution of dark energy in a precise and unambiguous manner.

A precision determination of spatial curvature on horizon scales would also provide a rare opportunity to test inflation. Current observations seem to point in the direction of flatness, with the most recent bounds from Planck finding $|\Omega_K| \lesssim 10^{-2}$ (95\% CL), consistent with the vast majority of inflation models, but if a detection of $|\Omega_K| \gtrsim 10^{-4}$ were made, the whole class of eternally inflating models would be put under pressure \citep{Kleban:2012ph, Guth:2012ww}.

\begin{figure}[t]
\includegraphics[width=\columnwidth]{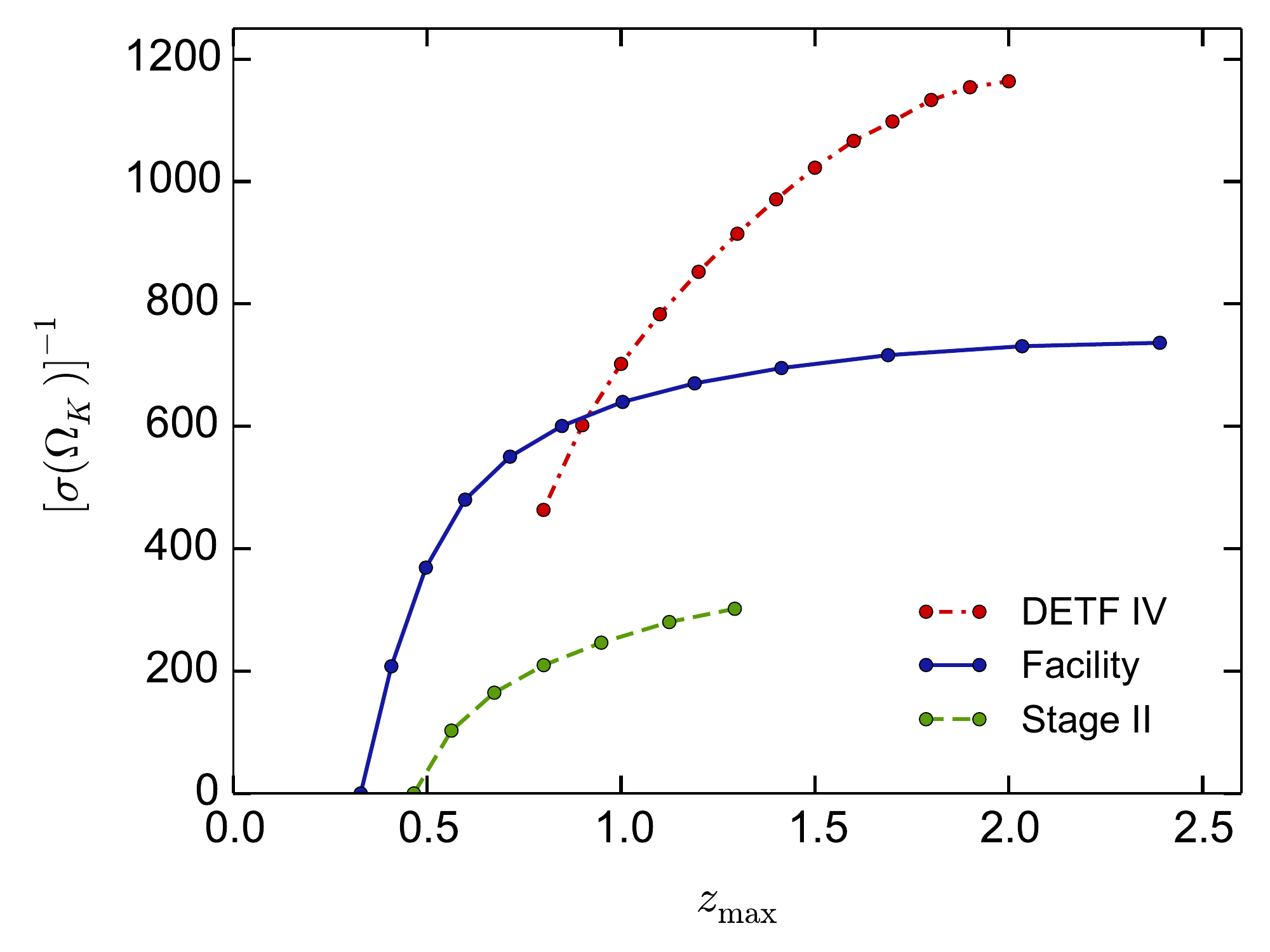}
\caption{Improvement in $\Omega_K$ constraints as a function of maximum redshift of the survey. We have marginalised over $w_0$, $w_a$ and $\gamma$ here.}
\label{pub-omegak-fnz}
\end{figure}

\begin{figure}[t]
\includegraphics[width=\columnwidth]{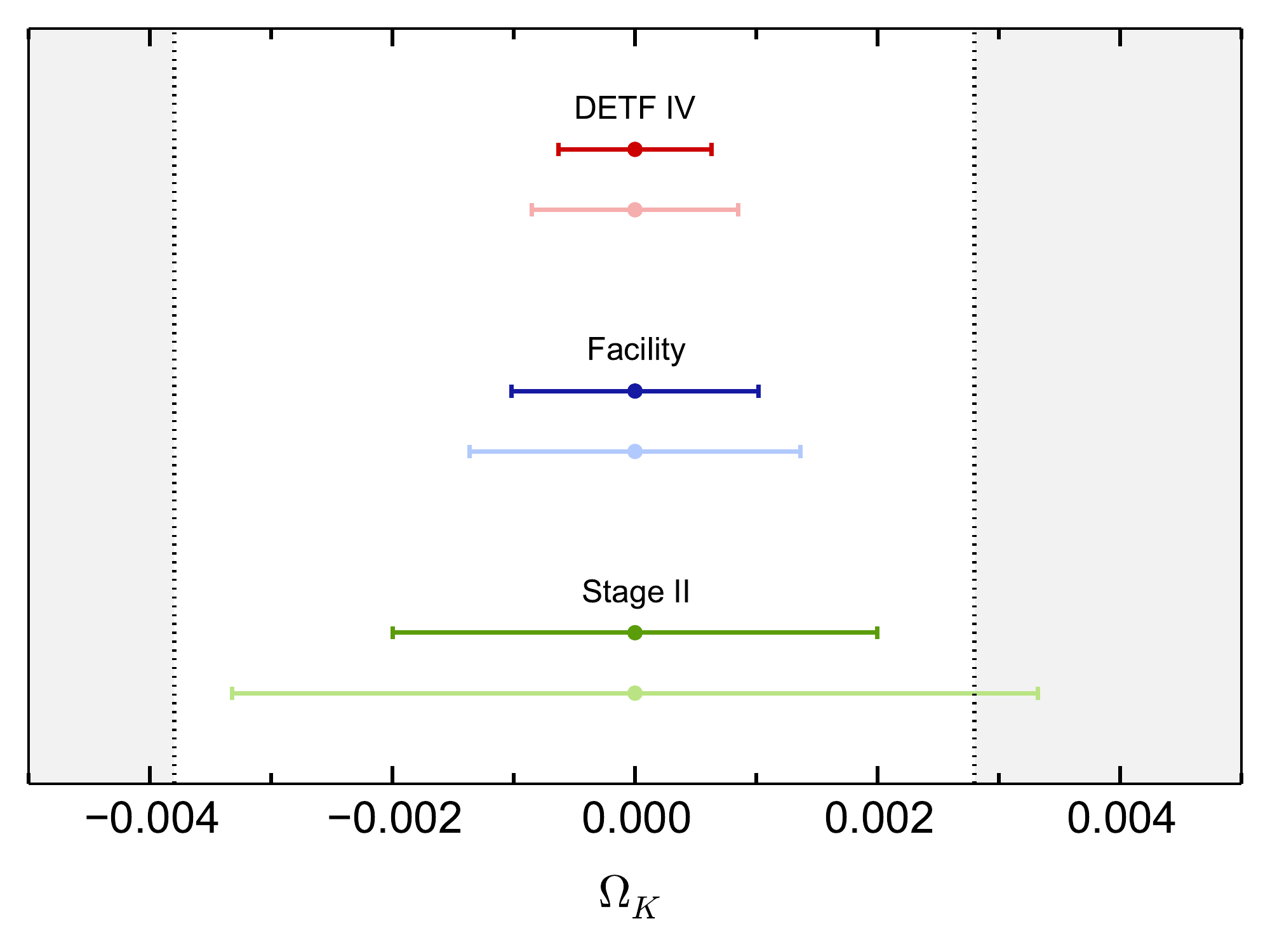}
\caption{Forecast marginal constraints on $\Omega_K$ (68\% CL) for the reference experiments plus Planck, with $(w_0, w_a)$ fixed to their fiducial values (upper errorbars), and marginalised over (lower errorbars). The shaded area shows the current best constraint on $\Omega_K$ ($w_0$, $w_a$ fixed) from Planck + WMAP + high-$\ell$ CMB + BAO \citep{2013arXiv1303.5076P}.}
\label{fig-omegak}
\end{figure}

The minimum curvature that can be detected unambiguously also happens to be at around the $10^{-4}$ level \citep{2009MNRAS.397..431V, 2013PhRvD..87h1301B}. Future CMB experiments should be able to approach this order of magnitude if $w$ is fixed to $-1$, and so too should a \StageThree-type IM experiment combined with Planck, as shown in Fig. \ref{fig-omegak}. There is little justification for putting such a strong prior on the equation of state, though -- any rigorous constraint on $\Omega_K$ must confront the geometric degeneracy head-on, and allow the full freedom of $w(z)$. Fig. \ref{fig-omegak} also shows the limits on curvature that can be achieved when the equation of state is left free. Though clearly worse than for fixed $w=-1$, the difference is relatively modest -- the combination of a \StageThree\ survey with Planck should still be able to measure $|\Omega_K|$ to around $10^{-3}$ without any particularly strong assumptions about the form of $w(z)$.

The effect of the geometric degeneracy runs both ways -- a lack of knowledge about $\Omega_K$ also degrades the reconstruction of the time evolution of the equation of state. Indeed, a percent level uncertainty in $\Omega_K$ can lead to a $\sim\! 100\%$ error on the recovery of more exotic forms for $w(z)$ \citep{2007JCAP...08..011C}, although it has been argued that current constraints can already mitigate this \citep{2013PhLB..719....1O}. As shown in Fig. \ref{pub-w0wa-priors}, the errors on the equation of state parameters do increase when we allow $\Omega_K$ to be free, albeit not substantially in the case of both \StageTwo\ and \StageThree; the combination of $H(z)$, $D_A(z)$, and $f(z)$ measurements from IM, plus the Planck prior, is enough to prevent any strong degeneracies from completely killing the $w_0 - w_a$ constraints. In fact, Fig. \ref{pub-w0wa-priors} shows that they are more sensitive to assumptions about $H_0$ than to curvature.

It is improved knowledge of the late-time expansion that most helps separate the effects of curvature and dark energy. We see this clearly in Fig. \ref{pub-omegak-fnz}, where $[\sigma(\Omega_K)]^{-1}$ is plotted as a function of the depth of each survey. There is an optimal point beyond which little new information is gained by the IM surveys, coinciding with the redshift at which the constraints on $f(z)$ and $D_A(z)$ start to degrade due to the limited angular resolution of the experiments (Fig. \ref{fig-redshift-fns}). This happens at higher $z_\mathrm{max}$ for the galaxy survey, which makes up for its lack of low redshift bins by having relatively flat fractional errors in $D_A(z)$, $H(z)$, and $f(z)$ out to $z \approx 2$. At even higher redshift, the dynamical effect of curvature is completely negligible, so little extra information could be gained anyway.


\begin{figure*}[t]
\includegraphics[width=2\columnwidth]{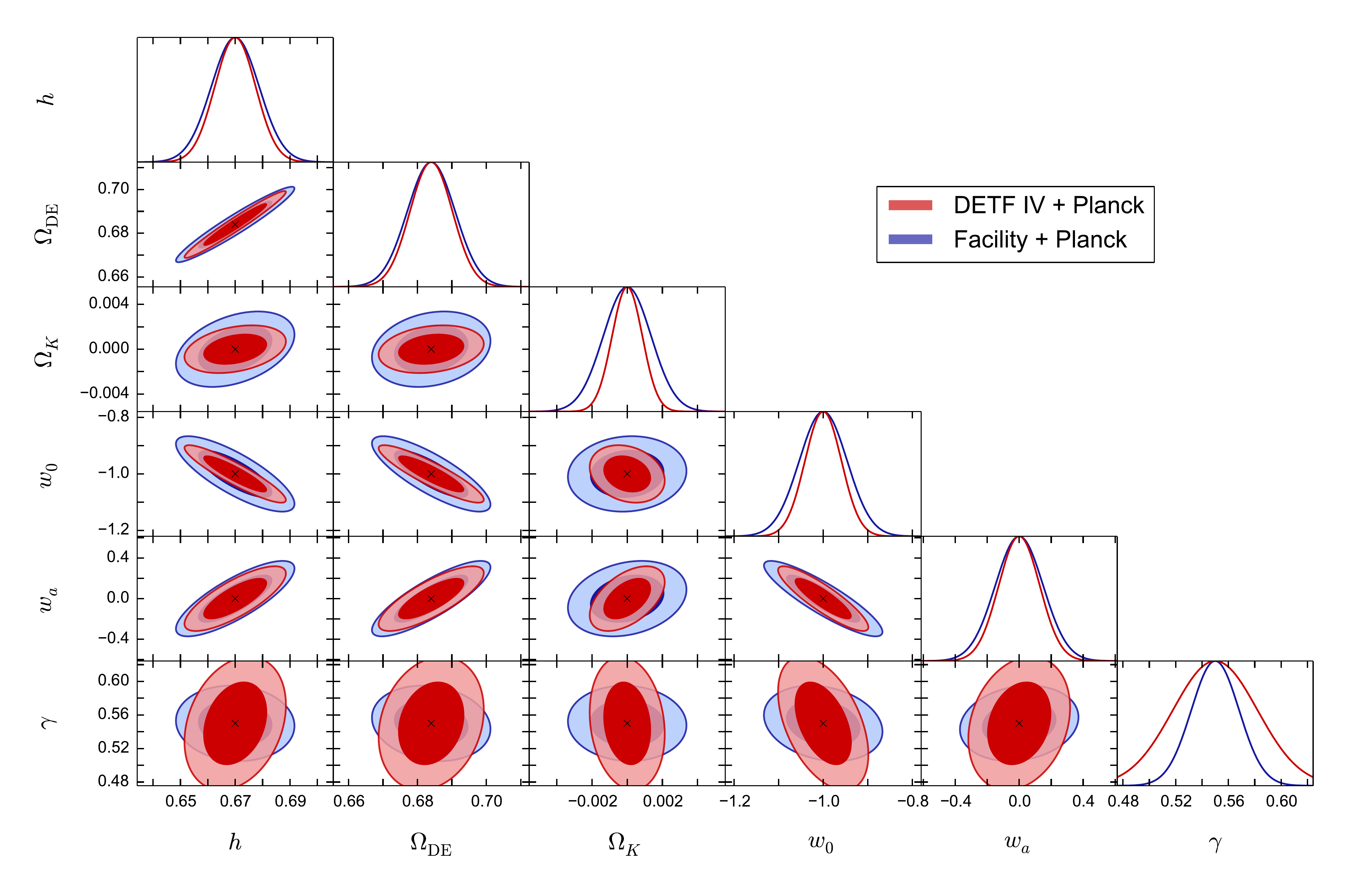}
\caption{Forecasts for dark energy and modified growth parameters for two of the reference experiments. Note the significantly different behaviour with respect to the growth index, $\gamma$.}
\label{fig-triangle-eos}
\end{figure*}

\subsection{Parametrised growth history} \label{sect-parametrised-growth}

The growth history of the Universe is a particularly powerful test of gravity. Modified theories of gravity generically alter the growth of structure from its GR behaviour, typically enhancing clustering on non-linear scales, and increasing peculiar velocities. Signatures of modified gravity in the non-linear regime are difficult to disentangle from less exotic astrophysical effects, leaving the linear velocity field as, in some sense, the `cleanest' modified gravity observable from large scale structure.

There is great variety in the effects that different modifications to gravity have on the linear growth history; the space of theories is complex, and has proved difficult to parametrise in a simple way \citep{PhysRevD.76.104043, Baker:2012zs, Battye:2012eu}. For the purposes of illustration, we will fall back on one of the simplest parametrisations of growth, using the growth index formulation of \citet{peebles1980}: $f(z) = \Omega^\gamma_M(z)$. Deviations from GR are captured, in part at least, by the difference in $\gamma$ from its $\Lambda$CDM+GR value, $\gamma_\mathrm{GR} \approx 0.55$.

Two notes of caution are necessary: Firstly, many modified gravity theories do not have growth histories that are well-described by a constant $\gamma$. Allowing $\gamma$ to be a function of redshift can help \citep{Linder:2007hg, Ishak:2009qs}, but even then there are many cases where the growth rate becomes scale-dependent, or is otherwise poorly described by this parametrisation. Secondly, dark energy models that require no modifications to GR can also modify the growth history, often in a way that is well-described by making the growth index a function of the equation of state parameter, $\gamma(w)$ \citep{Linder:2005in,2011ApJ...728L..46G}. We neglect this possibility here, and instead treat $\gamma$ as being independent of $w$.

Fig. \ref{fig-triangle-eos} shows forecasts for the various dark energy parameters for the DETF Stage IV galaxy redshift survey and the \StageThree\ reference experiment. The 1D marginal constraints from the galaxy survey outperform the IM experiment for all parameters, except one -- the growth index. Furthermore, the 2D constraints involving $\gamma$ are roughly orthogonal between the two surveys, despite this not being the case for other combinations of parameters.

At first, this may seem surprising. In Fig. \ref{fig-redshift-fns}, the galaxy redshift survey constrains $f\sigma_8(z)$ to around 1\% across most of its redshift range, while \StageThree's precision can only match this in the lowest redshift bins, increasing to $\sim 4\%$ at higher $z$. For $\gamma$, though, it is the very lowest redshifts that make the most difference. At low $z$, the growth factor evolves most rapidly, and is most sensitive to the value of $\gamma$ (i.e. $|df/d\gamma|$ increases as $z \to 0$), whereas at higher redshifts, matter begins to dominate, growth is slower, and the dependence on $\gamma$ is relatively weak. By virtue of its substantially lower $z_\mathrm{min}$, then, the \StageThree\ experiment captures more of the redshift range most sensitive to $\gamma$, and wins out over the galaxy survey.

\begin{figure}[t]
\includegraphics[width=\columnwidth]{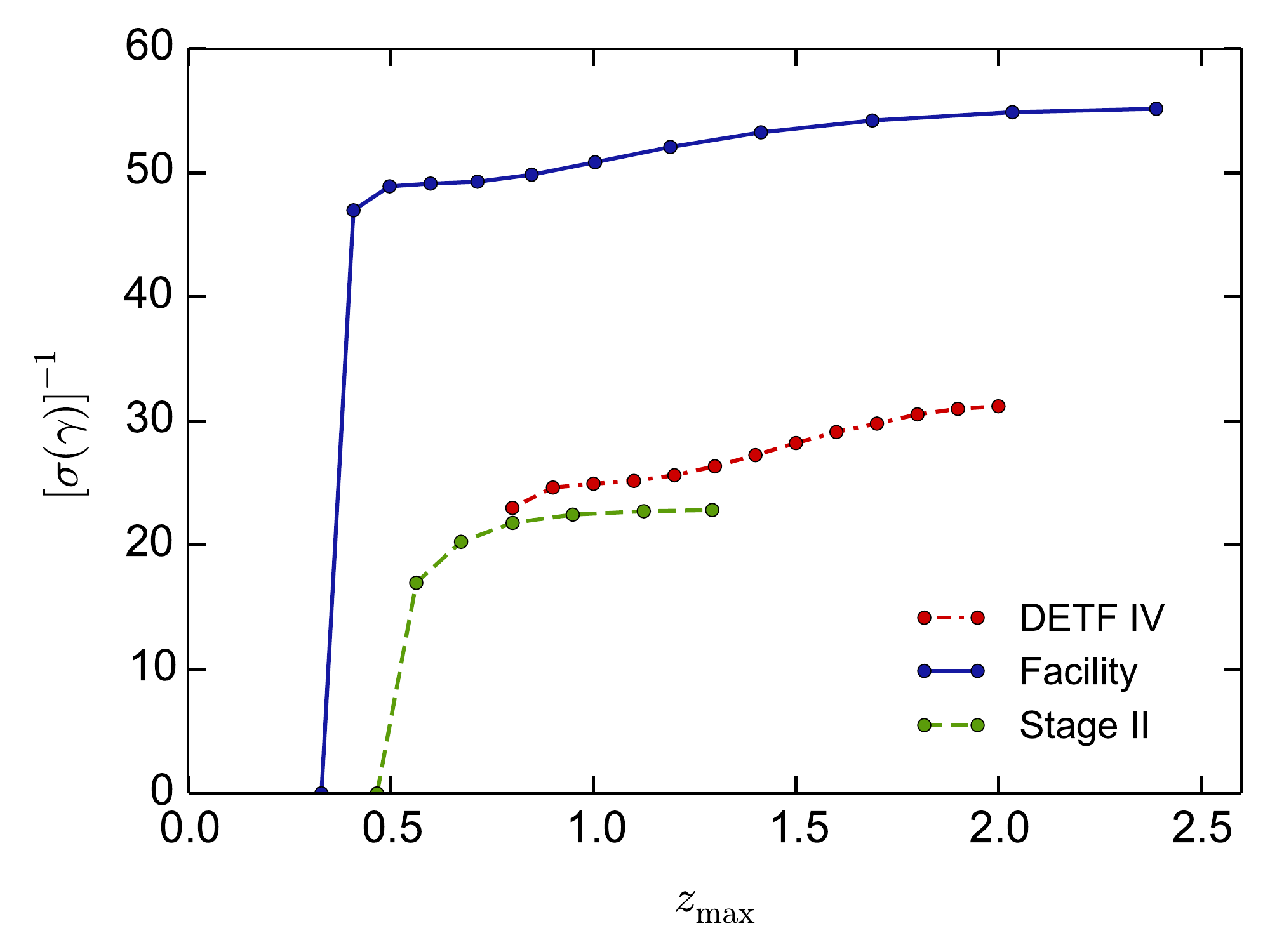}
\caption{Improvement in $\gamma$ constraints a function of maximum redshift of the survey. We have marginalised over $w_0$, $w_a$ and $\Omega_K$ here. The low value in the first redshift bin for the IM experiments is due to a degeneracy.}
\label{pub-gamma-fnz}
\end{figure}

Fig. \ref{pub-gamma-fnz} shows the effect of the lowest redshift bins on $\gamma$ more clearly. Even \StageTwo\ outperforms the Stage IV survey for $z_\mathrm{max} \lesssim 1$ -- again thanks to its lower $z_\mathrm{min}$ -- despite producing significantly worse constraints on almost every other parameter. This behaviour is also related to the choice of distance measures, and how the degeneracies between them get broken. As shown in Fig. \ref{fig-distance-terms}, the choice of measure can have a big effect on the $f(z)$ errors, so one might expect the strength of the constraint on $\gamma$, and its orthogonality to the galaxy redshift survey, to change if a different subset of measures was used.

\begin{figure}[t]
\includegraphics[width=\columnwidth]{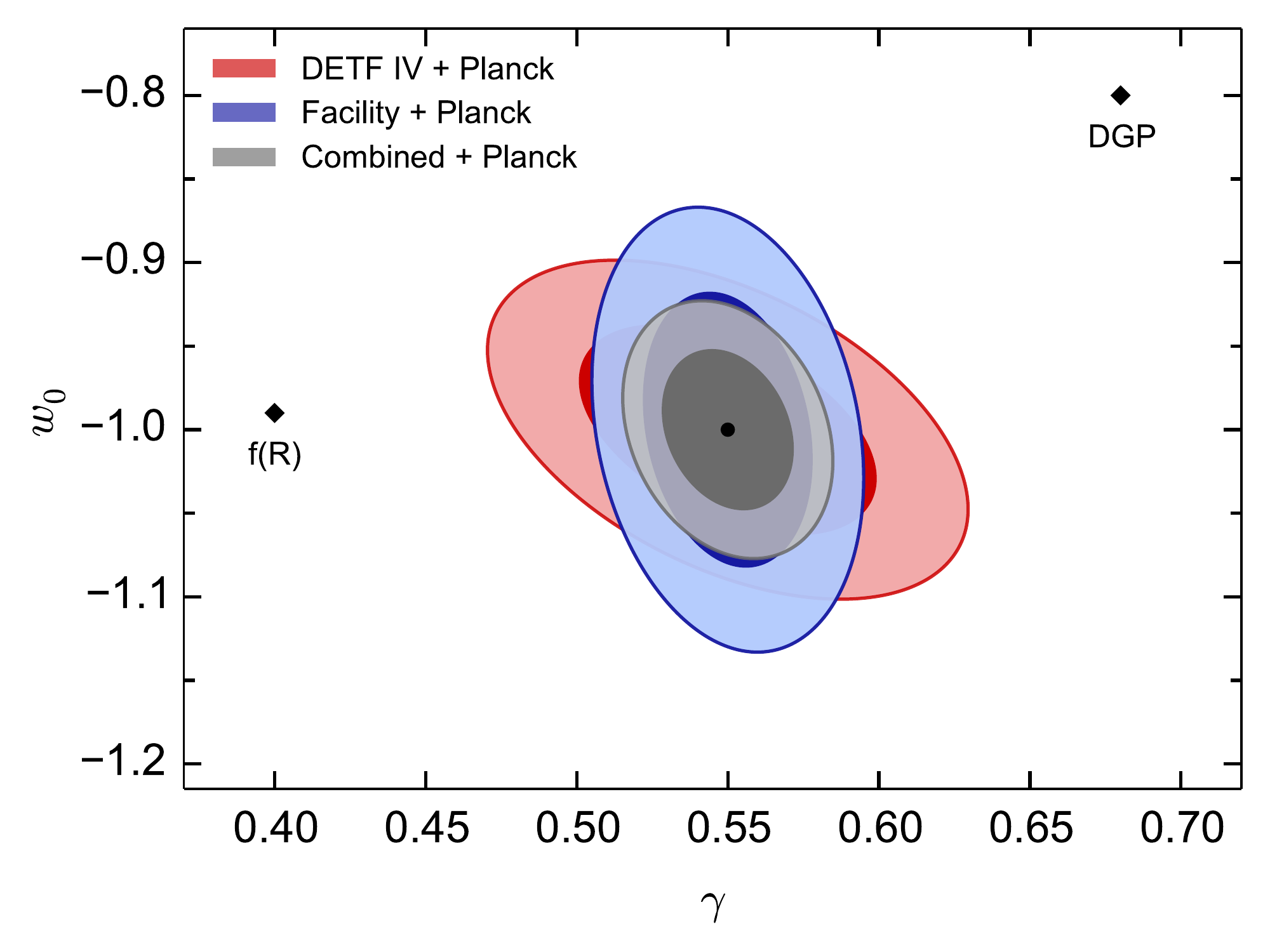}
\caption{Constraints on $(\gamma, w_0)$ for \StageThree, the DETF Stage IV survey, and the combination of the two, including Planck CMB priors. $w_a$ and $\Omega_K$ have been marginalised over, and the biases for both surveys, $b_\mathrm{HI}(z)$ and $b_\mathrm{gal}(z)$, are free in each bin. Also plotted are example $f(R)$ and DGP modified gravity models from \citet{2013LRR....16....6A}.}
\label{pub-w0gamma}
\end{figure}

Assuming the full set of distance measures, the complementarity between intensity mapping and galaxy redshift surveys can be used to significantly increase the precision of the constraint on $\gamma$, to the point where it becomes possible to clearly distinguish between many modified gravity models. Fig. \ref{pub-w0gamma} shows the result of combining the two surveys on the errors for $w_0$ and $\gamma$, along with some example predictions from modified gravity theories. The marginal error on $\gamma$ goes from $\sigma_\gamma = 0.024$ for \StageThree\ + Planck to $\sigma_\gamma = 0.015$ for \StageThree\ + DETF IV + Planck.

\section{Systematic effects} \label{sect-systematics}

Throughout this paper, we have compared the results from IM mapping experiments with those of a DETF Stage IV spectroscopic survey. In fact we have gone further and argued, in Section \ref{sect-zsurvey}, that we can think of an IM survey as a spectroscopic survey with an anisotropic $V_{\rm eff}(k)$, with very distinctive characteristics tied to the chosen mode of operation (single-dish or interferometric). While a useful comparison, IM experiments have their own, particular types of systematics that must be dealt with. In what follows we touch on these effects and attempt to quantify their impact on the target science.

\subsection{Evolution of the cosmological HI signal} \label{sect-HI-evol}

\begin{figure}[t]
\vspace{-0.5em}\includegraphics[width=\columnwidth]{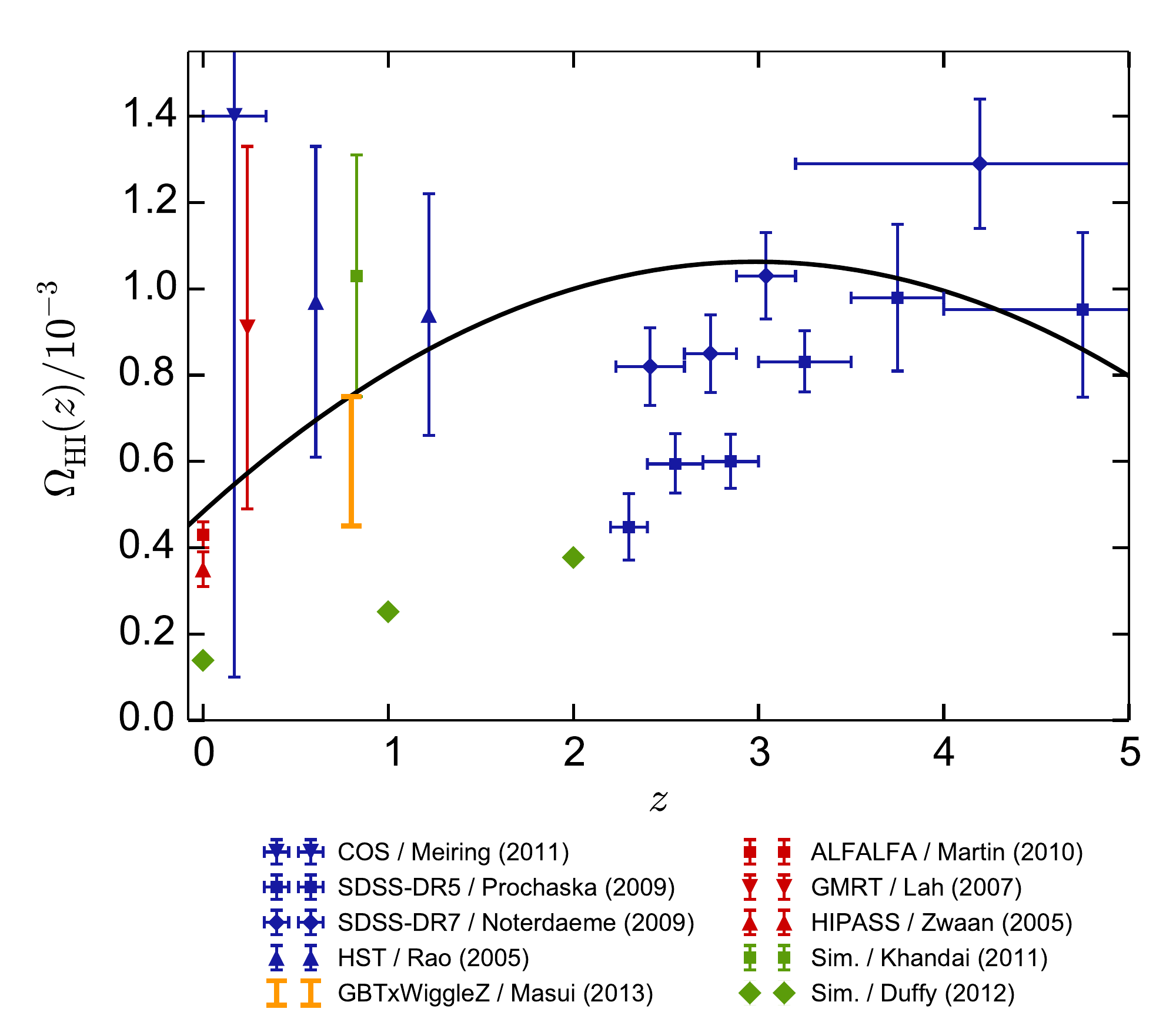}
\caption{Current constraints on the HI density fraction as a function of redshift, $\Omega_\mathrm{HI}(z)$ \citep{Meiring:2011pg, 2009ApJ...696.1543P, Noterdaeme:2009tp, Rao:2005ab, Martin:2010ij, Lah:2007nk, Zwaan:2005cz, 2011MNRAS.415.2580K}, partially based on the compilation in \citet{2012MNRAS.420.2799D} {\corr (see also \cite{Padmanabhan:2014zma}).} DLA observations are shown in blue, cross-correlations in yellow, other observations in red, and simulations in green. The thick black line shows the fiducial $\Omega_\mathrm{HI}(z)$ that we have adopted in this paper, {\corr which has $\Omega_\mathrm{HI}(z\!\!=\!\!0) = 4.86 \times 10^{-4}$.}}
\label{fig-omegaHI-evol}
\end{figure}

We have assumed a fiducial value of $\Omega_{{\rm HI}, 0} = 6.5 \times 10^{-4}$ throughout our analysis. Clearly our results will strongly depend on this value, as it is important in setting the ``signal-to-noise'' of the experiment -- the more neutral hydrogen there is, the more easily the cosmological signal can be detected. There are, however, large uncertainties in $\Omega_{\rm HI}(z)$ from current observations (Fig. \ref{fig-omegaHI-evol}). In particular, different tracers of the HI density give inconsistent results, so neither the normalisation nor the redshift evolution of $\Omega_{\rm HI}(z)$ are well understood.

The constraint of most relevance to us is from \citet{2013ApJ...763L..20M}, where IM measurements were cross-correlated with the WiggleZ galaxy redshift survey. It was found that
\bea
\Omega_{\rm HI}b_{\rm HI} \bar{r} = 4.3\pm0.7 \,({\rm stat.})\pm 0.4 \,({\rm sys.)} \times 10^{-4} \nonumber
\eea
at $z\!=\!0.8$, where the best theoretical estimates for the cross-correlation coefficient are $\bar{r} = 0.9-0.95$. This was obtained by restricting the analysis to $0.075 h \ {\rm Mpc}^{-1}<k<0.3 h \  {\rm Mpc}^{-1}$; extending it to $0.04 h \ {\rm Mpc}^{-1}<k<0.8 h \  {\rm Mpc}^{-1}$ lowers the constraint slightly to 
\bea
\Omega_{\rm HI}b_{\rm HI} \bar{r} = 4.0\pm 0.5 \,({\rm stat.})\pm 0.4 \,({\rm sys.)}\times 10^{-4}. \nonumber
\eea
The value of $\Omega_{\rm HI}$ is entangled with the bias which, from semi-analytic models combined with N-body simulations \citep{2011MNRAS.415.2580K}, is found to be consistently low, $b_{\rm HI}\approx 0.55-0.66$, although it can go up to unity for certain model choices. As a result, \citep{2013ApJ...763L..20M} proposes that one should assume $\Omega_{\rm HI}=4.5-7.5 \times 10^{-4}$ at $z=0.8$.

The lack of agreement between observations makes it difficult to reconstruct the redshift evolution of $\Omega_{\rm HI}$. At the upper end of the redshift range we are considering ($z \lesssim 3$), constraints from Damped Lyman$-\alpha$ (DLA) systems are scattered between $\Omega_{\rm HI} \approx 4 - 9 \times 10^{-4}$. At $z\sim 1$ there are discrepant results between theoretical models that find $\Omega_{\rm HI}\simeq 3\times 10^{-4}$ \citep{2012MNRAS.420.2799D} and observations of DLAs with HST that give $\Omega_{\rm HI}\simeq 9\times 10^{-4}$. At $z=0$, the ALFAFA and HIPASS surveys find $\Omega_{\rm HI}\simeq 4 \times 10^{-4}$, which is {\corr slightly lower} than our fiducial $\Omega_{{\rm HI}, 0}$.

For the forecasts in this paper, we tread the middle ground (solid line, Fig. \ref{fig-omegaHI-evol}). The $\Omega_{\rm HI}(z)$ redshift evolution is derived from a simulated HI halo mass function, as described in App. \ref{app-HI-signal}, and we choose its fiducial normalisation to be consistent with the GBT/WiggleZ cross-correlation measurement at $z=0.8$. The magnitude and redshift evolution of the HI bias, $b_{\rm HI}(z)$, are derived from the same mass function.

\begin{figure}[t]
\includegraphics[width=\columnwidth]{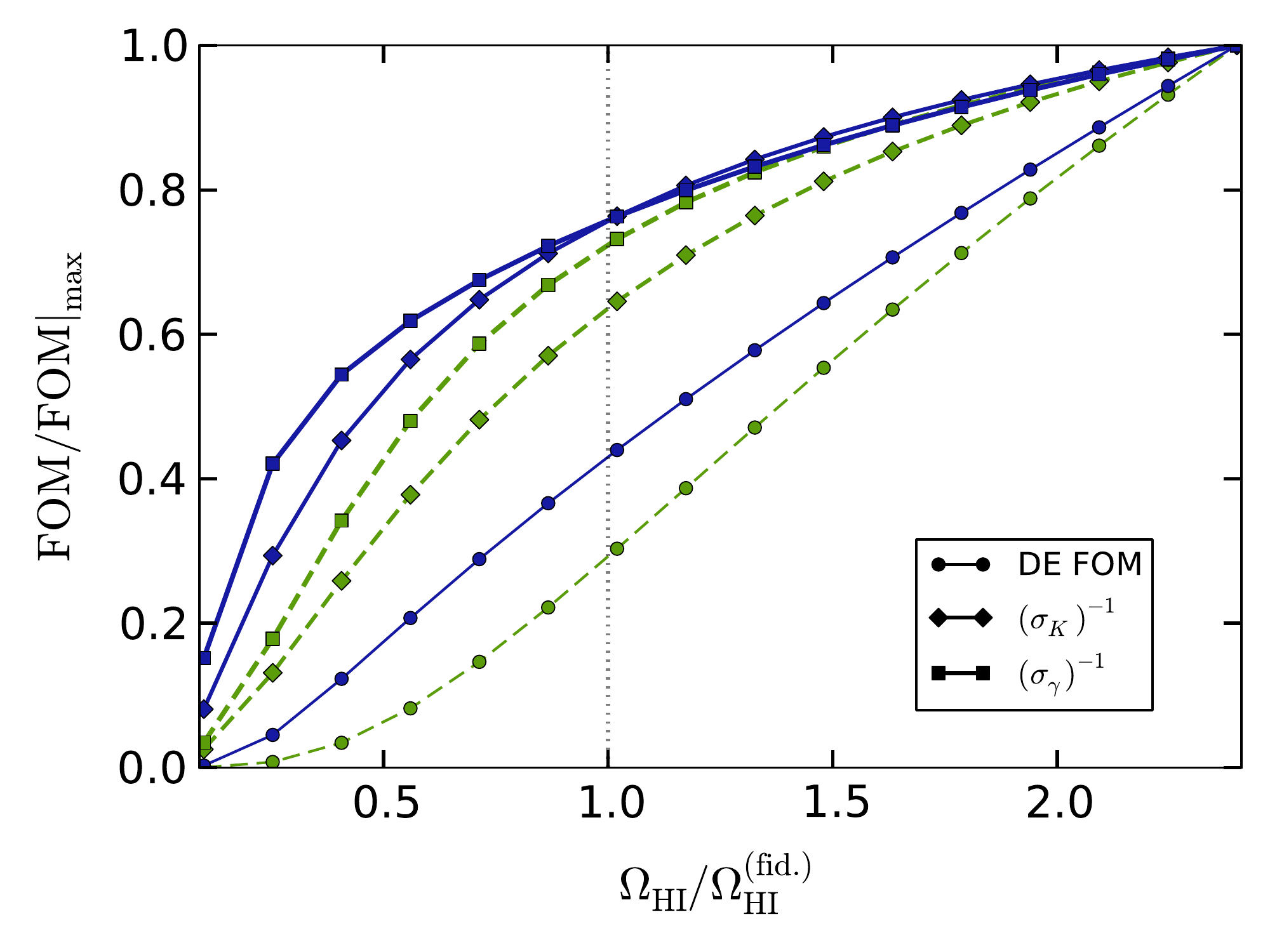}
\vspace{-2em}\caption{Normalised FOM and $\Omega_K$ and $\gamma$ marginal errors, {\corr as a function of $\Omega_{\mathrm{HI}}$ rescaled by a constant factor,} for the \StageThree\ (blue, solid) and \StageTwo\ (green, dashed) surveys.}
\label{fig-omegaHI}
\end{figure}

{\corr Fig. \ref{fig-omegaHI} shows the effect of rescaling $\Omega_{{\rm HI}}$ by a constant factor on the constraints for several observables. If $\Omega_{\mathrm{HI}}(z)$ was halved,} for example, the FOM for \StageThree\ would drop by a factor of 5. This highlights the sensitivity of cosmological constraints from IM to the HI density, and gives some idea of the degradation/improvement in performance that would be expected if $\Omega_{\rm HI}(z)$ substantially differs from what we have assumed.

\subsection{Non-linear scale} \label{sect-nonlinear} \vspace{-0.5em}

We have marginalised over the non-linear scale, $\sigma_\mathrm{NL}$, in all of our forecasts. As described in Section \ref{sect-zsurvey}, this parameter is responsible for setting the resolution in the radial direction; beyond this scale, non-linear peculiar velocities wash out all redshift information.

\begin{figure}[t]
\includegraphics[width=\columnwidth]{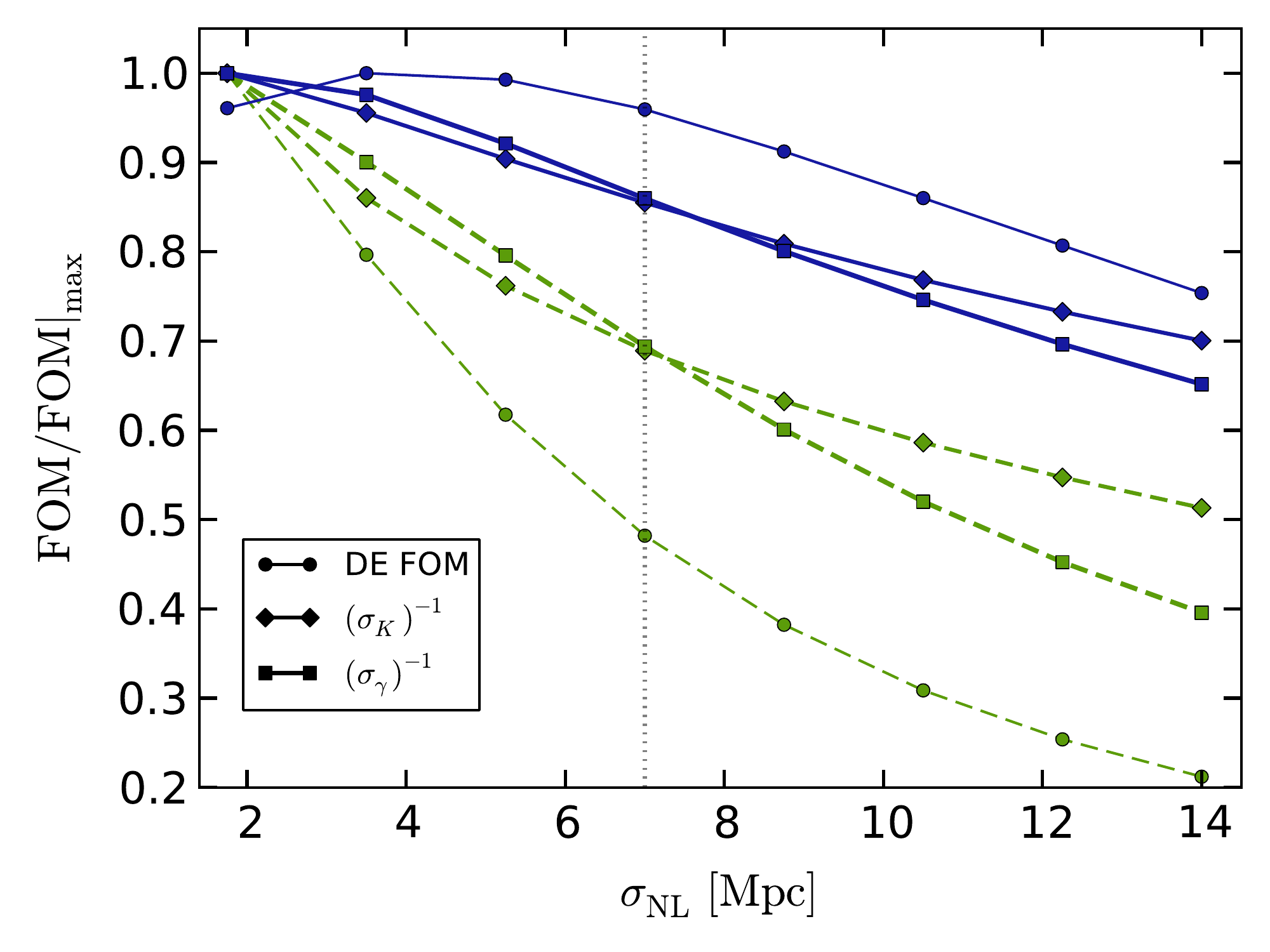}
\caption{Normalised FOM/marginal errors as a function of $\sigma_\mathrm{NL}$ (see Fig. \ref{fig-omegaHI} for key).} 
\label{fig-sigmaNL}
\end{figure}

Fig. \ref{fig-sigmaNL} shows the effect of changing the fiducial non-linear scale. As one might expect, increasing $\sigma_\mathrm{NL}$ degrades the various constraints, as information is lost at progressively larger scales. For the \StageThree\ experiment, which uses a combined single-dish and interferometer mode, the change in non-linear scale has a similar, relatively mild effect on all of the figures of merit, which change by less than a factor of two for a doubling of $\sigma_\mathrm{NL}$. This is not the case for the purely interferometric \StageTwo\ survey, which is more sensitive to smaller scales (especially at low $z$), so is hit harder by the loss of information there.

\subsection{Foreground contamination} \label{sect-fg}

The viability of intensity mapping as a cosmological probe vitally depends on the availability of accurate foreground removal techniques, as the contaminating signals have an amplitude of between 4 to 6 orders of magnitude greater than the cosmological HI signal (e.g. \citet{Alonso:2014sna}). In the previous sections we adopted a fiducial value for the residual foreground contamination amplitude of $\epsilon_\mathrm{FG} = 10^{-6}$, which is a reasonable target value for current foreground subtraction methods. In this section we quantify the sensitivity of our forecasts to the assumed removal efficiency, and discuss a number of potential problems surrounding foreground contamination.

Most of the proposals for how to subtract foregrounds from IM data rely on a simple qualitative assumption: that foregrounds have a smooth (coherent) frequency dependence over the observed frequency ranges\footnote{See \citet{2006ApJ...648..767M} for foregrounds affecting the Epoch of Reionisation.}. This is generally true, apart from in the presence of polarisation leakage, which we will discuss shortly. A simple common-sense approach is to assume that the foreground signal along the frequency direction is accurately modelled as a sum of low-order polynomials or similar, which capture what should essentially be a (very mildly) modulated power law behaviour. This is at the heart of the methods presented in \citet{2006ApJ...650..529W, 2008MNRAS.391..383G, 2008MNRAS.389.1319J, 2009MNRAS.398..401L}.

Another possibility is to be agnostic about the frequency dependence of the foregrounds, but decompose the total signal in some form of signal-to-noise eigenbasis. Since the foregrounds have such large magnitudes, the hope is that they will be contained only in the very high signal-to-noise part, and thus will be suitably segregated from the cosmic signal. This is the logic behind the methods used in \citet{Chang:2010jp, 2013arXiv1310.8144W}. When applied to real data in \citet{Chang:2010jp}, it was found that the foregrounds were not as strongly segregated from the cosmic signal as expected, so it was necessary to subtract a larger number of modes than originally planned (inevitably throwing-out some of the cosmological signal too). Even in the optimal case, this type of foreground removal method has been shown to leave a residual bias in (e.g.) the cosmological parameters that best-fit the recovered BAO \citep{2013arXiv1310.8144W}.

As described in Section \ref{forecast-dg}, we model the effects of foregrounds in terms of a residual noise term with an overall relative amplitude, $\epsilon_\mathrm{FG}$, and a minimum cutoff wavenumber, $k_\mathrm{FG}$, along the radial (frequency) direction. For any given foreground removal method, these two parameters are intertwined -- the more large-scale modes one uses to estimate the shape of the foregrounds (i.e. the larger $k_\mathrm{FG}$ is made), the better the removal efficiency, and so the lower $\epsilon_\mathrm{FG}$ should be. We have chosen $k_\mathrm{FG}$ to be a fixed fraction of the total bandwidth across {\it all} the redshift slices of a survey, and have implicitly absorbed the removal efficiency into $\epsilon_\mathrm{FG}$. (Note that this model is purely stochastic, and does not allow us to model the biases that were discussed in \citet{2013arXiv1310.8144W}.)

\begin{figure}[t]
\includegraphics[width=\columnwidth]{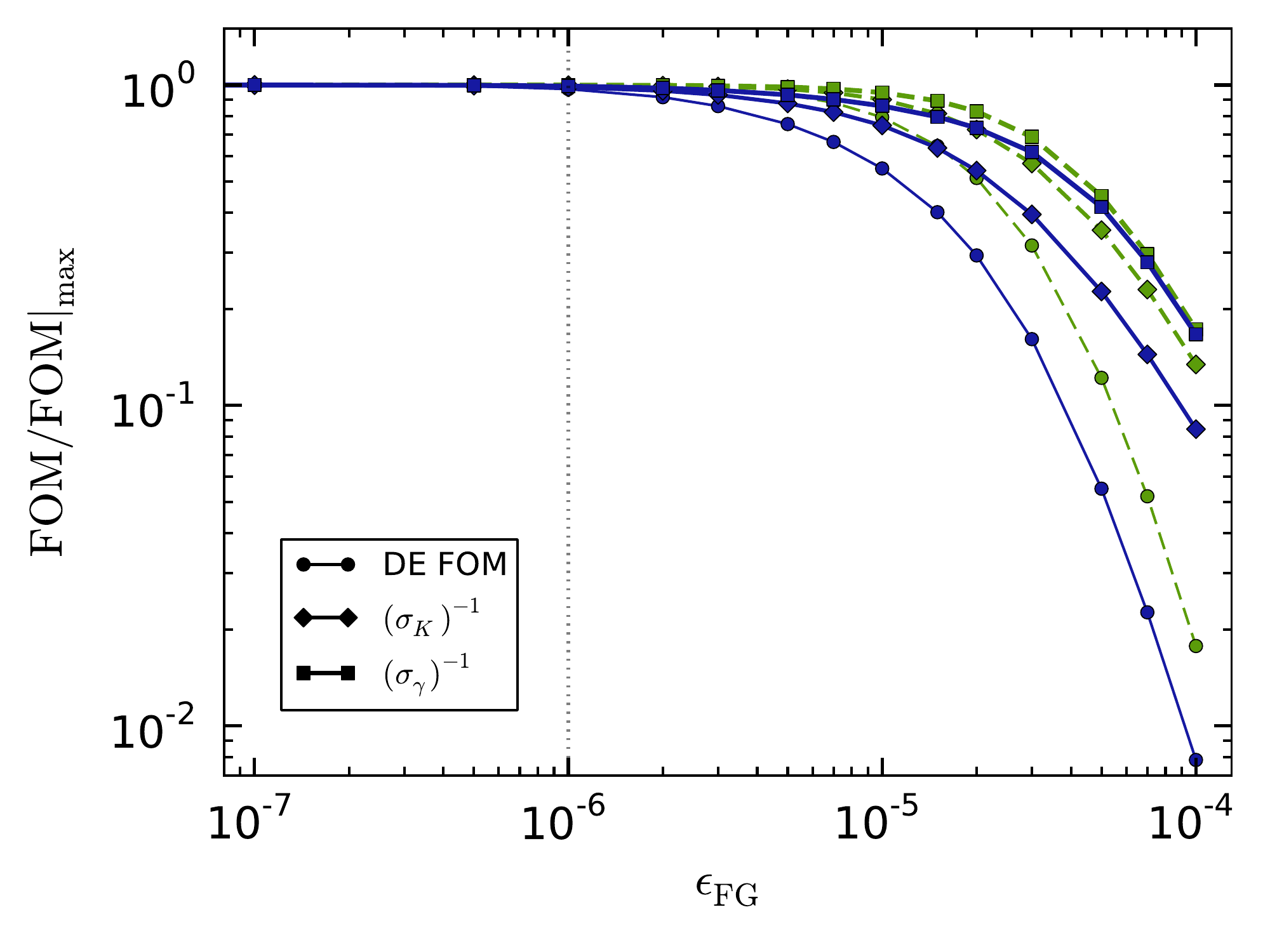}
\caption{Normalised FOM/marginal errors as a function of $\epsilon_\mathrm{FG}$ (see Fig. \ref{fig-omegaHI} for key).}
\label{fig-fg-efficiency}
\end{figure}

\begin{figure}[t]
\includegraphics[width=\columnwidth]{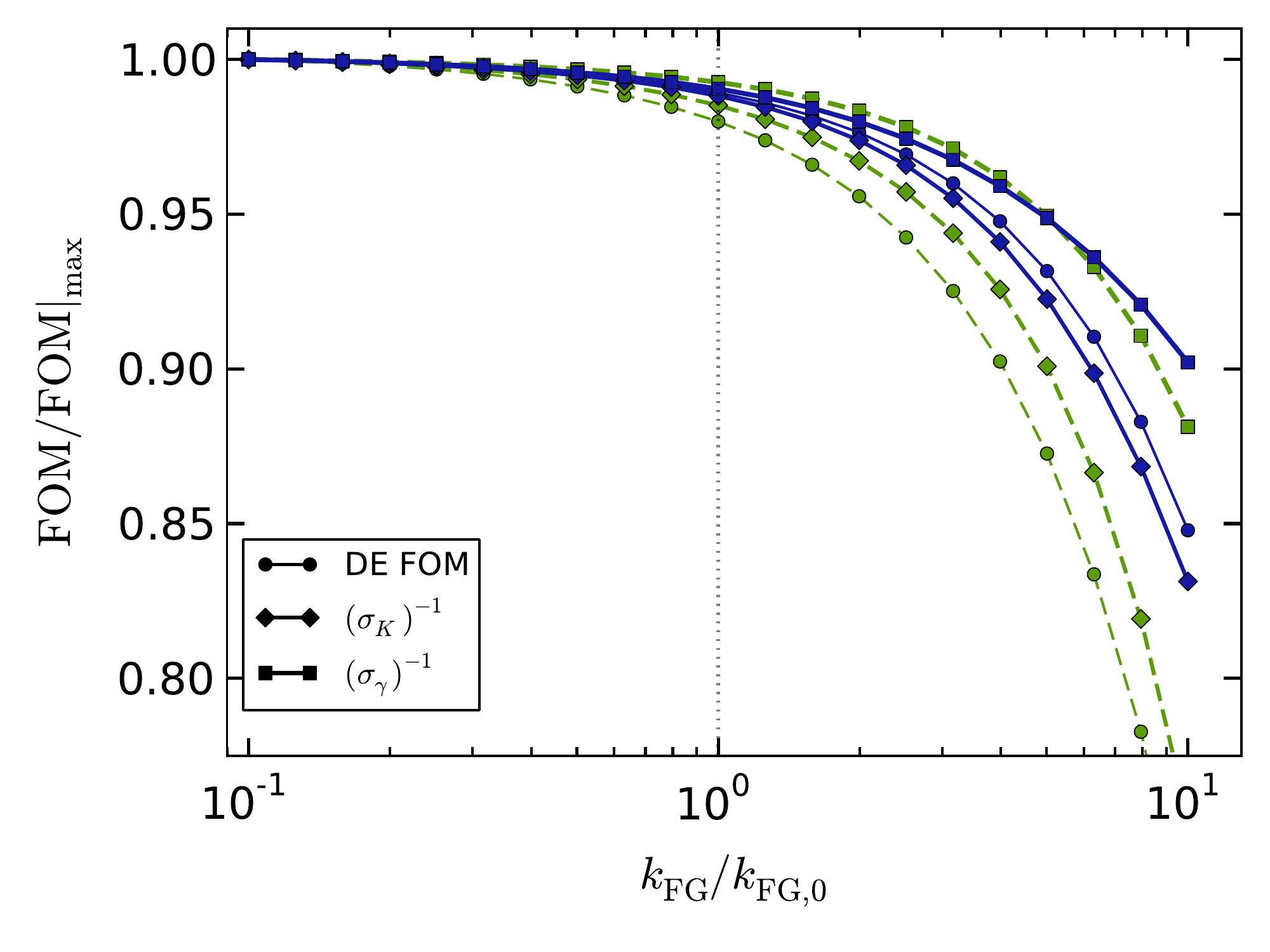}
\caption{The effect of rescaling the foreground cutoff scale, $k_\mathrm{FG}$, on the normalised FOM/marginal errors (see Fig. \ref{fig-omegaHI} for key). The base value of the cut-off scale is $k_{\mathrm{FG}, 0} = 2 \pi / (r_\nu \Delta{\tilde \nu}_{\rm tot})$.}
\label{fig-kfg}
\end{figure}

In Fig. \ref{fig-fg-efficiency}, one can clearly see that the optimal level of foreground subtraction is around $\epsilon_\mathrm{FG}\simeq 10^{-6}$ for the \StageThree\ configuration, but can be larger for other configurations (which have higher noise levels). The impact of changing $k_\mathrm{FG}$ is shown in Fig. \ref{fig-kfg}; while larger $k_\mathrm{FG}$ should yield better foreground subtraction, there is a trade-off involved in losing more data at small $k$.

While our analysis has so far focused on unpolarised foregrounds, problems may arise if one considers instrumental leakage from {\it polarised} foregrounds into the total intensity mode. For a typical receiver, one expects a cross-leakage of order a few percent, so given the large amplitude of the foregrounds, this can have a significant effect on the total signal. Synchrotron emission is the main polarised foreground, and has a non-trivial angular and frequency dependence due to Faraday rotation. To see this, consider the polarisation angle, $\phi$, which is rotated by the galactic magnetic field, $\bm{B}$, through
\be
\phi(r) = \phi_0(r) + c^2 \nu^{-2} \int_0^r n_e(r^\prime) \bm{B}(r^\prime) \cdot dr^\prime, \label{eqn-faraday}
\ee
where $\phi_0(r)$ is the initial angle, $n_e$ is the electron density, and $r$ is the distance along the line of sight. The galactic magnetic field is a non-linear superposition of an overall coherent mode, tied to the spiral structure of the Galaxy, with a turbulent stochastic mode on small scales \citep{2001SSRv...99..243B}. From (\ref{eqn-faraday}), one can see that the rotation of the polarisation vector depends on both frequency and the line-of-sight through the galaxy with some decoherence length (corresponding to the coherence of $\mathbf{B}$ along the line of sight). This leads to a more complex foreground signal that is considerably less smooth in frequency than unpolarised foregrounds -- an effect that increases in severity at low frequencies \citep{Alonso:2014sna}. While we have not explicitly included it in our foreground model, the effects of polarisation leakage can be partially accounted for by a larger fiducial $\epsilon_\mathrm{FG}$.

\subsection{Autocorrelation calibration (single-dish)} \label{sec:autocorr}

In this paper, we have advocated using some instruments as collections of single-dish experiments, i.e. in autocorrelation mode. This is common practice in CMB mapping experiments, and has been the leading method of producing large-scale, high-resolution maps. Its use with radio telescope arrays at lower frequencies is {\corr less common, however, and must be treated with some care because of a number of potentially serious systematics.}

Nevertheless, there is some precedent for using autocorrelation mode to detect both individual HI sources and unresolved emission. In \citet{2003A&A...406..829B}, the WSRT array was used in this mode to perform a wide-field survey, yielding a sample of $\sim\!150$ HI galaxies. A key difficulty of the analysis was in obtaining an accurate calibration of the autocorrelation mode -- while the average gain over all receivers was relatively stable, there were variations of up to 10\% for individual receivers. This was calibrated out by using cross-correlation data for known radio sources. For the first attempt at mapping the unresolved HI signal with the GBT telescope \citep{Chang:2010jp}, the flux calibration was controlled by fixing an intermittent noise source at the feed point, and by periodically monitoring a known source.

Drifts in the gain (e.g. due to instrumental temperature variations) are just one type of autocorrelation systematic. Another is due to spillover and sidelobe pickup, which can arise from a poorly characterised beam and ground contamination. This is not an insurmountable problem, and ground-based and balloon-borne CMB experiments commonly incorporate design features to mitigate these effects. One approach for HI instruments is that proposed by BINGO \citep{2013MNRAS.434.1239B}. There, the idea is to use a partially-illuminated aperture to reduce the effect of sidelobes, spillover, and RFI contamination. Another aspect of the BINGO design is the use of a fixed dish, which scans the sky simply by allowing it to drift through the beam as the Earth rotates. This bypasses various problems that arise with moving parts, and allows a more precise pointing calibration than is possible with dynamic ``raster'' scan strategies.\footnote{{\corr See also the `on-the-fly' technique \citep{2007A&A...474..679M}.}}

Another key obstacle in the analysis of autocorrelation data is the presence of $1/f$ noise, which is a coherent {\corr (correlated)} noise drift on long timescales. Ideally, one would be able to construct a receiver system such that the $1/f$ knee (i.e. the timescale beyond which correlations become important) is at very low frequencies -- a few $\times 10^{-3}$ Hz, for example. If this is possible, then the noise will drift over periods of several minutes which, for a drift scan, corresponds to angular scales of a few degrees (i.e. larger than the BAO scale). This is the method used by BINGO.

More traditionally, the way to deal with noise drifts or biases in CMB experiments has been to devise a scan strategy such that the modes one is looking at are scanned at frequencies higher than the knee frequency. The signal in a given pixel will be then be localised in particular (frequency) modes of the time series which are subject only to white (i.e. uncorrelated) noise. This requires that the instrument should have the ability to scan quickly across the sky, which can be challenging for large dishes. One can can further mitigate the effect of $1/f$ noise by devising a scan strategy with multiple cross-linking, i.e. in which each pixel is revisited a number of times on different time scales, from different directions. The inversion process for extracting the map from the time series (and estimating the noise) is then well-defined and numerically robust \citep{2000MNRAS.312...89F}. {\corr If the noise is smooth in frequency, it may also be possible to clean out any noise bias component, as one would a foreground component in the signal.}

In summary, there are clearly a number of issues that must be considered carefully when working in autocorrelation mode, but as we have shown in our analysis throughout this paper, the scientific potential of single-dish IM experiments is tremendous. {\corr Ideally, one would design the dishes, receivers, and other hardware of a given array specifically to mitigate the problems discussed above, but in some cases (e.g. with the SKA) this is not possible due to competing design constraints enforced by the need to operate primarily in an interferometric mode -- optimising the hardware for autocorrelation mode as well is simply too expensive. Any chance of controlling these (potentially critical) systematic effects to a sufficient degree is then left down to the choice survey strategy and data analysis methodology.}

{\corr Experience with autocorrelation CMB experiments suggests that reliance on non-hardware techniques to reduce important systematics is a risky strategy, but there are a number of features of forthcoming radio telescope arrays that may be helpful in making this feasible. For example, an experiment with hundreds of individual dishes can make use of the fact that some systematics will be uncorrelated between the dishes; by cross-correlating (detected) maps of the same volume produced by different dishes, many effects can therefore be expected to correlate out. Nevertheless, it remains to be demonstrated that an experiment as complex as Phase 1 of the SKA can successfully control its autocorrelation calibration down to the required level, and how the survey/analysis strategy affects its overall sensitivity. This must therefore be seen as an important caveat of our analysis.}

\subsection{Sensitivity to large scales (interferometers)}

A particular limitation to HI mapping with interferometers is the difficulty of sampling modes on large angular scales. In the simplest model of an interferometer -- as a collection of dishes -- the minimum measurable wavenumber is set by the minimum baseline, which cannot be smaller than the diameter of the dishes. From Fig. \ref{fig-resolution-z} we can see that arrays with large dishes will not adequately sample BAO scales at low redshift in interferometer mode, so most of the constraints must come from single-dish mode.

There are a few ways to mitigate this shortcoming. The simplest is the approach effectively taken by BAOBAB and dense aperture arrays -- to just use smaller dishes and pack them closer together. This results in smaller baselines and a larger field of view for the interferometer, but reduces its total effective collecting area (and thus its sensitivity). One can add more dishes to compensate, although this can substantially increase the cost of correlation hardware, which scales roughly like $\sim \! N^2_\mathrm{dish}$. GPU-based correlators, or correlating only a subset of receiver pairs, can reduce costs for large numbers of receivers.

Alternatively, one can use a more novel reflector design. For example, CHIME uses long cylindrical reflectors with many closely-spaced receivers installed along the cylinder \citep{Shaw:2014khi}. This provides a large number of short baselines, and a primary beam that is $\sim\!\!180^\circ$ in one direction but much narrower along the orthogonal direction. This leads to a very anisotropic sampling of transverse Fourier modes, and only the large angular modes that are exactly aligned with the cylinders will be properly sampled. Because the visibilities measured by the interferometer are convolved with a window function defined by the primary beam, however, modes larger than the shortest side of the FOV will be aliased, making them difficult to disentangle. This is only the case if the interferometer tracks a single patch of the sky, though; if one progressively scans over the patch with different pointing offsets, and has precise knowledge of the primary beam pattern, it is possible to remove the aliasing effect of the primary beam and thus independently measure modes larger than the instantaneous FOV by {\it mosaicing} \citep{1999ASPC..180..401H}. In the case of CHIME, drift scanning provides a continuous range of pointing offsets, and in principle the array can see the whole sky over a 24 hour observation period.



One can also make interferometric measurements over a number of separate pointings without mosaicing, simply to survey a larger area of sky \citep{1999ApJ...514...12W}. Drift scanning can be seen as a continuous limit of this. The advantage of such a method is that one can greatly reduce the sample variance of the smallest-baseline modes, simply by observing them on several independent patches of the sky. A crucial point is that simply patching together multiple fields does not allow modes larger than those defined by the minimum baseline to be measured, and therefore does not change the range of modes sampled, but {\it does} increase the sensitivity within that range. We have implicitly assumed that interferometers can handle multiple pointings by allowing $S_\mathrm{area} > \mathrm{FOV}$ in our forecasts.

\subsection{Combined mode}

{\corr Another possibility is to operate some experiments in a ``combined mode'', where both autocorrelation and cross-correlation data are collected.} The simplest way of doing this in practise is to split the total survey time into two chunks, using only one of the observing modes for each. From the previous two sections, we can see that each mode will have different systematics and hardware requirements, and it is likely that substantially different survey strategies would be needed for each mode.

The situation is considerably more difficult if one tries to collect data in both modes simultaneously. As discussed above, in single-dish mode one has to mitigate $1/f$ noise, typically by rapidly scanning across the sky. Conversely, interferometers require precisely-measured baselines and pointings to allow accurate phase calibration and reconstruction of the beam pattern; as dishes accelerate while scanning, even small distortions of the mounts can make this difficult. The combination of all these issues can potentially be overcome by drift scanning or through the use of novel mounts, but neither option appears to have been tested yet.

\section{An Ideal HI Survey?} \label{sect-survey-design}

In this section, we suggest what an `ideal' future HI intensity mapping experiment for late-time cosmology would look like. So far, we have assumed that Phase I of the SKA will represent the pinnacle of HI intensity mapping science for the coming decade. While its performance in terms of cosmological parameter constraints will indeed be impressive, it is worth remembering that the SKA is a general purpose facility, and is not specifically designed for IM. We propose that a cheaper purpose-built instrument, optimised for HI science, would be able to match, and perhaps even surpass, the SKA's performance.

\begin{figure}[t]
\includegraphics[width=\columnwidth]{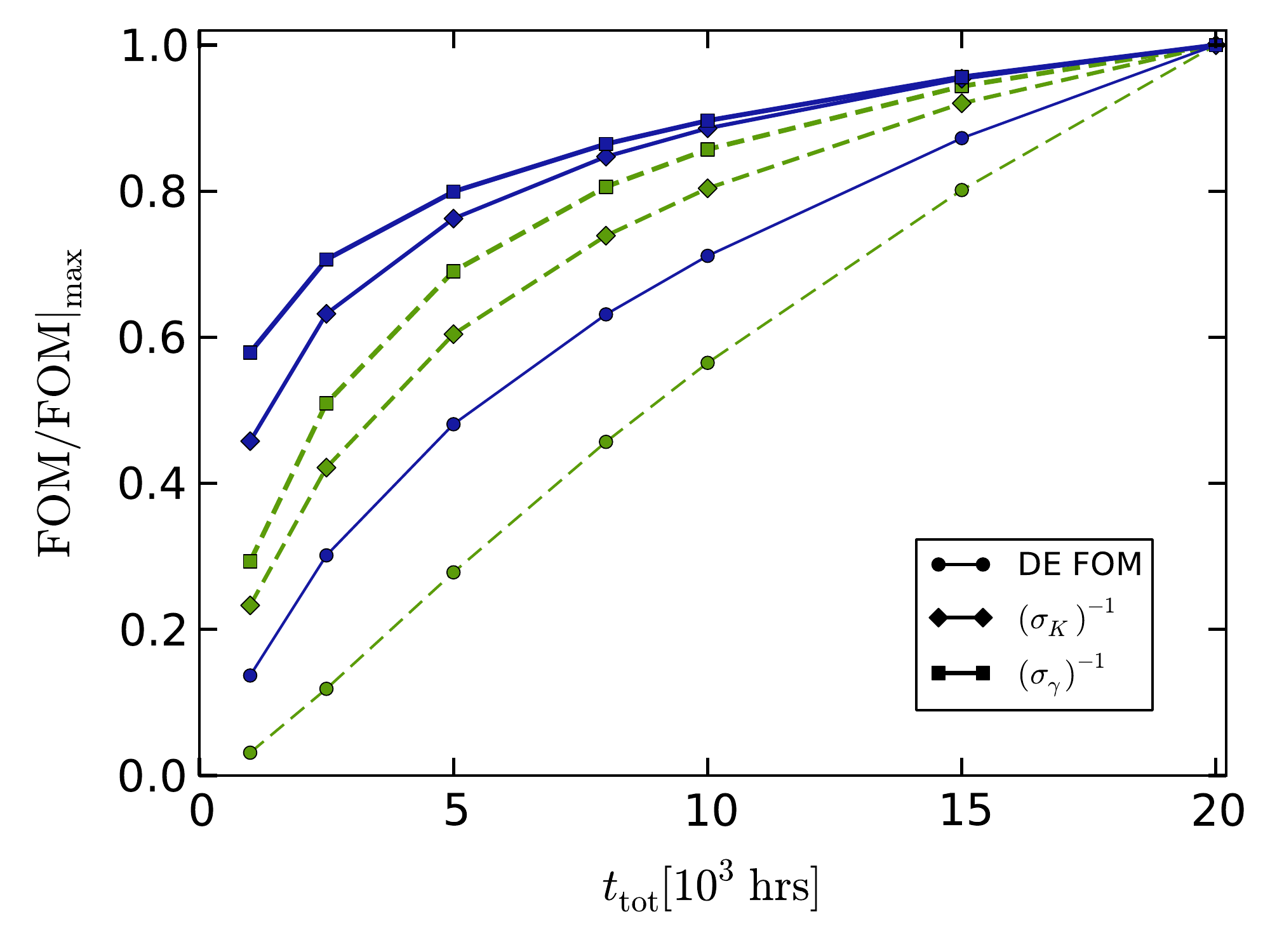}
\caption{Normalised FOM/marginal errors as a function of survey duration, $t_\mathrm{tot}$ (see Fig. \ref{fig-omegaHI} for key).}
\label{fig-ttot}
\end{figure}

For the sake of simplicity, we target the dark energy figure of merit (FOM) as the only parameter to be optimised for. Sensitivity to intermediate scales, where the BAO and other distance measures are most important, is therefore a priority, although distance information from RSDs and the overall shape of the matter power spectrum are also useful (see Sect. \ref{sect-dahz}). Dark energy is typically most important at low redshift, and so one might reasonably expect to focus the survey on the interval $0 \leq z \leq 2$. The lowest redshifts are likely to be subject to RFI, however, and pushing to higher redshifts brings in issues of limited angular resolution and increasing galactic foreground emission.

\begin{figure}[t]
\includegraphics[width=\columnwidth]{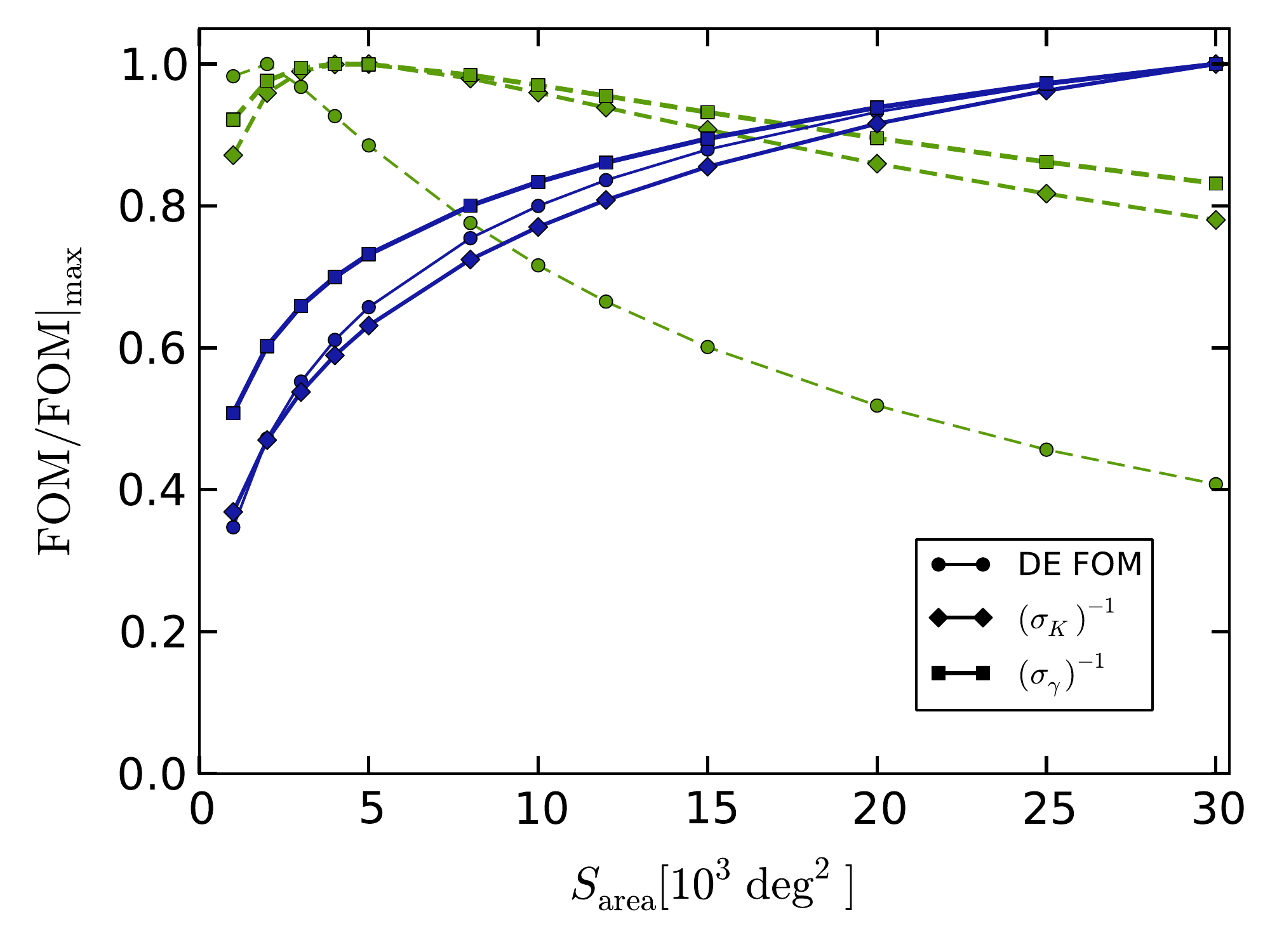}
\caption{Normalised FOM/marginal errors as a function of survey area, $S_\mathrm{area}$ (see Fig. \ref{fig-omegaHI} for key).}
\label{fig-sarea}
\end{figure}

In terms of the available technology, we assume that $T_\mathrm{inst} \approx 25$K wideband receivers can be built cheaply and in bulk, and that correlators for several thousand receivers will also be relatively affordable. Time allocation is not an issue for a purpose-built instrument and, as we can see from Fig. \ref{fig-ttot}, one only gains from increasing the amount of integration time. We assume that at least 10,000 hours of effective observing time can be used. This leaves only a handful of basic design parameters that can be varied:
\begin{itemize}
 \item Survey area, $S_\mathrm{area}$
 \item Dish size, $D_\mathrm{dish}$
 \item Array configuration (maximum and minimum baselines, $D_\mathrm{max}$ and $D_\mathrm{min}$, and filling factor)
 \item Frequency range (corresponding to redshift range, $[z_\mathrm{min}, z_\mathrm{max}]$)
\end{itemize}
The optimal survey area for a fixed amount of integration time depends on how quickly one can integrate down to the signal-dominated regime at each pointing. In Fig. \ref{fig-sarea} we show that, in the case of the \StageTwo\ experiment, increasing $S_\mathrm{area}$ above its optimal value will lead to a reduction in overall signal-to-noise and hence in the overall FOM. This is not the case for the \StageThree\ experiment, which already has sufficient time to reach signal-domination at each pointing; increasing $S_\mathrm{area}$ simply reduces the cosmic variance and therefore improves the FOM. In designing an optimal survey, one should pick $S_\mathrm{area}$ such that it is signal-dominated at each pointing, but only just, so as not to spend too much time integrating in a regime that is dominated by cosmic variance. This is the approach used when designing survey strategies for CMB experiments.

For interferometers, a high filling factor is desirable, as it equates to higher sensitivity. For a fixed number of dishes, increasing the filling factor amounts to increasing the dish diameter, or decreasing the maximum baseline length. Smaller dishes are useful for increasing the field of view, and thus the survey speed, however, and allow for smaller minimum baselines, which is important for resolving larger angular scales, especially at low redshift.

For single-dish experiments, smaller dishes also improve survey speed due to their increased field of view, but this now comes at the cost of angular resolution. A balance must therefore be found between resolving intermediate scales over as much of the redshift range as possible, and survey speed. There is also the issue that larger dishes cost more, but for a pure single-dish experiment this is offset by there being no need for expensive correlator hardware.

Under the design constraints that we imposed, it turns out that the SKA Phase I arrays are close to ideal for single-dish experiments, given the assumed foreground removal efficiency. This was to be expected; \StageThree\ (broadly representative of a single-dish SKA configuration) is already nearing the cosmic variance-limited constraints on $P(k)$ from the DETF Stage IV galaxy redshift survey, as we showed in Fig. \ref{fig-pk-constraints}. The only significant improvement to be had is around $k \sim 0.1$ Mpc$^{-1}$, which could be obtained by increasing the survey time or, even more effectively, by decreasing $z_\mathrm{min}$. Shifting the maximum frequency of \StageThree\ from 1100 to 1200 MHz while keeping the total bandwidth fixed to 700 MHz effectively matches its FOM to that of the galaxy redshift survey.

As interferometers, the SKA configurations have too small a field of view and too low a filling factor to achieve competitive dark energy constraints. A purpose-built interferometer operating over the desired redshift range would be better off having much smaller dishes, closely packed together. This is the approach that CHIME and BAOBAB are effectively taking. A 250-element array with 2.5m dishes distributed over a 44m core (giving a filling factor of 0.8) would surpass \StageThree's FOM for $\nu_\mathrm{max} = 1100$ MHz, and match the galaxy survey's for $\nu_\mathrm{max} = 1200$ MHz (where $\Delta \nu = 700$ MHz in both cases).

\begin{figure}[t]
\includegraphics[width=\columnwidth]{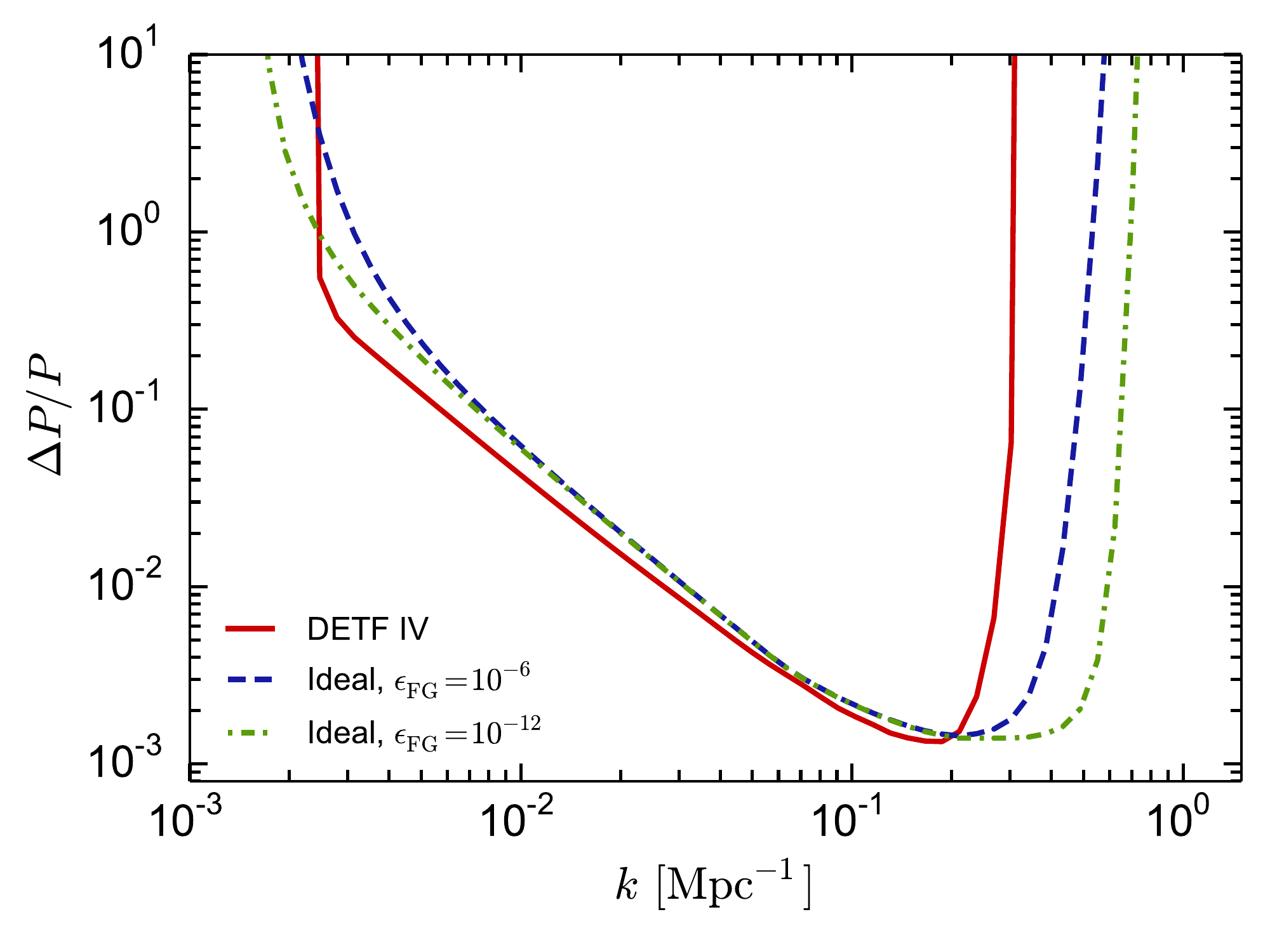}
\caption{Fractional constraints on $P(k)$ for an `ideal' (noise-free) 15m single-dish survey, covering the same redshift range and survey area as the DETF Stage IV reference experiment. The deviation from the cosmic variance limit on large scales is (partially) due to the $k_\mathrm{FG}$ cutoff.}
\label{fig-ideal}
\end{figure}

Is it possible for IM experiments to do better than the reference galaxy redshift survey? Yes, but not without relying on either a higher maximum frequency or small angular scales. On large scales, the single-dish SKA configurations are limited by residual foregrounds, as shown in Fig. \ref{fig-ideal}. Extending to higher redshifts increases the total volume being probed, but the sensitivity to dark energy decreases significantly above $z \gtrsim 2$, so in practise little is gained by doing this. As we have already seen, going to lower redshift (i.e. increasing $\nu_\mathrm{max}$) can have a big effect, as dark energy is most important here. The problem of RFI (radio interference) increases towards 1.4 GHz though, so this is also difficult. Besides, other sources of information on the matter density field are available at low-$z$ (e.g. existing galaxy redshift surveys), so it is not clear whether extending IM surveys into this region would be particularly useful.

Another option is to improve sensitivity on small angular scales, $k_\perp \gg 0.1$ Mpc$^{-1}$. This can be achieved by increasing the number of detectors, improving the single-dish angular resolution, reducing the instrumental noise, or performing longer surveys (Fig. \ref{fig-ideal} shows the ideal case). In theory this would provide extra distance information from the shape of the power spectrum, but this relies on being able to accurately model the non-linear power spectrum, which is also tricky. To significantly improve DE constraints past what a \StageThree-class experiment is capable of, one is probably better off focusing on combining IM with other probes, such as weak lensing.

\section{Discussion} \label{sect-discussion}

Neutral hydrogen (HI) intensity mapping looks set to become a leading cosmology probe during this decade. In this paper we have assessed its potential for constraining cosmological parameters, focusing on `late times', $z \lesssim 3$. We used a few reference experimental designs -- \StageOne, \StageTwo, and \StageThree\ -- that are inspired by up-and-coming experiments to assess how well we will improve our knowledge of the standard cosmological model, the nature of dark energy, the spatial curvature of the Universe, and the growth rate of structure. We have done so being mindful of the potential systematic problems that need to be faced.

Intensity mapping at radio frequencies has a number of advantages over other large scale structure survey methodologies. Since we only care about the large-scale characteristics of the HI emission, there is no need to resolve and catalogue individual objects, which makes it much faster to survey large volumes. This also changes the characteristics of the data analysis problem; rather than looking at discrete objects, one is dealing with a continuous field, which opens up the possibility of using alternative analysis methods similar to those used (extremely successfully) for the CMB. Thanks to the narrow channel bandwidths of modern radio receivers, one automatically measures redshifts with high precision too, bypassing one of the most difficult aspects of performing a galaxy redshift survey.

These advantages, combined with the rapid development of suitable instruments over the coming decade, look set to turn HI intensity mapping into a highly competitive cosmological probe in only a short space of time. In Sect. \ref{sect-parameters}, we showed that \StageThree-class experiments would be broadly competitive with DETF Stage IV galaxy redshift surveys such as Euclid, LSST, and WFIRST in terms of cosmological parameter constraints, in about the same timeframe. Indeed, the largest planned surveys, such as SKA1-SUR, may even be able to surpass the cutting-edge galaxy surveys, although this is contingent on the (currently poorly-known) HI density and the performance of foreground removal algorithms. Since the currently-planned \StageThree\ class surveys are not specifically designed for IM, we also considered what a large, purpose-built HI experiment would be able to achieve in Sect. \ref{sect-survey-design}. We found that little extra could be gained without pushing to higher frequencies or smaller (non-linear) angular scales; neither are free of problems.

\begin{figure}[t]
\includegraphics[width=\columnwidth]{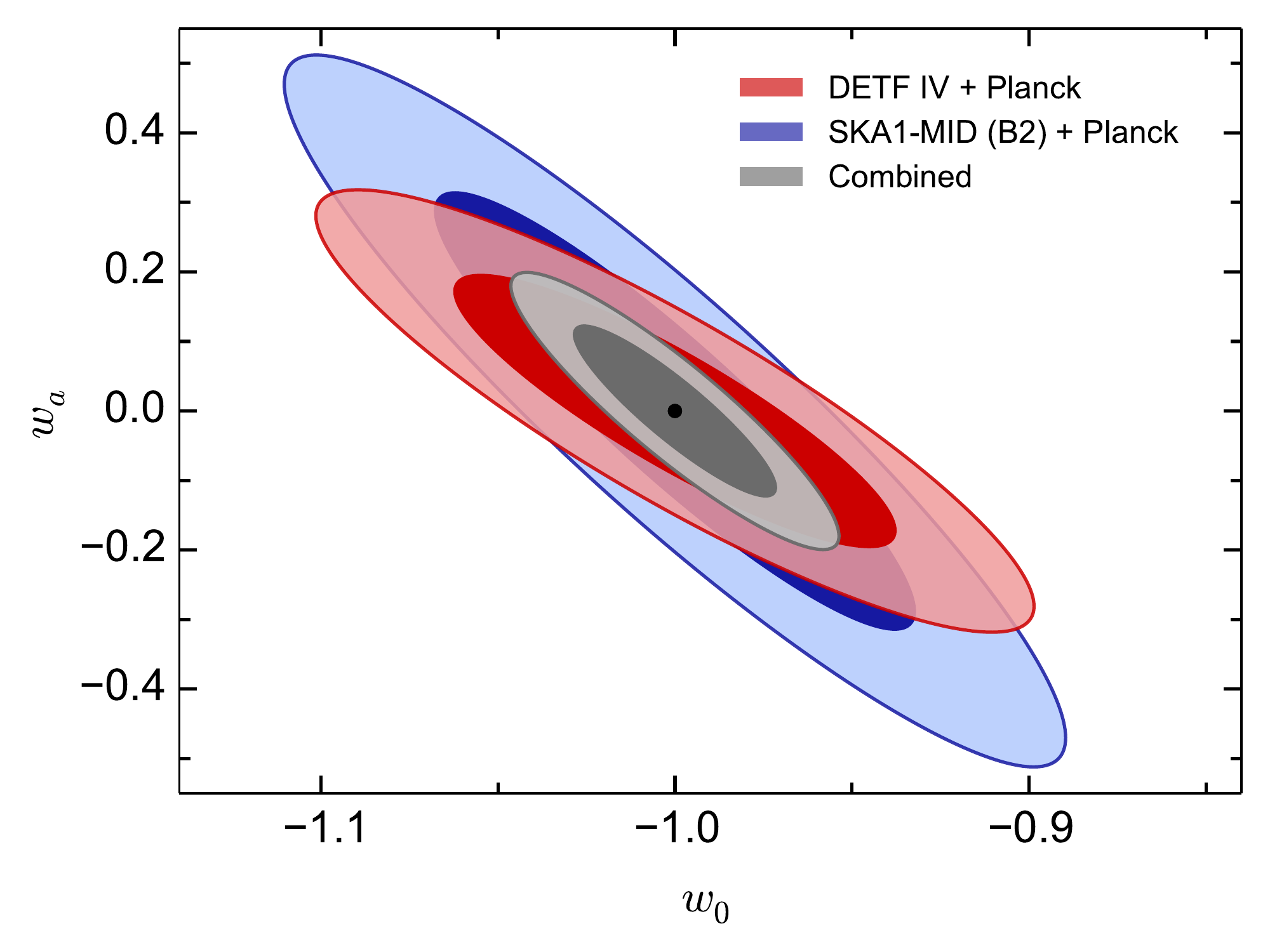}
\caption{Constraints on $w_0$ and $w_a$ from the combination of the DETF Stage IV galaxy survey and a combined-mode SKA1-MID configuration, compared with results for the experiments individually. The figures of merit are 427 (SKA1-MID), 438 (galaxy survey), and 2124 (combined). We have assumed that the survey volumes are independent; otherwise, cosmic variance would degrade the combined constraint.}
\label{fig-w0wa-combined}
\end{figure}

More important than their individual performance is what IM and galaxy redshift surveys can do in combination. In Fig. \ref{pub-w0gamma}, we showed that \StageThree\ and a DETF Stage IV survey give roughly orthogonal constraints on $w_0$ and $\gamma$ when combined with CMB data, mostly as a result of their complementary redshift coverage. Fig. \ref{fig-w0wa-combined} shows the joint constraints on $w_0$ and $w_a$ for the combination of DETF Stage IV and a combined-mode SKA configuration with a lower-redshift band; the resulting dark energy FOM is almost five times that of either survey individually. This large improvement is due to the increase in the total surveyed volume, as well as the complementary redshift coverage. One can also benefit from the ``multi-tracer'' effect, whereby the limits imposed by cosmic variance on some variables can be overcome by measuring several distinct populations of tracers of the cosmic density field \citep{McDonald:2008sh, 2013MNRAS.432..318A}. Combining HI intensity mapping and galaxy redshift surveys should therefore offer particularly stringent constraints on the dark energy equation of state and growth index parameters -- an absolute necessity for distinguishing between different dark energy and modified gravity models.

\begin{figure}[tb]
\includegraphics[width=\columnwidth]{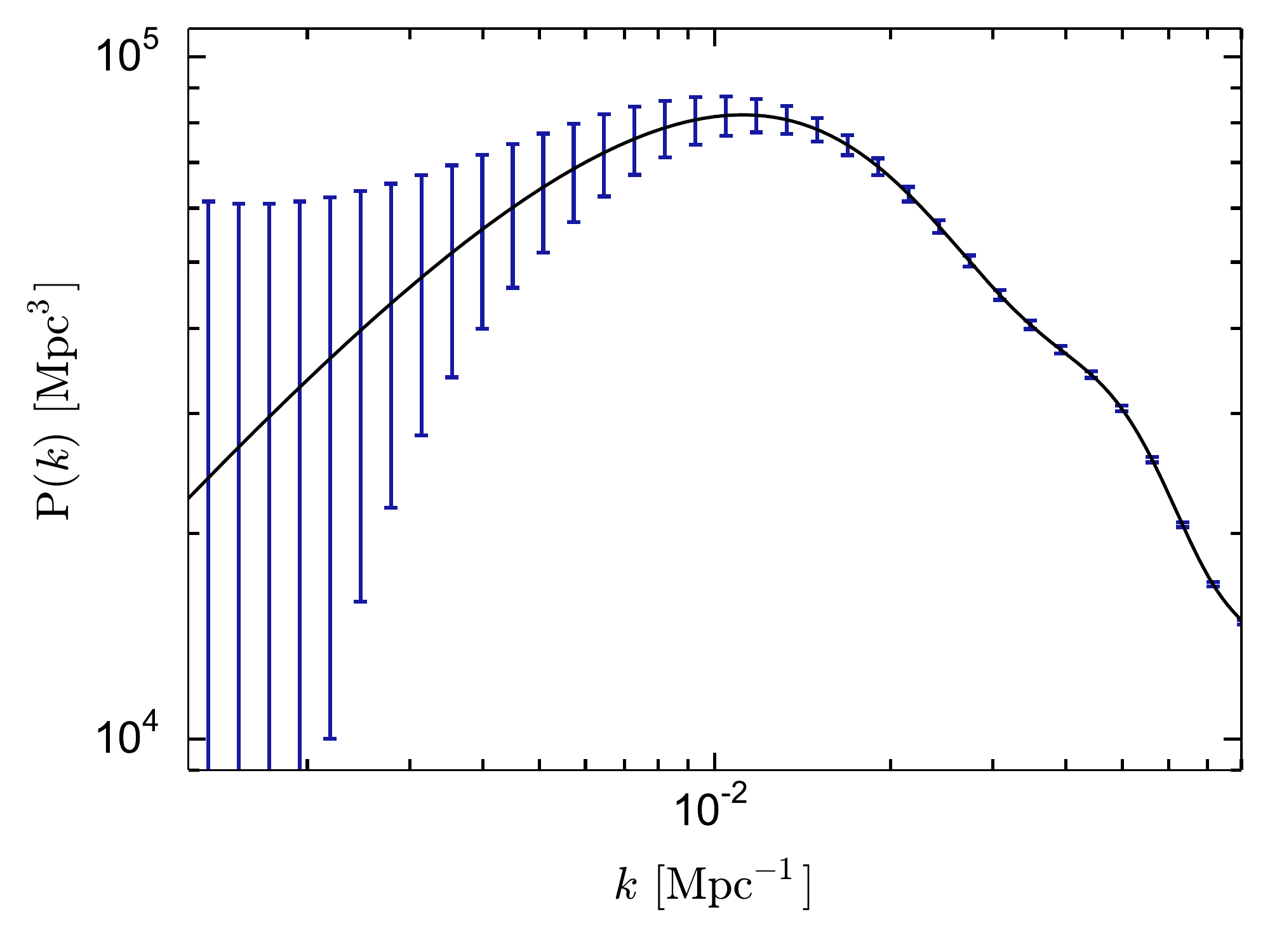}
\vspace{-2em}\caption{Forecast constraints on $P(k)$ from SKA1-MID (B1), assuming perfect foreground removal (i.e. $\epsilon_{\rm FG} = k_{\rm FG} = 0$). The turnover in the power spectrum should be clearly detectable.} 
\label{fig-pk-peak}
\end{figure}

HI intensity mapping experiments also offer some novel features -- for example, in their ability to probe ultra large scales in the late Universe. \StageThree-class arrays like Phase I of the SKA will be able to simultaneously survey an extremely wide range of redshifts over greater than half of the sky, covering volumes of several tens of cubic Gpc in one fell swoop. This is sufficient to detect physical effects beyond the matter-radiation equality scale (Fig. \ref{fig-pk-peak}), including non-Gaussianity, spatial curvature, and potential deviations from large-scale homogeneity and isotropy. As was shown in Fig. \ref{fig-resolution-z}, a sufficiently large HI survey could even probe beyond the horizon size at $z \gtrsim 1$, allowing us to access causally-disconnected regions long after recombination.

Before HI intensity mapping can contribute seriously to late-time cosmology, a number of potential pitfalls must be navigated. Chief amongst these is the overall magnitude of the HI density, $\Omega_{\mathrm{HI},0}$, and its evolution with redshift, both of which are currently poorly constrained. The lower the density, the harder the HI signal will be to detect (and the more aggressive the foreground cleaning will need to be). Fig. \ref{fig-omegaHI} showed the effect of changing $\Omega_{\mathrm{HI},0}$ on various figures of merit; a factor of two reduction in HI density from our fiducial value results in roughly a factor of five degradation in parameter constraints, which would be troublesome, although not catastrophic. Similarly, a higher density would make the signal much easier to detect. This situation is loosely analogous to the dependence of galaxy cluster surveys on the normalisation of the power spectrum -- when it was found that $\sigma_8$ was closer to 0.8 than 0.9, this vastly reduced expected cluster number counts, leading to a corresponding drop in forecast constraints from cluster surveys. For HI intensity mapping, all we can do is wait for better measurements of $\Omega_{\mathrm{HI}, 0}$ to see what the effect will be.

A confounding factor that is more directly under our control is the foreground cleaning efficiency, which we investigated in Sect. \ref{sect-fg}. The galactic foreground signal is around six orders of magnitude larger than the cosmological HI signal, but has a distinctive (and thus easy to separate) behaviour for the most part. Since they vary on similar angular/frequency scales to the galactic foregrounds, large-scale cosmological modes are likely to be hit harder by imperfect foreground cleaning. There is no reason why this cannot be overcome, however -- similarly large modes are routinely dealt with successfully in CMB analysis \citep{Ade:2013hta}. Of potentially more concern is the issue of polarisation leakage, which imprints a more variable signal on top of the cosmological one. Sufficiently sophisticated modelling, combined with a sustained effort to control or characterise leakage at the hardware level, should be able to deal with this.

Unsurprisingly, choosing the right survey strategy is vital; we investigated the effect of changing various survey parameters in Sect. \ref{sect-survey-design}. Throughout the paper, we have also considered the difference in performance between interferometer and single-dish observation modes. For purpose-built IM experiments (e.g. the `ideal' experiment described in Sect. \ref{sect-survey-design}), one will tend to prefer interferometry because of the comparative ease of controlling instrumental/atmospheric systematics, although this must be offset against the significant computational expense of correlating many baselines. For general-purpose instruments, whose design is likely to be set by other considerations, interferometry may be a poor choice -- if the array has large dishes, its interferometric field of view will be small, making it relatively insensitive to the intermediate scales that are most useful for detecting the BAO (see Fig. \ref{fig-resolution-z}). In this case, there is much to be gained by using a single-dish (or combined single-dish + interferometer) mode instead. This is the path that we advocate for Phase I of the SKA, at least for low redshifts -- at higher redshifts, the angular resolution in single-dish mode is actually too low, so the errors on quantities such as $D_A(z)$ get larger (see Sect. \ref{sect-dahz}). {\corr This choice brings with it a number of significant data analysis challenges however, as discussed in Section \ref{sec:autocorr}; how to precisely and consistently calibrate many hundreds of dishes operating in autocorrelation mode is currently an open problem, and the various advantages of single-dish operation will only be available if it can be solved.}

As we discussed at the start, our forecasting framework makes a number of approximations such as neglecting wide-angle effects and correlations between redshift bins. These simplifications (which were instrumental in allowing a direct comparison with galaxy redshift surveys) are only likely to have any material impact on our forecasts at the very largest scales, well away from where the strongest distance constraints come from. As such, the constraints on cosmological parameters that we have presented should not be expected to change appreciably under a more sophisticated treatment.

More important are the effects that we have accounted for that are sometimes neglected in other forecasts. By explicitly including non-linearities, unknown bias evolution, and foreground subtraction residuals, and systematically exploring parameter degeneracies, we have tried to be as comprehensive (and pessimistic) as possible in acknowledging possible adversities for IM surveys. This should be kept in mind when comparing results from this paper with those from elsewhere. Even so, there is always scope to disagree with the particular decisions that go into any set of forecasts, so we have made our full forecasting code publicly available, with documentation.\footnote{\url{https://gitlab.com/radio-fisher/bao21cm}} The interested reader is encouraged to use it to make their own forecasts. \\

\textit{Acknowledgments.---} We are grateful to Y. Akrami, D. Alonso, D. Bacon, T. Baker, B. Bassett, R. Battye, E. Calabrese, T.-C. Chang, H.~K. Eriksen, K. Grainge, M. Jones, Y.-C. Li, T. Louis, E. Macaulay, L. Miller, S. N{\ae}ss, U.-L. Pen, W. Percival, R. Shaw, J. Sievers, K. Sigurdson, O. Smirnov, K. Smith, A. Stebbins, A. Taylor, S. Torchinsky, P. Wilkinson, K. Zarb Adami and J. Zuntz for useful discussions. PB is supported by European Research Council grant StG2010-257080, and acknowledges Oxford Astrophysics, Caltech/JPL, and CITA for hospitality. PGF acknowledges support from Leverhulme, STFC, BIPAC and the Oxford Martin School and the hospitality of the Higgs Centre in Edinburgh. PP is funded by a NRF SKA Postdoctoral Fellowship. MGS acknowledges support from the National Research Foundation (NRF, South Africa), the South African Square Kilometre Array Project and FCT under grant PTDC/FIS-AST/2194/2012.

\appendix

\section{HI line intensity and brightness temperature} \label{app-HI-line}
Consider a clump of neutral hydrogen with number density $n_{\rm HI}=n_0+n_1$, where
$0$ and $1$ denote the lower and upper level of the hyperfine splitting respectively. The spin temperature, $T_S$, can be defined using
\bea
n_1=n_0\frac{g_1}{g_0}e^{-\frac{T_*}{T_S}}\simeq 3n_0=\frac{3}{4}n_{\rm HI}, \nonumber
\eea
where $T_*=h\nu_{21}/k_B=0.0682 K$, $g_1=3$, $g_0=1$, and we have assumed $T_S\gg T_*$.
The emissivity (energy per unit time, solid angle/volume, and frequency) of the clump is
\bea
j_{21}&=&\frac{A_{10}h\nu_{21}}{4\pi}n_1\phi(\nu), \nonumber
\eea
where $A_{10} \simeq 2.869 \times 10^{-15} \,\mathrm{s}^{-1}$ \citep{wilson2009tools} is the Einstein coefficient for spontaneous emission and $\phi(\nu)$ is the line profile, which is assumed to be very narrow, with width $d\nu$ (a simple approximation is $\phi\simeq 1/d\nu$). The clump's luminosity is then
\bea
dL = \frac{3}{4}{A_{10}h\nu_{21}}n_{\rm HI}\,\phi(\nu)\, d\nu\, dA\, dr \nonumber,
\eea
where $\nu$ is evaluated in the rest frame of the clump and $dA\, dr$ is the volume of the clump ($dr$ being the distance along the line of sight) in comoving units, if $n_{\rm HI}$ is the comoving number density. Absorption can be neglected if the spin temperature of the gas is much larger than the background temperature (usually the CMB), so that the total intensity against background just follows from the above luminosity.

The total flux (against the background radiation) from an object at redshift $z$ is then
\bea
dF = \frac{3 h \nu_{21} A_{10}}{16\pi(1+z)^2r^2(z)}n_{\rm HI}\,\phi(\nu)\,d\nu\, dA\, dr, \nonumber
\eea
where $\nu=\nu_{21}/(1+z)$. Defining the brightness, $I$, of the clump through $dF\equiv I d\Omega d\nu_{\rm obs}$ and multiplying by the Rayleigh-Jeans approximation factor, we obtain
\bea
T_b = \frac{3 h c^3 A_{10}}{32\pi k_B \nu^2_{21}}\frac{(1+z)^2}{H(z)} n_{\rm HI}, \nonumber
\eea
where we assume that the line width $d\nu/(1+z)$ is much smaller than the observed frequency interval $d\nu_{\rm obs}$ and the corresponding $dA=r^2 d\Omega$, $dr=\lambda_{21}(1+z)^2/H(z)\, d\nu_{\rm obs}$. The comoving number density is
\bea
n_{\rm HI}=\Omega_{\rm HI}\frac{\rho_{c,0}}{m_p}(1+\delta_{\rm HI}), \nonumber
\eea
where $\Omega_{\rm HI}$ is the comoving HI fraction, $m_p$ is the proton mass, $\delta_{\rm HI}$ is the HI density contrast, and $\rho_{c,0} = 3 H_0^2 / 8 \pi G$ is the critical density today.

\section{Redshift evolution of the HI signal} \label{app-HI-signal}

In this appendix, we derive the redshift evolution of the HI density, brightness temperature, and bias. We begin by assuming that the HI luminosity from a given spatial volume element (with solid angle $\Delta\Omega$ and frequency interval $\Delta\nu$) is proportional to the HI mass within the volume, $M_{\rm HI}$. If all the HI contributes to the signal, and the spin temperature is well above the background temperature, the observed brightness temperature of the volume element is
\be
T_{b}(\nu)=\frac{3.23\times 10^{-4}}{\Delta\Omega \Delta\nu}\frac{M_{\rm HI}}{(1+z)^2 D_A^2} \,\,{\rm mK\,Mpc^2\,Hz\,M_\odot^{-1}},\nonumber
\ee
and its proper volume is
\be
V_{\rm pix} = \Delta\Omega \Delta\nu \frac{ (c / \nu) D_A^2}{H + dv/ds}, \nonumber
\ee
where we are taking into account the effect of the peculiar velocity through $dv/ds$, the proper gradient of the peculiar velocity along the line of sight.

After reionisation, neutral hydrogen can only be found inside galaxies that are able to shield it from ionising UV radiation. The gas temperature and corresponding spin temperature should be much hotter than the CMB, so the approximation above can be used (there are a few cases where a strong background radiation source is capable of generating an absorption signal, but those are negligible given the low resolution we are considering here). We have also neglected HI self-absorption, or any other type of shielding of the HI emission. (Note that even if some re-absorption of the HI signal did happen, we would still expect a linear relation between the HI luminosity and mass, albeit with a smaller constant of proportionality.)

The next step is to connect the HI mass to the underlying halo mass, in order to relate the signal to the cosmological matter density field. We assume that a dark matter halo of mass $M$ contains one or more galaxies with a total mass $M_{\rm HI}$ that is only a function of the halo mass and redshift, i.e. $M_{\rm HI}(M, z)$. There may be some environmental dependence, which would make this a function of position as well. Some level of stochasticity can also exist in the relation between halo and HI mass, but given the low resolution pixels used in HI intensity mapping experiments, we expect a large number of HI galaxies per pixel, which should average-down any fluctuations and allow us to take the above deterministic relation for the mass function.

To detect the BAO scales at $z=1$, for example, one needs angular/frequency resolutions of around 1 degree and 5 MHz respectively, which translate into a comoving volume of $1.22\times 10^5$ $\mathrm{Mpc}^3$. In each volume element, we expect a total of around $10^6$ dark matter halos with mass between $10^{8} - 10^{15} M_\odot$, and $\sim\!31,000$ with masses between $5 \times 10^{9}$ and $1 \times 10^{12} M_\odot$ (where the latter range corresponds to halos expected to contain most of the HI mass). This supports our assumption of a position-independent HI mass function due to the averaging over many halos. Some level of stochasticity could still increase the shot noise of the signal, but this is expected to be quite small, as discussed below.

Given $M_{\rm HI}(M, z)$, we can then relate the signal to the underlying dark matter field. The number of halos of mass $M$ in the observed volume element is given by $\left[1+b(M,z)\delta_M(z)\right] \frac{dn}{dM} dM\, V_{\rm pix}$, where $\delta_M$ is the underlying dark matter fluctuation at that point in space (and time), $b(M,z)$ is the halo bias, and $dn/dM$ is the proper halo mass function. Integrating over all possible masses, the total observed temperature is then
\be
T_{b}(\nu) = \frac{\alpha}{(1+z)} \frac{\rho_{\rm HI}(z) \left[1+b_{\rm HI}\delta_M(z)\right]}{(H + dv/ds)(1 - v/c)}, \nonumber
\ee
where $\alpha = 2.21\times 10^{-27}\, {\rm mK\,Mpc^3\,M_\odot^{-1}\,s^{-1}}$, and the proper HI density and HI bias are
\bea
\rho_{\rm HI}(z) &=& \int_{M_{\rm min}}^{M_{\rm max}} dM \frac{dn}{dM}M_{\rm HI}(M) \nonumber\\
b_{\rm HI}(z) &=& \rho^{-1}_{\rm HI} \int_{M_{\rm min}}^{M_{\rm max}} dM \frac{dn}{dM} M_{\rm HI}\, b(z,M). \nonumber
\eea
Rewriting in terms of the fractional density\footnote{The $(1+z)^{-3}$ term shows up here because $dn/dm$ is the halo mass function in proper volume units.},
\be
\Omega_{\rm HI}(z)\equiv (1+z)^{-3} \rho_{\rm HI}(z) / \rho_{c,0}, \nonumber
\ee
and assuming that the peculiar velocity gradient and $v/c$ terms are small for these large pixels, we finally get
\bea
T_{b}(\nu,\Delta\Omega,\Delta\nu) &\approx& \overline{T}_{b}(z) \Big[1+b_{\rm HI}\delta_m(z)-\frac{1}{H(z)}\frac{dv}{ds}\Big] \nonumber\\
\overline{T}_{b}(z) &\approx& 566h\left(\frac{H_0}{H(z)}\right)\left(\frac{\Omega_{\rm HI}(z)}{0.003}\right)(1+z)^2\ {\rm \mu K}. \nonumber
\eea
Note that once $M_{\rm HI}(M, z)$ has been specified, we can calculate $\Omega_{\rm HI}$, the HI bias, and HI brightness temperature in a consistent manner. For the mass function, the most straightforward ansatz is to assume that it is proportional to the halo mass -- the constant of proportionality can then be fitted to the available data. Even in this case, however, we need to take into account the fact that not all halos contain galaxies with HI mass. {\corr Following \cite{Bagla:2009jy}, one can assume that only halos with circular velocities between $30 - 200$ km/s are able to host HI, which translates into a halo mass through
\be
v_{\rm circ} = 30 \sqrt{1+z} \left(\frac{M}{10^{10}M_{\odot}} \right)^{1/3} ~{\rm km/s}.
\ee
While reasonable, this model is unable to fit constraints on the HI density at high redshift. Possible refinements include allowing the minimum and maximum circular velocities to evolve with redshift, which could make a difference particularly at very low redshifts, or connecting the star formation rate to the halo mass, and then relating that to the HI mass. Alternatively, in \citet{2011ApJ...740L..20G} the relation between HI and halo mass was found using a non-linear function fitted to simulations from \citet{2009ApJ...703.1890O} (see their Table 1).}

{\corr In this paper, we have taken a different approach, using a redshift-independent power-law form for the mass relation,
\be
M_{\rm HI}(M) \propto M^\alpha.
\ee
An exponent of $\alpha \simeq 0.6$ provides a good fit to current low- and high-redshift constraints when we normalise the relation to the $z=0.8$ constraints from \cite{Switzer:2013ewa}. The resulting $\Omega_{\rm HI}(z)$ is shown in Fig. \ref{fig-omegaHI-evol}. }

Lastly, the shot noise power spectrum due to Poisson fluctuations in halo number is given by
\be
P_{\rm HI}^{\rm \,shot}(z) = \left(\frac{\overline{T}_{b}(z)}{\rho_{\rm HI}(z)}\right)^2\int_{M_{\rm min}}^{M_{\rm max}} dM \frac{dn}{dM}M_{\rm HI}^2(M). \nonumber
\ee
For the scales we are interested in, this quantity is rather small, although it would be increased somewhat if we allowed some level of stochasticity between each halo and the corresponding HI mass, as described before.

\begin{table}[b]
\begin{center}
{\renewcommand{\arraystretch}{1.2} \begin{tabular}{|c|c|c|c|c|c|}
\hline
Experiment & $\nu_\mathrm{min}$ & $\nu_\mathrm{max}$ & $\delta \nu$ [kHz] & ~$N_\mathrm{d}$~ & $D_\mathrm{dish}$ [m] \\
\hline
ASKAP & 700 & 1800 & 20 & 36 & 12.0 \\
CHIME & 400 & 800 & 1000 & $5 \times 256$ & $80 \times 20$ \\
JVLA (D) & 1000 & 2000 & 2000 & 27 & 25.0 \\
KAT7 & 1200 & 1950 & 50 & 7 & 12.0 \\
MeerKAT & 1200 & 1950 & 50 & 64 & 13.5 \\
SKA1-MID Base & 580 & 1015 & 50 & 190 & 15.0 \\
SKA1-MID Full & 580 & 1015 & 50 & 254 & 14.62 \\
Tianlai & 550 & 950 & 1000 & $8 \times 256$ & $100 \times 15$ \\
\hline
\end{tabular} }
\end{center}
\caption{Details of the array configurations for which $n(u)$ was calculated. $\nu_\mathrm{min, max}$ are in MHz, and $\delta\nu$ is the channel bandwidth. The efficiency factor for all dish arrays was taken to be $\eta=0.7$.}
\label{tab:msconfigs}
\end{table}

\section{Interferometer baseline density} \label{app-baselines}

In this appendix, we describe how to calculate the baseline density, $n(u)$, for a given array configuration. First of all, one must map-out the $uv$ coverage of the array. For a baseline with position components $(L_{X}, L_{Y}, L_{Z})$, the ellipse traced in the $uv$ plane is given by
\be
u^{2}+\left(v-\frac{L_{Z}/\lambda\cos\delta_{0}}{\sin\delta_{0}}\right)^{2}=\frac{L_{X}^{2}+L_{Y}^{2}}{\lambda^{2}}, \nonumber
\ee
where $\delta_{0}$ is the declination of the phase tracking centre. For an array with $N_{d}$ dishes, the total number of unique baselines is $N = N_\mathrm{d}(N_\mathrm{d}-1)/2$. Each baseline contributes one elliptical locus, given by the above expression. A full ellipse is traced for each baseline over the course of 24 hours of observation.

The baseline density $n(u)$ is just a histogram of $uv$ coverage in rings centred at the origin, i.e. the number of baselines per ring of radius $|u| = \sqrt{u^2 + v^2}$ and width $\Delta u$. The bin size is defined by the field of view (FOV), which depends on the effective area of a single array element, $A_{e}$, and the wavelength, $\lambda$, of the observation,
\be
\Delta u \sim \frac{1}{\sqrt{\textrm{FOV}}} = \sqrt{A_{e} / \lambda^{2}}. \nonumber
\ee
We neglect points with $|u| \le 1/\sqrt{\rm FOV}$, as these baselines are not independent (see next appendix).

\begin{figure}[t]
\centering{
\includegraphics[width=0.7\columnwidth]{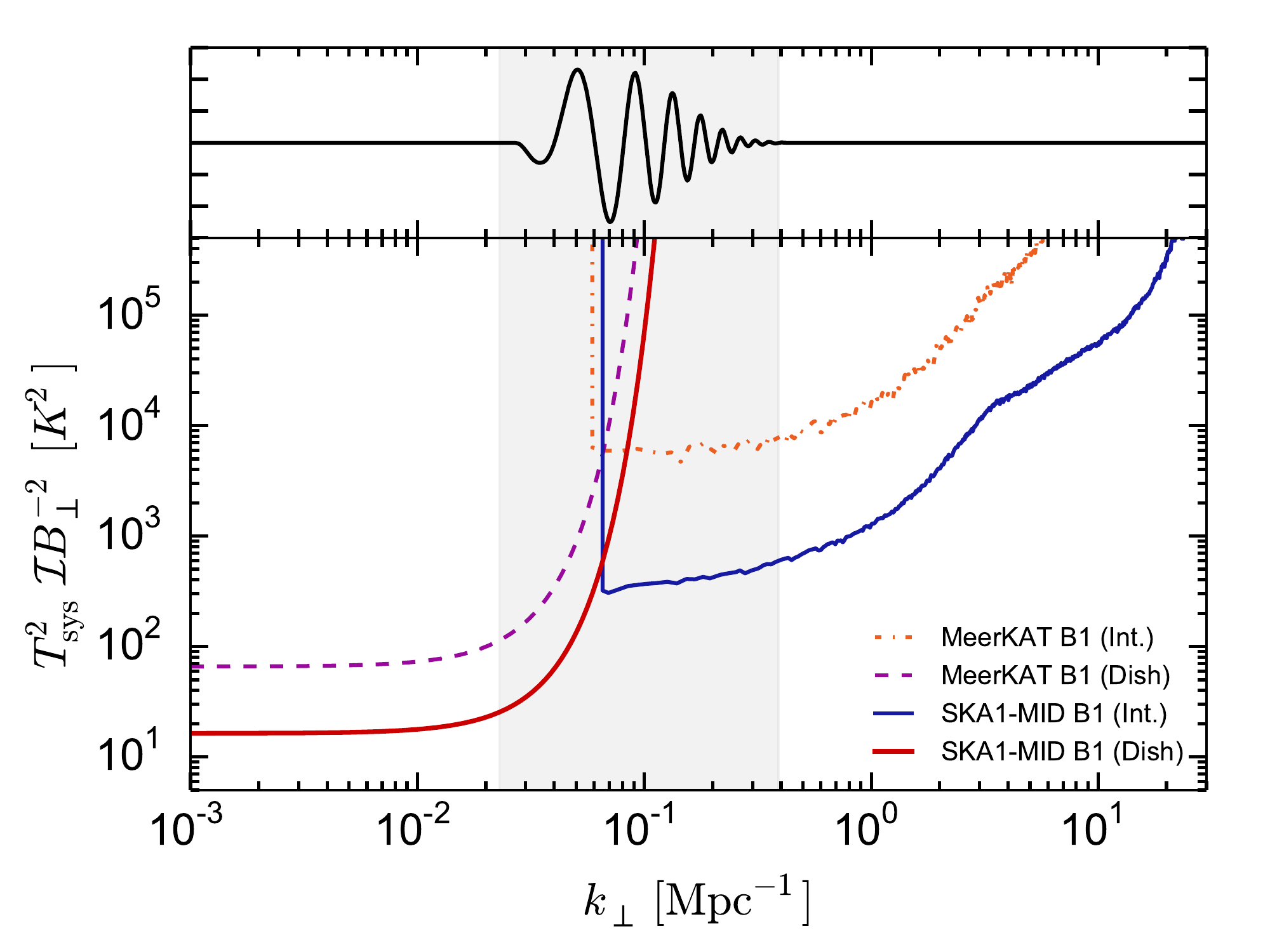} }
\caption{Noise sensitivity as a function of transverse wavenumber at $z=1$, for {\corr MeerKAT and SKA1-MID Band 1} (lower is better). The BAO scales (upper plot and grey band) are shown for reference.}
\label{fig-nu-beams}
\end{figure}

{\corr We have computed $n(u)$ for ASKAP, CHIME, JVLA, KAT7, MeerKAT, SKA1-MID, and Tianlai (only some of which have been used in interferometric mode in this paper). For the dish arrays, we generated $uv$ coverages for 24 hour observations, with 60 second integration time per visibility.} The $uv$ coverage was scaled depending on the observation frequency, and $n(u)$ computed as described above. As the different arrays operate in different bands, we simulated 10 frequency channels for each, uniformly spaced in each band. Details of the simulated configurations are given in Table \ref{tab:msconfigs}, {\corr and we have made the resulting $n(u)$ available online.\footnote{\url{https://gitlab.com/radio-fisher/bao21cm}} } {\corr The sensitivity for two of these experiments is shown as a function of transverse wavenumber (at $z=1$) in Fig. \ref{fig-nu-beams}. Note that, for the cylinder interferometers CHIME and Tianlai, the effective cylinder length is taken to be smaller than the geometric length, as the cylinders will likely be underilluminated to mitigate edge effects.}

\section{Derivation of noise expressions} \label{app:noise}

\subsection{Single-dish (autocorrelation)}

For a single dish with effective collecting area $A_e$, the noise associated with the measured flux is assumed Gaussian with an RMS (flux sensitivity) given by
\be
\sigma_S = \frac{2 k_B T_{\rm sys}}{A_e \sqrt{\delta\nu\, t_{\rm p}}},\nonumber
\ee
integrated over a frequency interval $\delta\nu$ and observation time per pointing $t_p$. Telescope sensitivities are often quoted in terms of the System Equivalent Flux Density\footnote{For a system with many dishes, both SEFD and $A_e/T_{\rm sys}$ are defined in terms of the total collecting area.}, $\mathrm{SEFD} \equiv 2 k_{\rm B} T_{\rm sys}/A_e$, or alternatively just $A_e/T_{\rm sys}$. The effective area $A_e$ is usually $\sim 70 - 80\%$ of the actual dish area, depending of the efficiency, $\eta$, of the system.
The total system temperature is $T_{\rm sys} = T_{\rm sky} + T_{\rm inst}$, where $T_{\rm sky}\approx 60 \,({300\, {\rm MHz} / \nu})^{2.55}$ K is the sky temperature, and $T_{\rm inst}$ is the instrument temperature (which is {\corr typically higher} than the sky temperature above 300 MHz). For typical instrumental specifications, the single-dish noise RMS can be written as
\be
\sigma_S = 18\, {\rm mJy} \left(\frac{T_{\rm sys}}{25\, \rm K}\right)\left(\frac{200\, \rm m^2}{A_e}\right)\left(\frac{0.1 \rm MHz}{\delta\nu}\right)^{1/2}\left(\frac{1 \rm h}{t_{\rm p}}\right)^{1/2}. \nonumber
\ee
Because we are interested in brightness temperature sensitivity (i.e. where the signal fills the primary beam), we need to look instead at the intensity of the signal. This is found by dividing the flux by the primary beam solid angle at FWHM, $\theta_{\rm B}^2$. In the Rayleigh-Jeans limit, {\corr the conversion from intensity to temperature then gives $\sigma_T \approx \lambda^2 T_{\rm sys} / (\theta_{\rm B}^2 A_e \sqrt{\delta\nu\, t_p})$.} 
For a given survey area, $\sim \!S_{\rm area}/\theta^2_{\rm B}$ pointings are needed, so the single-dish RMS noise temperature is
\be
\nonumber
\sigma_T\approx \frac{T_{\rm sys}}{\sqrt{\delta\nu\, t_{\rm tot}}}\frac{\lambda^2}{\theta_{\rm B}^2 A_e}\sqrt{S_{\rm area} / \theta_{\rm B}^2},
\ee
where $t_{\rm tot}$ is the {\it total} observation time for the survey. For an array with $N_\mathrm{d}$ identical dishes, the signals from each dish can be added incoherently, reducing $\sigma_T$ by a factor of $1/\sqrt{N_\mathrm{d}}$. Telescopes can also use focal plane arrays on each dish to increase the instantaneous FOV, effectively increasing the number of beams, $N_{\rm b}$, using multiple feeds or Phased Array Feeds (PAFs). {\corr Receivers can also support more than one polarisation channel, $n_\mathrm{pol} \ge 1$, and the channels can be added incoherently. Taking all of this into account, we can write}
\be
\sigma_T\approx \frac{T_{\rm sys}}{\sqrt{n_\mathrm{pol} \,\delta\nu\, t_{\rm tot}}} \frac{\lambda^2}{\theta_{\rm B}^2 A_e} \sqrt{S_{\rm area} / \rm \theta_{\rm B}^2} \sqrt{\frac{1}{N_\mathrm{d} N_\mathrm{b}}}, \nonumber
\ee
with the constraint that $S_{\rm area} \ge {\rm N_\mathrm{b} \theta_{\rm B}^2}$, since nothing is gained by pointing all the feeds in the same direction.

We are interested in a statistical detection of the HI signal. The 3D noise power spectrum associated with an autocorrelation measurement is just $P_N=\sigma_T^2\,V_{\rm pix}$, where $V_{\rm pix} = (r \theta_{\rm B})^2 \times (r_\nu\delta\nu/\nu_{21})$ is the 3D volume of each volume element. We can then obtain the expression for the noise covariance from Eq. (\ref{eqn-noise}) (ignoring the beams),
\bea
\nonumber
C^{N}({\bf q},y) \equiv \frac{P_N}{r^2 r_\nu}=\frac{T^2_{\rm sys} U_{\rm bin}}{n_\mathrm{pol}\, t_{\rm tot} \Delta\nu} \left (\frac{\lambda^4}{A_e^2 \theta^4_{\rm B}}\right)\, {\cal I},
\eea
where $U_{\rm bin}=S_{\rm area} \,\Delta{\tilde \nu}$ and ${\cal I} = 1 / N_{\rm b}N_{\rm d}$. {\corr For a dish reflector, $A_e \equiv \eta \pi (D_{\rm dish} / 2)^2$ and $\theta_{\rm B} \approx \lambda / D_{\rm dish}$, and so the factor in brackets in the expression above is $\mathcal{O}(1) / \eta^2$. For a dish equipped with a PAF, the beams begin to overlap below a critical frequency, $\nu_{\rm crit}$, and so there is a resulting loss of sensitivity, such that}
\bea
C^{N}({\bf q},y) \to C^{N}({\bf q},y) \times
\begin{cases}
    1, & \nu > \nu_\mathrm{crit}\\
    \left( \nu_\mathrm{crit} / \nu \right )^2, & \nu \le \nu_\mathrm{crit} \,. \\
\end{cases}
\eea

\subsection{Interferometer (cross-correlation)}

For an interferometer, a pair of elements separated a baseline of length $d$ measures a visibility $V({\mathbf u},\nu)$, where ${\mathbf u}$ is the vector in $uv$ space corresponding to that baseline\footnote{The projection of that baseline on the plane perpendicular to the line of sight (the telescope pointing) is what actually matters.}, and $u = |\mathbf{u}| = d/\lambda$. The $uv$-space resolution is set by the interferometer FOV, which for an array of dishes is given by the beam solid angle of a single dish, $\delta u\delta v \sim 1/{\rm FOV} \sim A_e/\lambda^2$. Visibilities separated by more than this distance in $uv$-space can be taken as independent.

The detector noise for a single complex visibility measurement, $N(\mathbf{u},\nu)$, is assumed Gaussian with variance
\be
\nonumber
\sigma_T^2\equiv\langle N(\mathbf{u},\nu_1) N^*(\mathbf{u},\nu_2)\rangle = \left(\frac{\lambda^2 T_{\rm sys}}{A_e\sqrt{\delta\nu t_\mathbf{u}}}\right)^2 \delta_{1,2},
\ee
where $\delta \nu$ is the channel bandwidth and $t_\mathbf{u}$ is the observing time for a given baseline. While each visibility measurement is independent in terms of instrumental noise, for the same sky signal the measurements will be strongly correlated for distances smaller than $\sqrt{A_e/\lambda^2}$, as explained above. One way of dealing with this is to average all visibilities falling into each $uv$-space resolution element of area $\delta u \delta v$. The noise will then be reduced by the number of points in that element, while the sky visibility stays essentially the same (assuming sufficiently high $uv$ resolution).

Let $t_s$ be the integration time for one visibility, and $N_s(\mathbf{u})$ the corresponding total number of visibilities falling into a given element after one ``snapshot''. $N_s$ will then be directly related to the baseline distribution (once we have factored in the observation angle), as explained in the previous appendix. The noise in each $uv$ resolution element is then
\be
\nonumber
\sigma_T^2(\mathbf{u},\nu) = \left(\frac{\lambda^2 T_{\rm sys}}{A_e\sqrt{\delta\nu\, t_s N_s(\mathbf{u})}}\right)^2.
\ee
Note that $t_s$ is usually just a few seconds, as longer integration times generate a ``smearing'' of the visibility in the $uv$ plane due to Earth's rotation. Smaller integration times allow more visibility points to be sampled, but each with larger noise.

The total observation time in a given patch of the sky, $t_p$, is usually more than the snapshot time $t_s$, and the telescope tracks the patch. This implies that the same baseline might produce different vectors in the $uv$ plane as the observation angle changes. To allow for different choices of the pixel area and integration time per visibility, one usually refers to the average number density of baselines averaged over a 24h period,
\be
\nonumber
n(\mathbf{u})=N(\mathbf{u})/(\delta u \delta v)/(24h/t_s).
\ee
$N(\mathbf{u})$ now corresponds to the total number of baselines falling into a given resolution element $\delta u \delta v$ in a 24 hour period, so that sky rotation is taken into account. For a given (square) resolution element, $(\Delta u)^2$, we can write
\be
\nonumber
\sigma_T^2(\mathbf{u},\nu) = \left(\frac{\lambda^2 T_{\rm sys}}{A_e\sqrt{\delta\nu\, n(\mathbf{u})\, (\Delta u)^2 t_p}}\right)^2,
\ee
where $n(\mathbf{u})$ is usually {\corr only a function of $u \equiv |\mathbf{u}|$ due} to the symmetrising effect of the rotation on the $uv$ coverage.

If we assume that $n(u)$ is constant on the $uv$ plane between some $u_\mathrm{min} = D_{\rm min}/\lambda$ and $u_\mathrm{max} = D_{\rm max} / \lambda$ (which is not the same as assuming an uniform distribution of antennas), we can write (see Eq. \ref{eqn-nu-const})
\be
\nonumber
(\pi u_{\rm max}^2-\pi u_{\rm min}^2) n(u) = N_\mathrm{d}(N_\mathrm{d}-1)/2.
\ee
This follows from noting that the {\corr integration of $n(u)$ over the $uv$ plane} should give the total number of baselines.

The variance of the noise in 3D Fourier space is related to the variance of the visibilities through
\be
\nonumber
\left<N_S(\mathbf{k}) N^*_S(\mathbf{k}) \right> \approx
(\delta\nu/\nu_{21}) \Delta{\tilde \nu} \left [ r^2 r_\nu \sigma_T(\mathbf{u},\nu) \right ]^2,
\ee
which is obtained by noting that the 3D Fourier component corresponds to a Fourier transform of the visibility along the frequency direction. The noise power spectrum {\corr for a single-pointing observation} is then given by
\be
P_N = \frac{\left<N_S(\mathbf{k}) N^*_S(\mathbf{k}) \right>}{(\mathrm{FOV} ~r^2) (r_\nu \Delta {\tilde \nu})}, \nonumber
\ee
{\corr which, using $(\Delta u)^2 \sim 1 / \mathrm{FOV}$,} reduces to an expression similar to Eq. (\ref{eqn-noise}),
\be
C^N(\mathbf{q}, y) \equiv \frac{P_N}{r^2 r_\nu} = \frac{T^2_\mathrm{sys}}{\nu_{21} t_p} \frac{\lambda^4}{A^2_e\, n(\mathbf{u})}. \nonumber
\ee
In the above, $t_p$ is the observation time for a single pointing of the interferometer. Once a signal-to-noise ratio of unity is achieved on the scales of interest, one can gain by moving to another pointing {\corr (i.e. increasing the survey area)}. The time spent at each pointing is then $t_p = t_\mathrm{tot} (\mathrm{FOV} / S_\mathrm{area})$, and the number of observed modes is increased by a factor of $S_\mathrm{area}/{\rm FOV}$. {\corr Taking this into account, as well as the possibility of having multiple beams, $N_\mathrm{b} \ge 1$, and polarisation channels, $n_\mathrm{pol} \ge 1$, we arrive at the expression}
\be \label{cn-interferom-full}
C^N(\mathbf{q}, y) = \frac{T^2_\mathrm{sys}}{\nu_{21} n_\mathrm{pol}\, t_\mathrm{tot}} \frac{\lambda^4 S_\mathrm{area}}{A^2_e \cdot \mathrm{FOV}} \frac{1}{N_\mathrm{b}\, n(\mathbf{u})}.
\ee
{\corr For a standard dish reflector, $\mathrm{FOV} \approx \theta_\mathrm{B}^2$. For an interferometer equipped with PAFs, the primary beam scales as}
\be \nonumber
\theta_B(\nu) = \theta_B(\nu_\mathrm{crit}) \times
\begin{cases}
    \left ( \nu_\mathrm{crit} / \nu \right ), & \nu > \nu_\mathrm{crit} \\
    1, & \nu \le \nu_\mathrm{crit}. \\
\end{cases}
\ee
{\corr For an aperture array, the primary beam scales as usual with frequency ($\theta_B = \theta_{\rm B}(\nu_\mathrm{crit})\times(\nu_\mathrm{crit} / \nu)$), but now the effective area picks up a correction as the array becomes dense below the critical frequency,}
\be \nonumber
A_\mathrm{eff}(\nu) = A_\mathrm{eff}(\nu_\mathrm{crit}) \times
\begin{cases}
    \left ( \nu_\mathrm{crit} / \nu \right )^2, & \nu > \nu_\mathrm{crit} \\
    1, & \nu \le \nu_\mathrm{crit}. \\
\end{cases}
\ee
{\corr The noise expression for cylinder interferometers (like CHIME) is more complicated. First of all, the primary beam is anisotropic; in the direction along the cylinder, the beam is $\sim 90^\circ$, while it is limited by the cylinder width, $w_{\rm cyl}$, in the perpendicular direction, giving $\mathrm{FOV} \approx 90^\circ \times \lambda / w_{\rm cyl}$ \citep{2014SPIE.9145E..4VN}. Secondly, many closely-packed feeds illuminate each cylinder, complicating the relationship between number of receivers and collecting area assumed in Eq. \ref{cn-interferom-full}. To fit the cylinder noise expression into this form, we write $A_{\rm e} = \eta \, l_{\rm cyl} \, w_{\rm cyl} / N_{\rm feed}$, the effective area {\it per feed}, where $N_{\rm feed}$ is the number of feeds per cylinder, and $l_{\rm cyl}$ is the cylinder length. There is also a restriction on the survey area; the beam cannot be steered and the telescope drift scans, fixing $S_{\rm area} \sim 30,000$ deg$^2$ (we choose $25,000$ deg$^2$ as an effective area).}

Finally, note that most of the quantities above depend on frequency. In particular, the FOV (and thus the minimum angular resolution) changes with frequency, so usually the maximum possible size for the $uv$-space resolution element is taken. Moreover, the final equation above is an approximation, and the middle of the frequency interval is taken in some of the expressions; otherwise we would need to consider an integral over the frequency when calculating the noise power spectrum.

\section{Derivatives used in the Fisher matrix} \label{app-derivs}

Most of the derivatives used in the Fisher matrix can be calculated analytically. It is advantageous to use analytic derivatives because their numerical behaviour can be regulated more easily. They can also offer insight into the behaviour of certain constraints.

The kernel of the Fisher integral consists of products of terms of the form
\be
\partial_{p_i} \log C^T = (\partial_{p_i} C^S) / C^T, \nonumber
\ee
where we recall that $C^T = C^S + C^N + C^F$, and the equality follows from the fact that only the signal covariance, $C^S$, is a function of the parameters $\{p_i\}$.

\paragraph{Basic parameters} The derivatives for most of the terms in the signal model, Eq. (\ref{sigcov}), are relatively straightforward:
\bea
\partial_A C^S&=& \frac{f_\mathrm{bao}(k)}{1 + A f_\mathrm{bao}(k)}C^S \nonumber \\
\partial_b C^S&=& \frac{2}{b + f\mu^2} C^S \nonumber \\
\partial_f C^S&=& \frac{2\mu^2}{b + f\mu^2} C^S \nonumber \\
\partial_{\sigma_8} C^S &=& (2 / \sigma_8) \, C^S \nonumber \\
\partial_{n_s} C^S &=& \log (k/k_\mathrm{piv}) \, C^S \nonumber \\
\partial_{\sigma^2_\mathrm{NL}} C^S &=& -k^2\mu^2 C^S \nonumber .
\eea
We have used the splitting of $P(k)$ into a smooth power spectrum plus a BAO feature from Eq. \ref{eqn-Abao} to calculate the derivative for the BAO amplitude, $A$. The derivative for $\sigma_8$ can be found by renormalising the power spectrum, $P(k) \to (\sigma_8 / \sigma^\mathrm{fid}_8)^2 P(k)$, {\corr and the derivative for $n_s$ by rewriting $P(k)$ in terms of the primordial power spectrum, $P(k) \propto T^2(k) \, (k/k_\mathrm{piv})^{n_s - 1}$, where $k_\mathrm{piv} = 0.05 \,\mathrm{Mpc}^{-1}$.}

\paragraph{Distance measures} The derivatives for the distance scales, $\{ \alpha_\perp, \alpha_\parallel \}$, are more complicated, but remain mostly analytic. For each $\alpha$, the derivative is
\bea
\partial_\alpha C^S=\left[\frac{n_\alpha}{\alpha}+\frac{2f}{b+f\mu^2}\partial_\alpha \mu^2+\partial_k \log P(k) \, \partial_\alpha k\right]C^S \nonumber
\eea
where $n_\perp=2$, $n_\parallel=1$. Only $\partial_k \log P(k)$ must be evaluated numerically; the other terms are given by
\bea
\partial_{\alpha_\perp}\mu^2 &=& - 2\, \alpha^{-1}_\perp \chi^2 \big / \left( 1+\chi^2 \right)^2 \nonumber \\
\partial_{\alpha_\parallel}\mu^2 &=& 2\, \alpha^{-1}_\parallel\chi^2 \big / \left(1+\chi^2\right)^2 \nonumber \\
\partial_{\alpha_{\perp}}k &=& \left(\frac{\alpha_\perp q}{ r}\right)^2 \big / \, \left (k \alpha_\perp \right ) \nonumber  \\
\partial_{\alpha_{\parallel}}k &=& \left(\frac{\alpha_\parallel y}{ r_\nu}\right)^2 \big / \, \left (k \alpha_\parallel \right ), \nonumber
\eea
where we have used the definitions
\bea
\chi &=& \frac{\alpha_\perp r_\nu q}{\alpha_\parallel r y} \nonumber \\
k^2 &=& {\left(\frac{\alpha_\perp q}{r}\right)^2+\left(\frac{\alpha_\parallel y}{ r_\nu}\right)^2} \nonumber \\
\mu^2 &=& \frac{y^2}{y^2+\left(\frac{\alpha_\perp r_\nu}{\alpha_\parallel r}\right)^2q^2}. \nonumber
\eea

\paragraph{Parameters from distance measures} Where we are interested in constraining cosmological parameters rather than functions of redshift, we project from $\{ D_A(z), H(z), f(z) \}$ into the parameters $\{ h, \Omega_K, \Omega_\mathrm{DE}, w_0, w_a, \gamma \}$ using Eq. (\ref{eqn-project-params}). To do this, we first assume the following forms for the functions of redshift, based on simple extensions to $\Lambda$CDM:
\bea
H(a) &=& H_0\sqrt{\Omega_M a^{-3} + \Omega_{DE}(a) + \Omega_K a^{-2}} \nonumber \\
{\corr r(a)} &=& {\corr \frac{c}{H_0} S \left ( \int \frac{da^\prime}{{a^\prime}^2 E(a^\prime)} \right )} \nonumber \\
f(a) &=& \Omega^\gamma_M(a), \nonumber
\eea
where for $\Omega_K = \{ \mathrm{-ve},~ 0,~ \mathrm{+ve}\}$ we have defined $S(x) = \left\{ \sin(x\sqrt{|\Omega_K|})/\sqrt{|\Omega_K|},~ x,~ \sinh(x\sqrt{\Omega_K})/\sqrt{\Omega_K} \right\}$, and 
\bea
w(a) &\approx& w_0 + (1 - a) w_a \nonumber \\
\Omega_M(a) &=& H_0^2 \Omega_M a^{-3} / H^2(a) \nonumber \\
\Omega_\mathrm{DE}(a) &=& \Omega_\mathrm{DE} \frac{\exp[3w_a(a-1)]}{a^{3(1+w_0+w_a)}}, \nonumber
\eea
$H(a)\equiv H_0 E(a)$, $H_0 = 100 h \,\mathrm{km s}^{-1}\mathrm{Mpc}^{-1}$, and $\Omega_M = 1 - \Omega_K - \Omega_\mathrm{DE}$.

Next, we need the derivatives of the original functions of redshift with respect to the new parameters. Most of them enter through the dimensionless Hubble rate, $E(a)$, for which the relevant derivatives are
\bea
\partial_{\Omega_k}E(a) &=& (a^{-2} - a^{-3}) \big / 2 E(a) \nonumber \\
\partial_{\Omega_\mathrm{DE}}E(a) &=& (1 - a^{-3}) \big / 2 E(a) \nonumber \\
\partial_{w_0}E(a) &=& -\frac{3}{2}\frac{\Omega_\mathrm{DE}}{E(a)}\log a \nonumber \\
\partial_{w_a}E(a) &=& -\frac{3}{2}\frac{\Omega_\mathrm{DE}}{E(a)}[\log a + (1-a)] \nonumber \\
\partial_\gamma E(a) &=& \partial_h E(a) = 0. \nonumber
\eea
For $\alpha_\parallel$, we then have (with $\beta$ being any of the parameters except $h$, and evaluating on $\Lambda$CDM)
\be
\partial_{\beta}\alpha_\parallel = \frac{\partial_\beta E}{E_\Lambda}. \nonumber
\ee
The expression for $\alpha_\perp$ is more complicated. For all but $\Omega_k$, the derivatives are given by
\bea
\partial_{\beta}\alpha_\perp &=& -\frac{\alpha_\perp}{r_\Lambda} \frac{\partial S(x)}{\partial x} \partial_\beta x \nonumber \\
\partial_\beta x &=& -\int \frac{c da^\prime}{{a^\prime}^2 E^2(a^\prime)} \partial_\beta E(a^\prime). \nonumber
\eea
Additional terms appear in the derivative w.r.t. $\Omega_K$,
\be
\partial_{\Omega_K}\alpha_\perp = -\frac{\alpha_\perp}{r_\Lambda} \left [ \frac{\partial S(x)}{\partial x} \partial_{\Omega_K} x + \frac{\Theta(\Omega_K)}{2 \Omega_K} \left ( x \frac{\partial S(x)}{\partial x} - S(x) \right ) \right ], \nonumber \\
\ee
where $\Theta=0$ for $\Omega_K=0$ and unity elsewhere. For both $\alpha_\parallel$ and $\alpha_\perp$, the $h$ derivative is $\partial_h \alpha = \alpha / h$.

For the growth function, the derivatives for all but $\{ \gamma, \Omega_K, \Omega_\mathrm{DE}\}$ are given by
\be
\partial_\beta f(a) = - \gamma f(a) \frac{\partial_\beta E}{E_\Lambda}. \nonumber
\ee
The derivatives with respect to the other parameters are
\bea
\partial_\gamma f &=& f(a) \log \Omega_M(a) \nonumber \\
\partial_{\Omega_K} f &=& - \gamma f \left [ \Omega_M^{-1} + \partial_{\Omega_{K}} \log E \right ] \nonumber \\
\partial_{\Omega_\mathrm{DE}} f &=& - \gamma f \left [ \Omega_M^{-1} + \partial_{\Omega_\mathrm{DE}} \log E \right ]. \nonumber
\eea


\bibliographystyle{hapj}
\bibliography{Bibliography_IM}

\end{document}